\documentclass{article}

\usepackage[round]{natbib}

\usepackage[margin=1in]{geometry}

\usepackage{epsfig,amsmath, amsfonts,amsthm,amssymb}
\usepackage{dsfont}
\usepackage{autonum}
\usepackage{esvect}
\usepackage{bm}

\usepackage[english]{babel}

\usepackage{stackrel}
\usepackage{verbatim}
\usepackage{setspace}
\usepackage{mathrsfs}
\usepackage{bm}
\usepackage{color}
\usepackage{epstopdf}
\usepackage{graphicx}
\usepackage{graphics}
\usepackage{multirow}
\usepackage[official]{eurosym}

\DeclareMathOperator*{\argmax}{arg\,max}

\usepackage[hyphens,spaces,obeyspaces]{url}

\urldef\berkeleyurl\url{https://www.econ.berkeley.edu/sites/default/files/course-homepage/2015-12-08/lecture-notes/Notes%20on%20Investment.pdf}
\urldef\morganstanleyurl\url{https://advisor.morganstanley.com/scott.altemose/documents/field/s/sc/scott-a--altemose/Case%20for%20Cryptocurrency_MAR%2017%202021.pdf}
	
\usepackage{colortbl}

\parskip = 0.0in
\setlength\parskip{\smallskipamount}%{\medskipamount}

\newtheorem{thm}{Theorem}[section]

\newtheorem{prop}[thm]{Proposition}
\theoremstyle{definition}
\newtheorem{defn}[thm]{Definition}

\newtheorem{rem}[thm]{Remark}
\numberwithin{equation}{section}
\setlength{\tabcolsep}{5pt}
\newcommand{\be}{\begin{equation}}
\newcommand{\ee}{\end{equation}}
\newcommand{\bq}{\begin{eqnarray}}
\newcommand{\eq}{\end{eqnarray}}

\newcommand{\half}{\frac{1}{2}}

\def\bbr{{\mathbb R}}
\def\bbe{{\mathbb E}}

\definecolor{Red}{rgb}{1.00, 0.00, 0.00}
\newcommand{\Red}{\color{Red}}
\definecolor{DRed}{rgb}{0.7, 0.3, 0.00}

\definecolor{Green}{rgb}{0.2, 0.5, 0.2}%{0.5, 0.00, 0.00}

\definecolor{Blue}{rgb}{0.00, 0.00, 1.00}%{0.00, 0.00, 1.00}

\definecolor{PaleGrey}{rgb}{.6, .6, .6}

\usepackage{listings}
\definecolor{mygreen}{RGB}{28,172,0} % color values Red, Green, Blue
\definecolor{mylilas}{RGB}{170,55,241}

\date{First version: June 16, 2021.}

\title{Proof-of-Work Cryptocurrencies: \\ Does Mining Technology Undermine Decentralization?}

\author{Agostino Capponi\thanks{Department of Industrial Engineering and Operations Research,
		Columbia University, New York, NY 10027, USA,
		\texttt{ac3827@columbia.edu}.} \quad
	Sveinn \'Olafsson\thanks{Department of Industrial Engineering and Operations Research,
		Columbia University, New York, NY 10027, USA,
		\texttt{so2570@columbia.edu}.} \quad
	Humoud Alsabah\thanks{Department of Industrial and Management Systems Engineering, Kuwait University, Kuwait City, Kuwait, \texttt{humoud.alsabah@ku.edu.kw}.}
	\footnote{This is a substantially revised version of a manuscript originally circulated under the title ``Pitfalls of Bitcoin's Proof-of-Work: R\&D Arms Race and Mining Centralization'', and co-authored with Humoud Alsabah. This new version models mining and manufacturing as distinct economic activities, and allow capacity constrained miners to invest in new mining hardware, developed by an exogenous manufacturing sector.}
}

\begin{document}		

\maketitle 

\doublespacing

\begin{abstract}

	Does the proof-of-work protocol serve its intended purpose of supporting decentralized {cryptocurrency} mining? To address this question, we develop a game-theoretical model where miners first invest in hardware to improve the efficiency of their operations, and then compete for mining rewards in a rent-seeking game. We argue that because of capacity constraints faced by miners, centralization in mining is lower than indicated by both public discourse and recent academic work. We show that advancements in hardware efficiency do not necessarily lead to larger miners increasing their advantage, %investment in new hardware contributes to decentralization 
	but rather allow smaller miners to expand and new miners to enter the competition. %\change{We calibrate our model to data from the Bitcoin network and study measures of centralization and network security. Our} 
	{Our calibrated model illustrates} that hardware efficiency has a small impact on the cost of attacking a network, while the mining reward has a significant impact. This highlights the vulnerability of smaller and emerging cryptocurrencies, {as well as of established cryptocurrencies transitioning to a fee-based mining reward scheme.}
	
\end{abstract}

\section{Introduction} 

%\noindent\note{Part 1: Relevant features of proof-of-work cryptocurrencies}

Since the introduction of Bitcoin in 2009, cryptocurrencies have gained widespread popularity and adoption, %gone from relative obscurity to mainstream adoption, and 
%even 
to the point of being acknowledged as a permanent asset class in diversified investment portfolios.
%\footnote{\note{Don't need to include a link, but see for example: \morganstanleyurl.}}
In June 2021, the market capitalization of cryptocurrencies was over \$1.5 trillion, after reaching \$1 trillion for the first time in January 2021. While Bitcoin is the best known and most widely used cryptocurrency, there are thousands of alternative cryptocurrencies in existence, collectively known as \emph{altcoins}. 
%The largest altcoins by market capitalization in June 2021 are 
Altcoins with a sizable market capitalization are Ethereum, Binance Coin, Tether, and Cardano Coin. %and Bitcoin Cash (BCH), which was created as a hard fork of Bitcoin.

A defining characteristic of cryptocurrencies is that no central authority supervises the integrity of transactions. Rather, a decentralized ledger technology, typically a blockchain, is used to maintain the transaction history, %of all transactions, 
and thus to keep track of coin ownership.\footnote{The ledger is public in the sense that anyone can download a copy of the blockchain, and it can be inspected to trace the path of coins from one transaction to another.} 
This blockchain is managed by a peer-to-peer network of computers, referred to as {nodes}, that validate transactions and host a copy of the blockchain. Validation of transactions is referred to as \emph{mining}, and nodes that participate in mining are referred to as miners. 

%A specific network protocol is used to achieve consensus about the correct record of transactions. 
A consensus protocol is used to achieve agreement about the state of the blockchain %validate transactions, 
and to produce new blocks. The most common protocol is \emph{proof-of-work}, 
%Miners adhere to a network protocol for  validating new block. 
%The most successful cryptocurrencies to date, including Bitcoin, use the proof-of-work protocol. 
%consensus about the correct record of transactions is most commonly achieved via the \emph{proof-of-work} protocol. 
where miners compete to solve a computationally costly problem, and the winner has the right to update the chain by appending a new block of transactions.  %(i.e., a collection of transaction) 
%A cryptographic hash function is used in the generation of blocks. %to maintain the difficulty of block generation.
Concretely, a miner generates a block by repeatedly modifying the input to a cryptographic hash function, until the output, referred to as a \emph{hash}, %starts with a given number of zeros, i.e., 
is small enough.
%\footnote{The Bitcoin protocol uses the SHA-256 hash algorithm, which maps an input consisting of the transactions in the proposed block, the hash of the most recent block, and a nonce (\emph{number only used once}), which is the component modified by the miner. The output of the function is a bit array, and a valid hash is one that starts with sufficiently many zero bits.} %Each output of the function is referred to as a hash, and the function evaluations as hashing. 
%\remove{A key characteristic of the hash function is that a small change to the input results in a completely different output, while it is simple to double-check the output generated by a specific input. In other words, finding a valid hash is difficult, and can best be done by a brute-force search of all possible inputs, but checking whether a given hash is valid is simple.}
A miner who successfully  generates a block is rewarded with %the \emph{block reward}, which consists of in the form of 
newly minted coins, known as the \emph{block reward}. Additionally, the miner gets to keep the fees paid by users to get their transactions processed. %The miner then broadcasts its block to the network. 
%\remove{Other nodes then verify the validity of the block, i.e., double check the ``proof-of-work'', and begin working on the next block.}
The mining process is also referred to as hashing, and the rate at which a miner can generate hashes is referred to as its hash rate. 

The original Bitcoin white paper of \cite{nakamoto2008bitcoin} envisions a world
%According to the vision described in the original Bitcoin white paper (\cite{nakamoto2008bitcoin}), mining were to be 
where mining is feasible to practically anyone using off-the-shelf computers. However, the increased adoption and price appreciation of Bitcoin %in addition to the emergence of alternative cryptocurrencies, CPU and GPU mining quickly became infeasible. The same can be said about so-called FPGA mining, and this 
led to the development of ASIC chips optimized for mining (\cite{Taylor}).\footnote{ASIC (\emph{application-specific integrated circuit}) is a silicon chip customized for a particular use. In the context of cryptocurrency mining, ASICs are optimized to evaluate hash functions efficiently.} %customized  ASIC chips, used exclusively for Bitcoin mining (\cite{Taylor}). %https://en.bitcoin.it/wiki/ASIC  	
Since the emergence of ASIC mining hardware in 2013, there has been a consistent increase in its efficiency  (see, e.g., Table 2 in \cite{Aste}).
%Bitcoin mining hardware has moved from CPU first to GPU (McNally et al., 2018) and later FPGA and ASIC but the principle behind the proof of work remains the same.
%After the initial ASIC chips had proven their value in Bitcoin mining, 
%\remove{A single firm, Bitmain, quickly emerged as a market leader in large-scale hardware manufacturing. Bitmain's flagship product is the AntMiner series, and its market share was estimated at around 65\% in December 2019} (\cite{CoinShare}). \remove{The fiercest competitor in the current landscape is another Chinese company, MicroBT, which has achieved a significant market share with its WhatsMiner series} (\cite{BitMex}).
%, estimated at 35\% in 2020, with Bitmain still controlling close to 50\% of the market (\cite{BitMex}).
These advancements in technology transformed Bitcoin mining into a highly specialized %both capital- 
and energy-intensive endeavor, making it largely unprofitable for individual miners using home computers. %\footnote{Mine altcoins and hope to hit the jackpot...} 
%\remove{Currently, the lion's share of the network's hash rate is in the hands of large mining organizations that build and maintain vast mining facilities and data centers all over the world.} 

%Large mining organisations build and maintain vast mining facilities and data centers all over the world. Individual and corporate miners alike point their hashing power towards the mining pools of their choice to increase the likelihood and frequency of finding new blocks and ``smooth earnings''.
According to cryptocurrency studies carried out by the Cambridge Center for Alternative Finance, %it is reported that 
a relatively small number of large mining organizations appear to cover the majority of the mining sector in terms of applied hash rate (\cite{HilemanRauchs,Rauchs}). 
%11 of the study participants are estimated to cover over 50\% of the total professional mining sector in terms of hash rate (\cite{HilemanRauchs}), and that while there are at least several dozen operators of mining facilities around the globe, a small number of large mining organizations appear to have a dominant position (\cite{Rauchs}).
%\note{Centralization risks if the PoW process is controlled mostly by big data centers instead of a decentralized network of hobbyists}
These studies also indicate that there is little vertical integration between hardware manufacturing and actual mining. 
%The mining value chain consists of hardware manufacturing, mining, and mining pools. While participants are involved in more than one mining activity, there is little vertical integration between miners and hardware manufacturers. A small number of large mining hardware manufacturers supply the
In fact, the official policy of prominent hardware manufacturers, such as the publicly traded {companies} Canaan and Ebang, is to not participate in {proprietary} self-mining.
{While the two largest manufacturers, Bitmain and MicroBT, are privately held, they seem to take a similar stance. Bitmain's IPO prospectus from September 2018 indicates that 95\% of its revenue comes from sales of mining hardware,}\footnote{The IPO application expired in March 2019, with ``plans to reapply at an appropriate time in the future''. The primary reason is believed to have been a prolonged Bitcoin bear market in 2018, which hurt mining hardware sales and the company's future outlook.} {and there is no evidence of MicroBT being involved in significant mining operations.}

Cryptocurrencies are only as secure as their networks. Most users of cryptocurrency take the actual network protocol and its implementation for granted, and are not necessarily aware of {the network's vulnerabilities.} 
%Most users of cryptocurrency take the actual network protocol and all of its implementation — mining, pools, validation, messaging, and more — for granted, and aren’t necessarily aware of all the ways these mechanisms might be attacked or fail.
%If a network is insecure, it can become unusable or users could lose their money. 
%Businesses and users who are deeply invested in cryptocurrency should understand how to evaluate and mitigate this risk.
For Bitcoin and other proof-of-work cryptocurrencies, %miners secure the network by using their computational resources to compete for block rewards. 
%In total, these resources are known as the network hashrate, which is the rate at which all the mining equipment operating in the network can execute the cryptocurrency’s hash function. 
%Bitcoin and other longest-chain protocols 
the security of the network relies on {\it  mining decentralization}, which is higher if the distribution of hash rate among miners is more even; if an attacker can gain control of a majority share of the network hash rate, they can perform double-spending attacks. 
%they can mine an alternative chain that will become the network's longest chain and they can double spend.
The rapid developments in mining technology bring up the question of whether the proof-of-work protocol will be able to serve its intended purpose of being the backbone of a decentralized, and thus secure, network. %This is not only relevant for Bitcoin, but for cryptocurrencies in general as they predominantly rely on the proof-of-work mechanism. 
Going forward, another concern is whether network security can be sustained as the mining reward gradually switches from being primarily based on new coins to relying on transaction fees, which so far have been a small component of the mining revenue. %\note{Is this last statement for all proof of work cryptocurrencies or only for Bitcoin?}%\note{\Red The majority of them. To save space I removed the following from above: Although the implementation details differ, most cryptocurrencies also have a hard cap on the total number of coins that will be generated, as there is not a central authority to adjust the money supply. For Bitcoin, the number of coins awarded per block is halved every four years, and over 95\% of the 21 million maximum will have been mined by 2030.}
%Historically, transaction fees have been a small component of mining revenue, 
Answering these questions is critical to assess the ability of the proof-of-work protocol to support decentralized cryptocurrency networks.

We consider a cryptocurrency ecosystem consisting of a finite number of miners. Each miner is characterized by its cost of hashing, which depends on the efficiency of its mining hardware and its price of electricity. %In reality, miners not only have a finite amount of mining hardware, but of equal importance is the fact that they do not have unlimited 
Typically, miners have limited access to low-cost electricity to run their operations.  
Because of the large amounts of energy required, mining organizations need to secure wholesale agreements with energy providers, which specify a fixed production cap.\footnote{In China, it is common for miners to migrate between different parts of the country depending on where they can access cheap electricity (\cite{CoinShare}).} %See, e.g., \url{https://news.bitcoin.com/chinese-government-crackdowns-and-cheap-hydropower-miners-migrate-from-north-to-south-china/}.} % to access cheap electricity, 
Miners {in our model} thus face a capacity constraint which prevents them from increasing their hash rates unboundedly. %without bounds.

We formalize the competition between miners via a two-stage game. In the first stage of the game, miners decide {whether and how much to invest} in state-of-the-art mining equipment to improve the efficiency of their operations.
%can take advantage of advancements in technology by investing in state-of-the-art mining hardware.
	%This improves theefficiency of their operations, i.e., lower their cost of hashing.}%Miners first invest in hardware to improve the efficiency of their mining operations. Subsequently, they compete in a mining game by exerting costly hashing power. 
%\footnote{In late 2020, the two largest manufacturer of mining hardware, Bitmain and MicroBT, were confirmed to be sold out of new mining machines until May 2021. See: \url{https://www.coindesk.com/secondary-markets-asic-bitcoin-mining-manufacturing-delays}.}
In the second stage, miners simultaneously decide on hash rates to exert in the mining competition in order to maximize their profits. %in order to maximize their profits. %The payoff to a firm consists of the second-stage mining profits net of the first-stage R\&D expenditure. Investing in R\&D results in more efficient mining equipment, and thus allows firms to mine bitcoins at a smaller cost. 
Our model of the mining competition is designed to capture two key properties of the proof-of-work protocol: (i) the total reward from mining is independent of the network hash rate, and (ii) the expected share of reward attained by a miner is proportional to its \emph{hash rate share}, i.e., the fraction of the network hash rate it contributes. % of its computational power (relative to the aggregate). 
The first property is a consequence of the adaptive difficulty level of the hashing problem, which ensures a fixed rate of mined blocks. The second property 
%is intrinsic to the hashing function used, where a higher hash rate implies a higher probability of successfully mining a block.
captures the nature of cryptographic hashing functions, which are chosen to guarantee that a miner's probability of generating a valid hash is linearly increasing in its hash rate.

%\noindent\note{Part 5: Mining competition. Active miners. Reward vs.\ hash rate + empirical} 

{We show that positive hash rates are applied only by a subset of miners}. A miner is \emph{active} if and only if its cost-per-hash is smaller than the equilibrium reward-per-hash, which is consistent with existing practices. This result also suggests that centralization is, to some extent, intrinsic to the mining competition, because inefficient miners are forced out of the market. Miners can be thought of entering the competition in order of increasing cost. In the absence of capacity constraints, a miner is unable to enter if its cost is only slightly greater than the average cost of more efficient miners, i.e., those who are already active. However, the entry condition becomes weaker if we account for the capacity constraint faced by miners. Limited capacity prevents the most efficient miners from taking full advantage of their lower costs. {This, in turn, allows smaller miners to expand and new miners to enter.}
%\note{(a) and (b) are two options to finish this paragraph}
%the second term, which arises because of the capacity constraint, significantly relaxes this condition. Intuitively, limited capacity prevents the most efficient miners from taking full advantage of having lower costs, which in turn allows smaller miners to expand, and miners who otherwise would be inactive to have positive hash rates.}
%\add{(b) However, this centralization effect is relaxed by the capacity constraint faced by miners. 
%In addition to analyzing the number of active miners, we study in detail the mining equilibrium for a fixed set of miners. Such analysis is of particular significance as establishing a new mining facility requires significant amount of time and planning. In particular, we show that 

{We show that there exists} a \emph{sublinear} relation between the mining reward and the network hash rate. {Active miners are not able to increase their hash rates at the same rate as the mining reward,} because they become increasingly capacity constrained. {We provide empirical support for this implication by identifying a statistically significant sublinear relationship between mining reward and the network hash rate for the Bitcoin network.} 
%We use data from the Bitcoin network to study the empirical relationship between mining reward and the network hash rate, which supports our model result. 
Furthermore, our result is supported by the study of \cite{Rauchs}, where it is reported that miners are commonly operating at full capacity and thus limited in how much they can increase their hash rates in response to an increase in mining reward.

%\noindent\note{Part 6: Effect of investment} \note{We are emphasizing (italicizing) adjustment costs in the first part of the next paragraph. Check if they enter into the discussion of that paragraph} 
%\note{This paragraph is very long}

%Bidding war, shortages, discounts:
%https://www.theblockcrypto.com/post/88151/bitcoin-mining-hardware-bidding-war
%https://www.theblockcrypto.com/post/90060/bitcoin-price-mining-hardware-renaissance
%yahoo article from referee

In the first stage of the game, miners {invest in new, more efficient mining hardware to replace a portion of their old, less efficient hardware}. When choosing their investment levels, miners account for the fact that more efficient hardware lowers the cost of mining and thus impacts the outcome of the subsequent mining competition. {Miners face} \emph{adjustment costs}, {which prevent them from quickly improving their efficiency by upgrading their stock of hardware through investment.}\footnote{In cryptocurrency mining,  adjustment costs can both be internal (e.g., cost of installing and optimizing the performance of new equipment, and drop in hash rate during the transitional period), %, as well as extending and enhancing cooling systems, technical expertise) 
	and external (e.g., scarcity of new hardware and long time elapsing between the ordering and delivery of mining hardware). For example, due to overwhelming demand, Bitmain has doubled its prices and pre-sold mining hardware months ahead of expected shipping dates (see: \url{https://www.coindesk.com/bitcoin-asic-mining-shortage-bitmain-sold-out}). We refer to \cite{Eisner} and \cite{Lucas} {for early studies on capital adjustment costs in the context of investment, and} to \cite{Hamermesh} {for a review of existing literature on models of adjustment costs.}}\label{f1}
 We show the existence of a unique equilibrium investment, and characterize explicitly its effect on the mining competition. %\remove{under the assumption of high adjustment costs, which translate into low levels of investment. We then verify numerically that our characterization results provide good approximations to the exact result when investment levels are not low.}
Whether a miner is able to capture the benefits of its investment depends on the miner's initial {cost of hashing}. {Specifically, the change in a miner's hash rate share and profit due to hardware investment are both increasing in the miner's initial cost of mining.} {This implies that smaller miners are able to overcome the negative externality imposed by the investment of other miners, and increase their share of the network's hash rate.}
%, as the cost advantage of more efficient miners as smaller miners is decreased. %increase their hash rate shares at the expense of more efficient miners. Bounded mining capacity is again a factor in this mechanism. The aggregate welfare also increases, as capacity-constrained miners improve their efficiency.\note{Should mention here or somewhere the effect of entry}
Larger miners are more efficient to begin with, relative to smaller miners, %already incur low costs relative to small miners, 
and therefore gain less from investment. %This outweighs the fact that larger miners are also more capacity constrained, and in that sense gain more from investment. 
%\remove{We also show that if heterogeneity among miners is not too great, then welfare in the mining ecosystem increases with investment. The primary reason is that miners are capacity constrained, which allows them to take advantage of higher mining efficiency - in the special case of unbounded capacity, the effect of lower mining costs is negated by a higher network hash rate, and the profit of each miner is the same before and after investment.}
{Hence, the effect of investment is to push the mining network towards decentralization, because} a larger number of miners is required to control any given fraction of the network hash rate. Although Bitcoin mining has undeniably turned into an activity dominated by relatively large mining operations,
%rather than being in the hands of a large number of individuals, 
%the blockchain's validation process became more centralized as more and more hashing power was consolidated into a handful of mining companies 
%instead of a large network of hobbyists, 
our model implies that advancements in hardware technology do not necessarily lead to a consistently higher level of centralization. This is because less efficient miners are able to catch up by upgrading their hardware, and new miners can enter the mining competition through investment. These findings are consistent with the fact that Bitcoin mining appears to be less concentrated, both geographically and in terms of hash rate ownership, than commonly indicated by public discourse (see Section 7 in \cite{Rauchs}).\footnote{The study in \cite{Rauchs} was undertaken near the end of 2018, when efficient ASIC mining hardware had already been dominating the market for a few years. While there have been significant improvements in ASIC hardware since then, to the best of our knowledge there is no evidence of the mining industry having become more centralized.} %a new technology has not emerged to transform the mining industry as the emergence of ASIC did

%\noindent\note{Part 7: Mining as rent-seeking}

%Also, is the standard in the sense that resources are wasted? Beneficial for the network for miners to be homogeneous, both because of centralization, but also because of a higher hash rate? Pay $R$ for security and get more bang for the buck? We don't have a result for $H$ as a function of $c^{(n)}$. Well, for a fixed $c^{(n)}$ with varying components, $H$ is fixed. So, more homogeneity is better for a fixed $c^{(n)}$. 

%The mining competition resembles a rent-seeking contest where miners trade valuable resources for the chance of earning rent in the form of block rewards.\footnote{See \cite{Nitzan} for an overview of rent-seeking games.}

{It is well known that proof-of-work mining can be modeled as a rent-seeking contest where miners trade valuable resources for the chance of earning rent in the form of block rewards; see for instance} \cite{Budish}.\footnote{See \cite{Nitzan} for an overview of rent-seeking games.}
%That Bitcoin mining can be modeled as a rent-seeking contest is now widely known; see for instance Kroll, Davey and Felten (2013) pg. 8; Huberman, Leshno and Moallemi (2017) Proposition 1; Easley, O’Hara and Basu (2017) equation (1); Chiu and Koeppl (2017) Lemma 1; and Ma, Gans and Tourky (2018) equation (7). 
%https://faculty.chicagobooth.edu/eric.budish/research/Economic-Limits-Bitcoin-Blockchain.pdf
However, in the works referenced therein, rent is completely dissipated, as in the classical rent-seeking model of \cite{Tullock}. In our framework, miners {use cost advantages to extract profits in a Cournot-Nash type equilibrium} - %from the contest - 
in a model with an infinite number of identical miners, profits would be driven to zero.
%Other papers include \cite{dimitri2017bitcoin}, \cite{arnosti2019bitcoin}, and \cite{biais2017blockchain}. However, in those papers the effect of changes in mining efficiency through investment is not considered, and all equilibrium quantities are independent of the mining reward. 
%Budish and other papers talk about rent being dissipated. Here that does not happen because of homogeneity and finite miners. So, unlike that literature... fixed cost paper of UCSB... Budish does not consider the effect of miner homogeneity on e.g. security. In the limit of a dynamic model, we would get a Budish case... 
A key insight {from} the rent-seeking contest is that the mining difficulty increases when a miner increases its hash rate, which imposes a negative externality on other miners. To the best of our knowledge, we are the first to explicitly characterize %\remove{the nature of these externalities, and how heterogeneity in costs translates into heterogeneity in hash rate shares.}
{how these externalities impact the equilibrium hash rates.}
%\note{does this heterogeneity translation has to do with externalities?} 
%Unlike in a standard Cournot competition, the mining reward is independent of total production, i.e., the network hash rate, so a monopolistic miner would exert zero hash rate. In the general case, miners are forced to exert costly hash rates to compete for the reward. Miners therefore impose externalities on each other by contributing to the network hash rate, which in turn impacts the marginal gain of hashing. 
Specifically, we show that the strategic behavior of a miner depends on its size, i.e., its share of the network hash rate. A miner generally takes advantage of cost increases of other miners to increase its own hash rate, and thus manages to capture a larger share of the mining reward. However, a large enough miner \emph{reduces} its hash rate as other miners become less efficient, while still managing to capture a larger share of the mining reward. This means that miners generally behave aggressively and fill in the void left by other miners, but a large enough miner manages its hash rate passively to fend off competition.

%\noindent\note{Part 8: Model calibration and centralization/security}%\note{In this part we discuss the model calibration. { Don't we also have an empirical analysis? Why aren't we describing it (we mentioned it briefly earlier on) in more detail?}}

We calibrate our model using data from the Bitcoin network, and further study measures of mining centralization and network security. {We argue that} if the market capitalization of a coin increases, mining is not deemed to become increasingly centralized, i.e., exhibit ``the rich getting richer'' phenomenon. The rationale offered by our model is that existing mining facilities are not able to increase their hash rate indefinitely in response to a higher mining reward, which allows smaller miners to grow and new miners to enter. 
{This result can also be understood by considering the nature of the mining competition. Specifically, the proof-of-work protocol implies that there are limited economies of scale when it comes to opening new mining facilities - doubling the hash rate simply doubles of the expected reward. This means that as the market capitalization of a coin grows, existing miners do not have an inherent advantage over new miners when it comes to opening new mining facilities.}

%For a fixed mining reward, we also observe the same thing with respect to the effect of advancements in mining technology. Investment leads to decentralization in mining, as a larger number of miners is required to control any given fraction of the network hash rate. In particular, any kind of collusion among miners is more difficult, as a larger fraction of miners is required to participate. 
%\note{Extrapolating from our model, an extremely high adjustment cost of smaller miners could cause a disrupt, but there is not really evidence of that; limited economy of scale - opening a new mining facility costs the same as opening a first mining facility}
%A primary measure of network security is the degree of decentralization. %, which indicates the number of miners required to control any given fraction of the network hash rate. 
As decentralization increases, the network becomes more secure. Potential collusion between miners aimed at capturing a given hash rate share would require the participation of a larger number of miners. The network security also increases with the \emph{cost of attacks}, i.e., the {cost} of capturing a large hash rate share. 
%An additional measure of network security is the \emph{cost of attack}, i.e., the {cost} of capturing a large hash rate share. 
We show that {the degree of investment in mining hardware} has a limited impact on the cost of attacks, because of two opposing effects. On the one hand, investment leads to a larger network hash rate, which makes a given fraction of the hash rate larger in absolute terms, and thus more expensive to generate. On the other hand, investment makes the cost of hashing lower, which largely neutralizes the first effect.  

Unlike hardware investment, {which we have argued to only have a mild effect on the cost of attacks,} we show that mining reward has a significant impact on such costs. This is consequential for smaller and emerging cryptocurrencies, which have smaller market capitalizations and thus attract lower hash rates. {A small proportion of miners from a large coin's network is sufficient to successfully attack a smaller coin's network and thus hinder its growth.}
%{It is sufficient that a small proportion of miners from the network of a coin with a large market capitalization switch to a coin with a small market capitalization to successfully attack the small coin's network and hinder its growth.}
Sufficient hash rate to attack smaller coins can also be rented from cloud-mining services, allowing non-mining entities to implement attacks.\footnote{{Miners can easily move their hash rate between any cryptocurrency using the same hashing function. The theoretical cost of renting hash rate to implement a 51\% attack on various smaller cryptocurrencies can be seen at:} \url{https://www.crypto51.app/}.}
This result also stresses that a significant threat to security is built into the network protocol itself, i.e., its gradual transition to a fee-based mining reward system. 
The network can only remain secure during this transition if a robust transaction fee market develops {to compensate miners for their efforts}, and thus keep them committed to the network. Such a market has yet to materialize for Bitcoin.

{The rest of the paper is organized as follows.} Section \ref{sec:literature_review} reviews existing related literature. In Section \ref{sec:model}, we present our model and define the equilibrium of the game. In Section \ref{secMiningI}, we study the competition between miners {given a prespecified {hardware} investment profile}. In Section \ref{secMiningII}, we determine the {optimal investment profile} and its impact on the mining competition. In Section \ref{secParams}, we calibrate the model and {study} measures of mining centralization and network security.
In Section \ref{secEmpirical}, we study stylized features of cryptocurrency mining, and test the empirical implications of our model. Section \ref{secConclusions} concludes. In Appendix \ref{secNumInvest}, we verify the accuracy of approximations introduced to analyze the equilibrium of the game. All proofs are deferred to Appendix \ref{appProofs} and Appendix \ref{appOtherProofs}.

\section{Literature Review}\label{sec:literature_review}

{The Bitcoin white paper by} \cite{nakamoto2008bitcoin} was the first to outline the conceptual and technical details of decentralized digital currencies, {also referred to as} cryptocurrencies. 
%\remove{In January 2009, Nakamoto released the original Bitcoin source code and launched the network by mining the \emph{genesis block}, i.e., the first block of the chain.} 
Since then, %Bitcoin and other cryptocurrency networks 
{the study of cryptocurrencies has attracted significant} attention from academics, practitioners, and policy makers. We refer the reader to \cite{bohme2015bitcoin}, \cite{halaburda2016beyond}, and \cite{narayanan2016bitcoin}, %for an overview of Bitcoin and its operations. 
for comprehensive introductions to %coverview of the operations of 
cryptocurrencies, the underlying technology, and the economic incentives driving their creators and users. 

%design principles and properties of cryptocurrency platforms 
%conceptual and practical foundations
%Beyond Bitcoin explores the economic forces underlying the design of their features and their potential. Halaburda and Sarvary argue that digital currencies are best understood by considering the economic incentives driving their creators and users

Cryptocurrencies are a relatively recent invention, and their design principles are not set in stone. A number of studies have considered the optimal design of cryptocurrencies as well as variations and alternatives to the widely used proof-of-work protocol (see, e.g., \cite{chiu2017economics} %, \cite{hinzen2019proof}, 
and \cite{saleh2017blockchain}). 
Cryptocurrency as an asset class has also received attention. \cite{Fang} provide a comprehensive survey of 126 papers studying cryptocurrency trading and risk management. %\remove{, including ones studying the effect of news and social media.} %\add{Some studies have investigated prediction and formation of cryptocurrency prices.} %trading and price prediction. 
Methodological contributions in this direction include \cite{biais2018equilibrium} and \cite{pagnotta2021bitcoin},
%Contributions towards understanding the formation of cryptocurrency prices include equilibrium models 
%\add{Noticable contributions in this direction 
who study equilibrium models for pricing of Bitcoin and other decentralized network assets. %, such as those proposed in the studies of . 
%\cite{pagnotta2018bitcoin}, and \cite{pagnotta2018equilibrium}). 

%A separate stream of literature
%there has been a significant interest %in the analysis of Bitcoin operations, by
%analyzing the incentives faced by key participants in the system. 
%incentives faced by key participants in the system;  
%especially the incentives faced by miners %To study the dynamics of the blockchain, we analyze the behavior of its key participants: the miners
%operation of cryptocurrencies

%Our work takes the network mechanism as given and focuses on the operational side of the proof-of-work protocol, analyzing the economic incentives and strategic behavior of miners

%studying the operational side of the proof-of-work protocol, analyzing the economic incentives and strategic behavior of miners. 

In recent years, there has been significant interest in studying the operational side of proof-of-work %cryptocurrency 
blockchains by analyzing the incentives %the interaction 
of its key participants: miners and users seeking transaction processing. 
\cite{biais2017blockchain} and \cite{abadi2018blockchain} study the emergence of \emph{forks}, i.e., competing branches of the blockchain.  \cite{huberman2017monopoly} and \cite{Easley} study the determinants of transaction fees paid by users to %gain priority and 
avoid transaction-processing delays. 
\cite{Garratt} propose a model where %\change{miners incur a fixed cost for setting up mining operations}
the initial equipment investment of miners is a sunk cost, and study how this impacts their response to fluctuations {in mining reward}. Their study is partly motivated by the work of \cite{prat}, who calibrate a structural model and find that fixed costs account for a significant portion of the total mining cost.

Essential to the results in the papers surveyed above is free entry of miners - for homogeneous miners this is equivalent to the expected profit of each miner being zero. % separate set of studies has 
A separate stream of literature analyzes the economic incentives of %behind decisions of 
profit-maximizing miners in a Cournot competition framework. \cite{dimitri2017bitcoin} models Bitcoin mining as a static game where each miner decides on the hash rate to exert. He characterizes the equilibrium hash rate profile, and shows that the special structure of the Bitcoin system prevents the formation of a monopoly. In a similar framework, \cite{biais2017blockchain} show that if miners are homogeneous, {the network equilibrium hash rate is higher than the socially optimal level, i.e., the one which maximizes the aggregate mining profit}.  %The study of 
Building on the framework of \cite{dimitri2017bitcoin}, \cite{arnosti2019bitcoin} %, contemporaneous to our work, 
assume that each miner is also a producer of mining hardware. In a two-stage game, they show that production of mining hardware is dominated by a single miner who supplies all other miners; the subsequent mining equilibrium is the same as in \cite{dimitri2017bitcoin}. The model of \cite{arnosti2019bitcoin} closely resembles the setup in \cite{Ferreira}, who study the formation of ``conglomerates'' in the blockchain ecosystem, capturing the governance of the blockchain. Their model implies that a single firm dominates the market for mining equipment and employs the same pricing strategy as in \cite{arnosti2019bitcoin}.

%The study by \cite{arnosti2019bitcoin}, contemporaneous to our work, adopts the framework of~\cite{dimitri2017bitcoin} and show that small differences in firms' hash costs can {increase concentration in the} mining industry. Different from their model, firms' hash costs are endogenous in our framework. Using a calibrated model, we demonstrate that R\&{D} contributes to driving the Bitcoin mining industry towards centralization, and that the extent of this phenomenon strongly depends on Bitcoin rewards. 

Our study extends the above stream of literature on oligopolistic mining in a number of important ways. First, mining and hardware production are separate economic activities in our model -  %make a clear distinction between miners and producers of hardware - we 
we endogenize the decision of miners to invest in new hardware, which is developed by an exogenously {specified} manufacturing sector.
%and endogenize the investment of miners in new hardware to improve the efficiency of their operations. %Unlike the aforementioned studies, which only focus on the equilibrium hash rate profile, we additionally endogenize the investment profile. 
Second, miners incur a fixed cost for setting up their operations, {as in} \cite{Garratt}, and account for this cost when they deciding whether to invest in hardware to enter the mining competition. Third, we account for the fact that mining facilities have bounded capacity. These modeling choices yield a mining equilibrium  drastically different from that obtained in \cite{dimitri2017bitcoin}, \cite{biais2017blockchain}, and \cite{arnosti2019bitcoin}. In their studies, equilibrium quantities such as the set of active miners and their hash rate shares are \emph{independent} of the mining reward, which is in stark contrast to what is observed empirically. Such an independence result also implies that the mining reward has no impact on mining centralization.
In contrast, our work sheds light on the crucial role of mining reward in determining equilibrium hash rates. It also provides insights on the impact of gradual technological advancements on the mining equilibrium, %mining centralization, 
above and beyond what can be deduced from a static Cournot competition between miners. Additionally, we are able to assess the sustainability of the equilibrium outcome by considering the effect of both investment and mining reward on centralization and network security.

%Mining pools are another important component of the mining value chain where concentration of mining power can occur. 
{In addition to centralization at the mining level, concentration of mining power can also occur among mining pool operators.}
\cite{Cong} {study \emph{mining pool centralization} where the initial pool sizes are determined by ``passive hash rates'' that may not be optimally allocated, for example due to miners' inattention, and the transfer of ``active hash rate'' between pools is interpreted as growth. They argue that a single mining pool will not dominate, as a larger size allows the pool operator to charge higher fees, which in turn hinders the pool's growth. However, their results do not have direct implications for actual mining centralization. Specifically,} \cite{Cong} {assume a homogeneous mass of infinitesimal miners, and are not concerned with the distribution of mining power among actual miners. In particular, they do not capture the important fact that mining pools allow for the inclusion of small miners that would otherwise find mining infeasible. In regards to mining pool centralization, their study does not account for the fact that miners have an incentive to leave large mining pools for the good of the network, because their income and net worth is tied to the network's health. Miners' ability to switch between pools thus alleviates the potential adverse effects of mining pool centralization. Additionally, it should be noted that significant restrictions have been placed on the ability of mining pool operators to control the hash rate directed to their pools.} 
In contrast to \cite{Cong}, {we study centralization at the mining level. We model the initial hash rates is terms of exogenously determined mining costs, and interpret changes in hash rates due to investment as growth.}

	\section{Model of Cryptocurrency Mining}\label{sec:model}
		
	{We model cryptocurrency mining as a two-stage game. First, $N \geq 2$ miners choose their levels of investment in new mining hardware to increase the efficiency of their operations.} Then, they decide on hash rates to exert during the mining competition. 
	
	%We consider a two-stage game. In the first stage, the manufacturers simultaneously decide their R\&D levels, i.e., how much they try to advance their hardware efficiency, and their pricing schemes, i.e., how much they charge miners for new hardware. \note{If I understand correctly from all discussions we had, we no longer have this first stage, so we should remove it.} 
	
	%\begin{rem}
	%(i) Miners generally have an upper bound on capacity. Miner $i$ should have initial capacity $H_i$, and without an update should either not mine or use hash rate $H_i$. There should be correlation between size and efficiency of a mining operation.
	%(ii) Ferreira does not have R\&D and equipment upgrades.
	%(iii) In an equilibrium with $n\geq 2$ miners, the profit per hash of each miner has to be the same. Is there an intuitive reason for why it is equal to $1/(n-1)$? That is:${\pi_i}/{h_i} = {1}/{n-1}$.
	%\end{rem}
	
	%\remove{We consider a set of $N\geq 2$, and assume $N\geq 2$ to exclude the trivial case of a single miner}. 
	We use $\tilde c_i>0$ to denote the initial cost-per-hash of miner $i$, 	
	%\change{With advancements in mining hardware, new equipment is introduced that has cost-per-hash}
	and $\tilde c_0\leq\min_{1\leq i\leq N}\tilde c_i$ to denote the cost-per-hash of the most recently introduced hardware. Each miner can invest in the latest hardware to replace a fraction $0\leq\beta_i\leq 1$ of its hardware stock. 
	%\note{The value of $\tilde c_0$ is not of fundamental importance for the results, but we can discuss how the value of $\tilde c_0$ can be based on how much hardware efficiency has improved per year historically. There exist reports and papers covering how hardware efficiency has evolved, e.g., \url{https://medium.com/meetbitfury/bitcoin-mining-and-electricity-price-a-new-paradigm-b65ff46d7e96}}. 
	%Denoting by $0\leq\beta_i\leq 1$ the proportion of miner $i$'s stock of hardware that gets replaced, 
	The cost-per-hash of miner $i$ is 
	\begin{align}\label{c_beta}
	c_i := c_i(\beta_i) := \tilde c_i - \beta_i(\tilde c_i-\tilde c_0) %+ \beta_i p_i 
	+ \frac{\eta_i}{2}\beta_i^2, % =: c_i - \beta_iu_i + \frac{\eta_i}{2}\beta_i^2,
	\end{align}
	where the linear component quantifies how the cost-per-hash of miner $i$ is reduced with investment. The %\remove{polar} 
	case $\beta_i=0$ corresponds to no upgrading, in which case the cost of miner $i$ stays unchanged, and the upper bound %other extreme %\remove{case} 
$\beta_i=1$ corresponds to full upgrading, which pushes the cost of miner $i$ down to $\tilde c_0$. 
	The final term in (\ref{c_beta}) captures adjustment costs faced by miners - %\remove{, and the parameter $\eta_i\geq 0$ quantifies the magnitude of this cost for miner $i$}. 
	we follow existing literature and assume it to be convex and quadratic in the level of investment (\cite{Hamermesh}). % and \cite{Hamermesh}).
	{For a fixed level of investment, the parameter $\eta_i\geq 0$ quantifies the magnitude of adjustment costs for miner $i$.  Smaller miners, i.e., those who face higher initial cost-per-hash, typically also incur higher adjustment costs (see Remark} \ref{remEta}).\footnote{The size of a miner refers to its hash rate. We show in Proposition \ref{propH} that the equilibrium hash rate of a miner is decreasing in its cost-per-hash.} To capture this stylized feature, we use the following parametric specification,
	%\remove{Smaller miners, i.e., those who face higher initial costs, typically incur higher adjustment costs (see also Remark} \ref{remEta}). \remove{To capture these stylized patterns, we use the following parametric specification,}
	\begin{align}\label{eta}
	\eta_i := \eta (\tilde c_i-\tilde c_0),
	\end{align}
	where $\eta\geq 0$. {Such a specification also captures} potential price discrimination, with larger miners being charged on average less because of bulk discounts and greater bargaining power.\footnote{Large hardware orders are usually privately negotiated, and specific terms %of the deals 
		such as prices are typically not disclosed. As an example, in the article ``The bidding war for bitcoin mining hardware is heating up'', posted on The Block (see \url{https://www.theblockcrypto.com/post/88151/bitcoin-mining-hardware-bidding-war}), the section ``Big order negotiations'' states: ``U.S.-based bitcoin mine operator Core Scientific said Thursday that it has executed agreements with Bitmain to facilitate the purchase of \ldots Core Scientific didn't disclose details such as the prices \ldots but the execution of agreements suggests the two parties could have price lock-in deals that would lower the cost below the market average.''}

	\medskip 
	
	\begin{rem}\label{remEta}
		In the \emph{Global Cryptocurrency Benchmarking Study} (\cite{HilemanRauchs}) {conducted by The Cambridge Center for Alternative Finance (CCAF)} the section on ``Mining'' is based on a sample of 48 mining organizations, classified as ``small miners'' and ``large miners'' {based on their hash rate levels.} %, that participated in a voluntary study. 
		%\change{In particular, the study listed}
		{The study lists} a number of operational risk factors that can have a negative impact on both the operational functioning and profitability of mining activities. Miners were asked to rate those factors according to the risk they might pose to their daily operations. 	
		The largest discrepancy between small and large miners was observed with regards to ``insufficient availability of capital that is needed to continually upgrade and/or replace mining equipment''. This {represents a serious concern for small miners, {unlike} large mining organizations {which} have easier access to capital to invest in their mining infrastructure}. 
		The CCAF's \emph{2nd Global Cryptoasset Benchmarking Study} (\cite{Rauchs}) further reinforces this point. Therein, the section on mining is structured in the same way, based on public and private data on 128 mining facilities around the globe. Small miners raise concerns over growing difficulties in accessing state-of-the-art hardware equipment in a timely fashion.\hfill\qed %, an issue that large miners seem to be less concerned by. \hfill\qed 
	\end{rem}
	 
	Denote by $\beta:=(\beta_i)_{1\leq i\leq N}$ the investment level of each miner, and by $h:=(h_i)_{1\leq i\leq N}$ the hash rates exerted at the mining stage. The hash rates may depend on the investment profile $\beta$, and we emphasize this dependence by writing
	\[
	h := h(\beta), \qquad h_i := h_i(\beta) = h_i(\beta_i,\beta_{-i}).
	\] 
	The hash rate of miner $i$ is nonnegative, i.e., $h_i\geq 0$, and we denote by 
	\begin{align}\label{A}
	A:= A(\beta) := \{1\leq i\leq N: h_i>0\}
	\end{align}
	the set of active miners, i.e., the subset of miners with strictly positive hash rates. The objective function of miner $i$ is then given by %\note{I suggest we change $K$ with a greek letter, say $\mu$} 
	%\begin{align}\label{pi_i}
	%\pi_i := \pi_i(\beta_i,h_i;\beta_{-i},h_{-i}) := \frac{h_i}{H}R-c_ih_i - \frac{\gamma}{2}h_i^2 - K{\bf 1}_{\{i\notin A(0),\,h_i>0\}}, 
	%\end{align}
	\begin{align}\label{pi_i}
	\pi_i := \pi_i(\beta_i,h_i;\beta_{-i},h_{-i}) := 
	\renewcommand*{\arraystretch}{1.3}
	\left\{\begin{array}{ll}
	\frac{h_i}{H}R-c_ih_i - \frac{\gamma}{2}h_i^2 - K{\bf 1}_{\{i\notin A(0),\,\beta_i>0\}}, &\; H>0,\\
	0, &\; H=0,
	\end{array}\right.
	\end{align}
	where $R>0$ is the total reward from mining, {and $H:=\sum_{j=1}^Nh_j$ is the aggregate hash rate of all miners.} The quadratic cost term $(\gamma/2) h_i^2$ captures the limited hashing capacity of miner $i$, for instance due to a bounded supply of low cost electricity %. The parameter $\gamma\geq 0$ quantifies the capacity of {a miner}, 
	- {a larger value of $\gamma\geq 0$ corresponds to} smaller capacity. It is worth observing that a convex cost function has a qualitatively similar effect to {imposing} a capacity constraint. However, the constraint is soft, because production can be increased at an increasing marginal cost.\footnote{The quadratic form for the cost function buys us tractability, but the main results in the paper would {carry through} with an {arbitrary convex cost} function.} 
	{The last term in} (\ref{pi_i}) is the fixed cost $K>0$ incurred by new miners when setting up mining operations. {In our framework, miner $i$ is said to have set up new a mining operation if it is not active without investment, i.e., $i\notin A(0)$, but invests in new mining hardware, i.e., $\beta_i>0$.} Therefore, entry through investment in new mining hardware does not occur unless the revenue from mining exceeds the {incurred cost} (cf.\ \cite{Garratt}).

	{We conclude {this} section by defining the Nash equilibrium of the investment and mining stages of the game.}
		
	\medskip 
	
	\begin{defn}\label{def1} \hfill
		\begin{itemize}
		\item[(i)] For a fixed $\beta\in[0,1]^N$, an {\it equilibrium hash rate profile} is a vector $h^*\in[0,\infty)^N$ such that, for $1\leq i\leq N$,
		\begin{align}\label{eq1}
		\pi_i(\beta_i,h_i^*,\beta_{-i},h_{-i}^*) &= \sup_{h_i\geq 0}\pi_i(\beta_i,h_i;\beta_{-i},h_{-i}^*). %\\
		%\pi_i(\beta_i,h_i^*(\beta),\beta_{-i},h_{-i}^*(\beta)) &= \Big.\sup_{x\geq 0}\pi_i(\beta_i,h_i(\beta);\beta_{-i},h_{-i}^*(\beta))\Big\lvert_{h_i(\beta)=x}
		\end{align}
		\item[(ii)]	An {\it equilibrium investment} is a vector $\beta^*\in[0,1]^N$ such that, for $1\leq i\leq N$, %$0\leq\beta_i^*\leq 1$ and 
		\begin{align}\label{eq2}
		\pi_i\big(\beta_i^*,h_i^*(\beta^*);\beta_{-i}^*,h_{-i}^*(\beta^*)\big) = \sup_{0\leq\beta_i\leq 1}\pi_i\big(\beta_i,h_i^*(\beta_i,\beta_{-i}^*);\beta_{-i}^*,h_{-i}^*(\beta_i,\beta_{-i}^*)\big),
		\end{align}
		where $h^*(\beta^*)=(h_i^{*}(\beta^*))_{1\leq i\leq N}$ is an equilibrium hash rate profile corresponding to the investment $\beta^*$.
		\hfill\qed
		\end{itemize}
	\end{defn} 
	
	\medskip 
	
	Note that in the definition of an equilibrium investment, each miner internalizes that investment lowers the cost of mining and thus impacts the reward captured in the subsequent mining competition. A full Nash equilibrium of the game is a tuple
	\begin{align}\label{tuple}
	(\beta^*,h^*):=(\beta^*,h^*(\beta^*)), 
	\end{align}
	satisfying equations (\ref{eq1})-(\ref{eq2}), and
	\begin{align}
	\pi_i^* := \pi_i^*(\beta_i^*,h_i^*;\beta_{-i}^*,h_{-i}^*),
	\end{align} 
	is the corresponding equilibrium profit of miner $i$. 
	
	\begin{rem} 
		{Block generation is a random process, and $h_i/H$ is the expected share of reward earned by miner $i$. The objective function} (\ref{pi_i}) is thus based on the assumption that miners are risk-neutral. %Block generation is a random process and results in an uneven revenue stream for miners. 
		An alternative interpretation can be obtained by observing that, in practice, the vast majority of hash rate belongs to \emph{mining pools} {that combine the hash rate of a large number of miners, and allow them to earn a steady rather than an uneven revenue stream from mining blocks at random times.}
		%\add{combine the hash rate of a large number of miners and absorb the risk associated with not mining a block for prolonged periods of time.} 
		%\note{Alternative: combine the hash rate of a large number of miners and allows them to earn a steady revenue stream rather than the revenue from mining blocks at random times.}
		%absorb the \add{risk of randomness}\note{risk of randomness is weird, risk of a risk} in the mining process. 	
		The objective function (\ref{pi_i}) can therefore also be obtained without making assumptions about the risk aversion of miners, and instead % obtained by 
		assuming that all miners {direct} their hash rate to mining pools, and thus earn a reward in proportion to their hash rate share. In this case, the parameter $R$ can be considered to be the mining reward net of fees charged by pool operators. 
		\hfill\qed
	\end{rem}
	
	%Note from Definition \ref{def1} that the equilibrium hash rate of a miner can be equal to zero. We say that miner $i$ is active in equilibrium if $h_i^*>0$. 
	
	%\medskip 
	
	%\begin{defn} 
	%	An {\it equilibrium investment} is a vector $\beta^*\in[0,1]^N$ such that for $1\leq i\leq N$, %$0\leq\beta_i^*\leq 1$ and 
	%	\begin{align}\label{eq2}
	%	\pi_i\big(\beta_i^*,h_i^*(\beta_i^*);\beta_{-i}^*,h_{-i}^*(\beta_i^*)\big) = \sup_{0\leq\beta_i\leq 1}\pi_i\big(\beta_i,h_i^*(\beta_i,\beta_{-i}^*);\beta_{-i}^*,h_{-i}^*(\beta_i,\beta_{-i}^*)\big),
	%	\end{align}
	%	where $h^*(\beta^*)$ is an equilibrium hash rate profile. \hfill\qed 
	%\end{defn}
	
	%\medskip 

	\section{Mining Competition}\label{secMiningI}
	
	In this section, we study the second stage of the game where miners compete for rewards by solving a computationally costly hashing problem. 
	In Section \ref{secH}, we establish the existence of a unique mining equilibrium. % and study the associated hash rates and profits. 
	%{\Red In Section \ref{secHalf}, we examine the mining system in the special case of two miners.} 
	In Section \ref{secNumMiners}, we {characterize} the set of miners who are active in equilibrium. 
	%In Section \ref{secDeltas}, we study the dependence of  equilibrium hash rates on mining reward.\note{Changed?} 
	In Section \ref{secStatics}, we analyze the sensitivities of equilibrium hash rates to changes in the model parameters.

	\subsection{Equilibrium Hash Rates and Profits}\label{secH}
	
	 We take the miners' investment profile $\beta$ as given{, and thus treat the cost-per-hash $(c_i)_{1\leq i\leq N}$ of all miners as exogenous}.  Without loss of generality, we assume miners to be sorted in order of increasing cost-per-hash, i.e., $c_i\leq c_{i+1}$\footnote{{Throughout the paper, we say the sequence $(x_i)_{1\leq i\leq N}$ to be increasing if $x_i\leq x_{i+1}$, and decreasing if $x_i\geq x_{i+1}$. %that are nondecreasing and nonincreasing, respectively. 
	 		%We use the terms strictly increasing and strictly decreasing if the inequalities are strict. 
	 	}}, and denote by 
	\begin{align}\label{cn}
	 c^{(n)} := c^{(n)}(\beta) := \sum_{i=1}^n c_i(\beta_i) % = \sum_{i=1}^n c_i,
	\end{align}
	the cumulative cost of the first $n$ miners. The following proposition {establishes} {the existence of} a unique equilibrium hash rate profile.

	\medskip 
	
	\begin{prop}\label{propH}
		For any $\beta\in[0,1]^N$, there exists a unique equilibrium hash rate profile $h^*$. 
		{The equilibrium} hash rate of miner $i$ satisfies
		\begin{align}\label{opth}
		h_i^* = \left\{\begin{array}{ll} \frac{H^*(R-c_iH^*)}{R+\gamma (H^*)^2}, &\; 1\leq i\leq n, \\
		0, &\; n<i\leq N,\end{array}\right.
		\end{align}
		for some $2\leq n\leq N$, and the equilibrium {aggregate} hash rate is 
		\begin{align}\label{optH}
		H^* = \left\{\begin{array}{ll} \frac{\sqrt{(c^{(n)})^2+4(n-1)R\gamma}-c^{(n)}}{2\gamma}, &\; \gamma>0, \\
		\frac{(n-1)R}{c^{(n)}}, &\; \gamma=0.
		\end{array} \right.
		\end{align}
		%Moreover, 
		%\begin{align}\label{cRH}
		%h_i^*>0 \quad\Longleftrightarrow\quad c_i<\frac{R}{H^*}.
		%\end{align}
		%The value of $n$ is such that
		%\begin{align}\label{cRH}
		%c_i < \frac{R}{H^*} \quad\Longleftrightarrow\quad 1\leq i\leq n. 
		%\end{align}
		%In the general $\gamma_i\geq 0$ case, the overall hash rate can be bounded from above and below, 
		%	\begin{align}
		%f_n(\bar\gamma,R) \leq H \leq f_n(\underline{\gamma},R).
		%\end{align}
		%Also, for any $\gamma\geq 0$, the overall hash rate satisfies 
		%\begin{align}\label{HbigO}
		%H = f_n(\gamma,R) + O\big(\max_{1\leq i\leq n}|\gamma_i-\gamma|\big).
		%\end{align}
	\end{prop}

\medskip 
	
	The set of active miners consists of the $n$ most efficient miners, and their hash rates are decreasing in the cost parameter $c_i$. {Hence, miners with lower costs are larger in the sense that they have higher hash rates.} 
	It follows that the marginal active miner, i.e, miner $n$, is both the least efficient miner and the miner with the smallest nonzero hash rate. 
	{Note that there are \emph{at least} two active miners. This can be understood by observing that if no miner exerts a positive hash rate, then each miner has an incentive to marginally increase its hash rate and earn a positive profit. Similarly, if only a single miner has a positive hash rate, then a marginal reduction in its hash rate will increase its profit.}
	
	The equilibrium hash rates are obtained using the first-order condition {for} the objective function (\ref{pi_i}), {which equates marginal gain to marginal cost.} For an active miner $i$, this condition is given by
\begin{align}\label{1st}
\frac{R}{H^*}\Big(1-\frac{h_i^*}{H^*}\Big) %- \frac{h_i}{H}\frac{R}{H} 
= c_i + \gamma h_i^*.
\end{align}
Observe that the {equilibrium} marginal gain is smaller than the equilibrium reward-per-hash, $R/H^*$. {This is because  the} marginal probability of earning the reward is decreasing in the exerted hash rate, i.e., each additional unit of hash {rate} {has a smaller impact} on the probability of earning the reward (see Appendix \ref{appOtherProofs} {for details}). %which is due to the concavity of the function $h_i^*/H^*$.
{The next proposition {provides a decomposition of} the marginal cost of miners.} 
%The left-hand side shows that the marginal gain of miner $i$ is equal to the reward-per-hash, $R/H$, minus an implicit cost term $h_iR/H^2$.\note{What does implicit cost mean?} \add{This cost captures the decreasing marginal probability of earning the reward for each additional unit of applied hash, and is due to the concavity of the function $h_i/H$.}\remove{fact that  appears because the hash rate $h_i$ also contributes to the denominator of the probability of earning the reward, $h_i/H$: Since $h_i/H$ is a concave function of $h_i$, the increase in  probability of earning the reward gets smaller for each additional hash.}

\medskip 

	\begin{prop}\label{propMC}
		The marginal cost of miner $i$ {in equilibrium} is given by 
				\begin{align}\label{MC}
		MC_i^* := c_i + \gamma h_i^*, %\left\{\begin{array}{ll} c_i + \gamma h_i^*, &\; 1\leq i\leq n, \\
		%c_i, &\; n<i\leq N,\end{array}\right.
		\end{align}
		and $(MC_i^*)_{1\leq i\leq N}$ is a increasing sequence. Furthermore,
		\begin{align}\label{active}
		MC_i^* < \frac{R}{H^*} \quad\Longleftrightarrow\quad c_i<\frac{R}{H^*} \quad\Longleftrightarrow\quad 1\leq i\leq n.
		%\qquad \frac{R}{H^*} < 2MC_i^* \;\;\Longleftrightarrow\;\; \frac{h_i^*}{H^*}<\frac{1}{2}.
		\end{align}
		The sequences $({c_i}/{MC_i^*})_{1\leq i\leq N}$ and $({\gamma h_i^*}/{MC_i^*})_{1\leq i\leq N}$ %given by\note{remove the below equation?}
		%\begin{align}\label{MCdecomp}
		%	\frac{c_i}{MC_i^*} = \frac{c_i}{c_i+\gamma h_i^*}, \qquad \frac{\gamma h_i^*}{MC_i^*} = \frac{\gamma %h_i^*}{c_i+\gamma h_i^*},
		%\end{align}
		{are increasing and decreasing, respectively.} 
	\end{prop}
	
	\medskip 
	The marginal cost of miners consists of two components. The first component, $c_i$, depends on the miner's efficiency and is independent of its hash rate. The second component, $\gamma h_i^*$, reflects that a higher hash comes with a larger cost due to the limited miner's capacity. 
	For smaller (and inactive) miners, the first component accounts for a larger fraction of the marginal cost. These miners have a high cost of hashing {and a less binding capacity constraint because they exert low hash rates.} For larger miners, instead, the cost of hashing is lower but the capacity constraint is more binding. These {features} are formalized through the monotonicity properties of the sequences given in  Proposition (\ref{propMC}).
	
	{We conclude this section by analyzing the profits of miners in equilibrium.}
	
	\medskip 
	\begin{prop}\label{propProfit}
		The equilibrium mining profits $(\pi_i^*)_{1\leq i\leq N}$ form a decreasing sequence, such that $\pi_i^*>0$ for $1\leq i\leq n$, and $\pi_i^*=0$ for $n<i\leq N$. Moreover, the profit-per-hash of active miners, $(\pi_i^*/h_i^*)_{1\leq i\leq n}$, is a decreasing sequence.  
	\end{prop}
	\medskip 
	
	{As expected, an active miners makes a profit that is increasing in its mining efficiency. Interestingly, the same holds true for its profit-per-hash. That is, not only do more efficient miners apply larger hash rates (see} Prop.\ \ref{propH}), {but their profit-per-hash is also higher. 
	Observe that, in this system, profitability arises because of miners' heterogeneity and the oligopolistic nature of the mining competition. {In a system with homogeneous costs, free entry of miners would make total expenditures equal to the total reward, as in the classical rent-seeking model of} \cite{Tullock}. {Hence, the profit of each miner would be driven to zero.}

	\subsection{Set of Active Miners}\label{secNumMiners}
	
	We study the set of miners who are active in equilibrium.  It follows from Equation (\ref{active}) that miner $i$ is active if and only if its cost-per-hash is lower than the reward-per-hash. Note that \emph{given} the equilibrium reward-per-hash, the set of active miners can be deduced from the miners' cost-per-hash parameters, i.e., it is independent of their mining capacity.
	%Note that this condition is independent of the parameter $\gamma$, which determines the hashing capacity of miners. %\note{This is misleading. At the end of the section we comment on the fact that the break-even point of mining depends on $\gamma$. Clarify} 
	This is because a miner decides to be active based on the cost it incurs from applying an infinitesimal hash rate,} in which case limited capacity is not an impediment. % in such circumstances. 
	That is, if $h_i^*=0$, the marginal cost in (\ref{MC}) is equal to the cost-per-hash $c_i$.

	Observe that the aggregate hash rate $H^*$ depends on the number of active miners, {thus the reward-per-hash is} not an exogenous quantity. In the following proposition, we provide a condition to determine the set of active miners only in terms of the model parameters. 
	
	\medskip
	
	\begin{prop}\label{propN}%\hfill 
		%\begin{itemize} 
		%	\item[(a)]
			{The number of active miners in equilibrium is} given by the largest $2\leq n\leq N$ such that
			\begin{align}
			%\frac{R\gamma}{c_n^2} + \frac{c^{(n)}}{c_n} > n-1.\\
			c_n < \frac{c^{(n)}+R\gamma/c_n}{n-1}.
			\end{align} 
			The value of $n$ is increasing in $R$ and $\gamma$. 
	\end{prop}
	
	\medskip 
	
	%\note{It is not clear the relation of (4.2) with the condition in Proposition 4.3. It seems to suggest that what matters is how close the cost of a miner is to the average, but then 4.5 seems to be a condition which concerns a miner in isolation (not really because $H^*$ is the aggregate hash of all miners. Rather than presenting the two as ``contradictory'', we should rephrase and present the two conditions as complimentary and looking at different aspects of the mining game.)}\note{\Red Added a line before proposition.}
	
	%\remove{Condition} (\ref{active}) \remove{indicates that miners with a high cost of mining will not be competitive.} 
	{The above proposition states} that whether a miner is active depends on its cost relative to that of other miners. 
	Specifically, the higher the average cost of the first $n-1$ miners, the weaker the participation condition for miner $n$ becomes. {Hence, less heterogeneity in mining costs leads to a higher number of active miners. If miners are fully homogeneous, they are all active regardless of their cost.} 
	
	The value of $\gamma$ plays a key role in determining the number of active miners. If $\gamma=0$, {then the cost of the marginal miner satisfies} 
	\[
	%c_n < \frac{n-1}{n-2}\frac{c^{(n-1)}{n-1}.
	c_n < \frac{n}{n-1}\frac{c^{(n)}}{n}. 
	%\frac{c^{(n)}}{n} \leq c_n<\frac{c^{(n)}}{n-1}.
	\]
	This means that in equilibrium, the cost of the marginal miner can only be slightly greater than the average cost of all active miners. {This stringent condition highlights} how vulnerable the mining competition is to centralization. For instance, if the number of active miners is $n=20$, then the marginal miner's cost is at most 5\% higher than the average cost of all active miners, because $n/(n-1)\approx 1.05$.
	{Observe that the above participation condition is independent of $R$. This is intuitive because if miners have unbounded capacity, the marginal cost of hashing is independent of the exerted hash rate, and the number of active miners becomes independent of the reward. For example, if the reward doubles, the most efficient miners simply expand their operations by a factor of two while less efficient miners remain inactive.} 
	
	If $\gamma>0$, the number of active miners increases relative to the case $\gamma=0$. This is because the capacity constraint prevents the most efficient miners from expanding their operations infinitely, as the marginal cost of hashing grows with the exerted hash rate.  %\note{Why is it a similar argument? If $R$ is larger, and there are capacity constraints why would the number of active miners increase?}\note{\Red Added:}
	{The effect of a larger mining reward $R$ can be similarly analyzed.} A larger reward increases the marginal gain of hashing, so active miners increase their hash rates. However, their ability to do so is limited from the capacity constraint, which thus allows a greater number of miners to apply positive hash rates. %{Hence, for a fixed $\gamma$, more miners exert positive hash rates if the reward $R$ is larger.}
	%\remove{To summarize, miners face a tradeoff between the cost and reward of mining}. %\remove{, and the mining difficulty, which is directly related to the overall hash rate.}\note{Cost of minining and mining difficulty are very similar concepts. We should just keep one of them.}\footnote{\add{The mining difficulty is a measure of the computing power required to mine a block.} The difficulty is adjusted by the Bitcoin protocol to keep a fixed rate of mined blocks, and is therefore directly related to the overall hash rate. See: \url{https://www.blockchain.com/charts/difficulty}.} 
	 {This finding is consistent with {events} observed after the Bitcoin halving event in 2020 - when the reward for mining a block was halved, many smaller and less efficient mining farms went out of operation (see Remark \ref{remR} for additional details).\footnote{{Reports from China show that smaller mining operations were forced to either shut down or switch to mining alternative cryptocurrencies where competition is smaller} (see, e.g.,  \url{https://news.bitcoin.com/a-number-of-small-bitcoin-mining-farms-are-quitting-as-older-mining-rigs-become-worthless}).}
	
	At the beginning of this section, we stated that miner $i$ is active in equilibrium if and only if its cost $c_i$ is smaller than the {\it break-even cost of mining} $R/H^*$. {It follows from} Proposition~\ref{propN} that larger values of $R$ and $\gamma$ raise the break-even cost of mining, making these two quantities key determinants of {\it mining centralization}. In particular, if $\gamma$ is large enough, limited mining capacity becomes such a restraint %\change{concern}{restraint} 
	that even the least efficient miners have become active. %\change{are active}{have positive hash rates}. 
	If, instead, $\gamma$ is sufficiently small, the capacity constraint is so soft that a small set of efficient miners manages to crowd out all less efficient miners.
	%only the most efficient miners \change{prevail}{}.\note{Not really prevail; rather grow a lot to push others out} 

	\medskip 
	
	\begin{rem}\label{remR}
		In the Global Cryptocurrency Benchmarking Study (see Remark \ref{remEta}), 
		large miners considered ``fierce competition among miners of the same cryptocurrency'' to pose the highest risk to their operations, while small miners deemed a ``sudden large price drop (e.g., 25\%)'' to be the primary risk factor. %of greater risk than the constant arms race between miners. 
		The latter is representative of small miners {being concerned that low mining rewards would prevent them from being able} to mine at all, consistent with {the} analysis in this section. %The former indicates that large miners are less {concerned} about not being competitive, {and more concerned} about earning lower profits due to high competition. 
		\hfill\qed
	\end{rem}

\subsection{Comparative Statics of Mining Equilibrium}\label{secStatics}

{We perform a comparative statics analysis of the mining equilibrium, i.e., analyze its dependence on the underlying model parameters. We consider states of the system where small changes in model parameters do not alter the number of active miners.}\footnote{It follows from Proposition \ref{propN} that the number $n$ of active miners is a piecewise constant function of the model parameters. As the mining reward $R$ increases, the number of active miners remains the same until reaching the threshold at which it increases by one. Mathematically, the number of active miners is a left-continuous with right limits (LCRL/c\`agl\`ad) function of $R$ and $\gamma$, and a right-continuous with left limits (RCLL/c\`adl\`ag) function of $c_i$.} The signs of all sensitivities are summarized in Table \ref{table0}.

\medskip 

\renewcommand*{\arraystretch}{1.2}
\begin{table}[h!]
	\begin{center}
		%\begin{tabular}{|c|c|c|c|c|} \hline % <-- Alignments: 1st column left, 2nd middle and 3rd right, with vertical lines in between
		\begin{tabular}{|>{\centering}p{0.1\textwidth}|>{\centering}p{0.1\textwidth}|>{\centering}p{0.1\textwidth}|>{\centering}p{0.1\textwidth}|p{0.1\textwidth}<{\centering}|} \hline 
			& $c_i$ & $c_j$ & $\gamma$ & $R$   \\ \hline
			$H^*$        & $-$   & $-$   &  $-$     & $+$   \\
			$h_i^*$      & $-$   & $+/-$ &  $+/-$   & $+$   \\ 
			$h_i^*/H^*$  & $-$   & $+$   &  $+/-$   & $+/-$  \\ \hline
		\end{tabular}
	\end{center}
	\vspace{-10pt}
	\caption{\small Sensitivities of equilibrium hash rates to model parameters. The first row shows the signs of the aggregate hash rate's sensitivities to $c_i$, $c_j$, $\gamma$, and $R$. The second and third rows, respectively, show the signs of the sensitivities of miner $i$'s hash rate and hash rate share to these parameters. The sign ``$+$'' (``$-$'') indicates that the derivative of the equilibrium quantity with respect to the model parameter is positive (negative). The sign ``$+/-$'' means that the sign can be either positive or negative. Explicit expressions for the sensitivities are provided in equations (\ref{derivsH}), (\ref{derivsh}), and (\ref{derivsHh}) of the Appendix.}
	\label{table0}
\end{table}

%\note{To be done: We should have three sections. I have already written the titles of these three sections. The first section would be on aggregate hash rate vs mining reward. The second section would be on individual hash rates $h_i^*$ to cost $c_i$ and $c_j$. Here, a separate section on the sensitivity of profits to costs would be stated and proven in the appendix. The third section would be on the sensitivity of individual hash rates to $\gamma$ and $R$.}

\subsubsection{Sensitivities of Aggregate Hash Rate {to Model Parameters}}\label{secDeltaH}

%\medskip
%\begin{prop}\label{propR}
%If $\gamma=0$, the aggregate hash rate $H^*$ is proportional to the mining reward $R$. 
%If $\gamma>0$, $H^*$ increases like the square root of $R$. 
%The aggregate hash rate $H^*$ increases like the square root of the mining reward $R$. 
%There exists an increasing sequence $(r_k)_{1\leq k\leq N}$ {with $r_1=0$ and $r_{N}=\infty$}, {such that for} $r_{k}<R\leq r_{k+1}$, the {aggregate} hash rate $H^*$ is given explicitly by (\ref{optH}) with $n=k+1$. 
%\end{prop}
%\medskip 

The sensitivities of the aggregate hash rate $H^*$ to the model parameters are consistent with intuition. First, $H^*$ gets smaller if mining becomes more costly, either due to an increase in the cost-per-hash of a miner, or because of the capacity constraint becoming tighter. Second, $H^*$ gets larger if the mining reward $R$ increases. 

We next analyze in more detail the relationship between $H^*$ and $R$, which takes a different form depending on the miners' capacity constraint.
If miners have unbounded capacity, i.e., $\gamma=0$, the aggregate hash rate $H^*$ is {directly} proportional to $R$. This is because an increase in the mining reward leads to a proportional increase in each miner's hash rate, as discussed in Section \ref{secNumMiners}, and thus also in the aggregate hash rate. Formally, the fact that $H^*/R$ is constant can be seen from Proposition \ref{propH} and that the number of active miners $n$ is independent of $R$ (see Prop.\ \ref{propN}). 
%As argued in Section \ref{secNumMiners}, an increase in the mining reward would then result in a proportional increase in each miner's hash rate, and thus in the aggregate hash rate. \change{T}{Formally, t}his result follows from Proposition \ref{propH}, which states that the aggregate hash rate satisfies $H^*/R=(n-1)/c^{(n)}$, along with Proposition \ref{propN}, which implies that the number of active miners $n$, and thus the proportionality constant $(n-1)/c^{(n)}$, are independent of $R$.
If $\gamma>0$, %the number of active miners is no longer independent on $R$, as established in Proposition \ref{propN}. However, 
and the increase in $R$ is small enough to leave the number of active miners unchanged, the aggregate hash rate $H^*$ increases like the square root of $R$ (see Proposition \ref{propH}). The reason for this sublinear growth is that active miners are not able to increase their hash rates at the same rate as the mining reward, because they become increasingly capacity constrained.

\subsubsection{Sensitivity of Individual Hash Rates to Mining Costs}\label{secDeltaC}

The sensitivity of a miner's \emph{hash rate} to its own cost-per-hash parameter consists of two terms. The first term captures the direct dependence on the parameter, while the second term captures an indirect dependence through the aggregate hash rate. The indirect dependence is also equal to the sensitivity of the miner's hash rate to the cost-per-hash parameter of other miners. Analogous decompositions hold true for the sensitivity of a miner's \emph{hash rate share} to the cost of mining. These relations are formalized in the following proposition; explicit expressions for all components are provided in equations (\ref{derivsh}) and (\ref{derivsHh}) of Appendix \ref{appProofs}. 

\medskip 
	\begin{prop}\label{propDeltaC}
	For active miners $i$ and $j$, the sensitivities of $h_i^*$ to $c_i$ and $c_j$ are of the form
	\begin{align}
	\frac{\partial h_i^*}{\partial c_i} &= \Delta_{i,1} + \Delta_{i,2} <0,  \qquad 
	\frac{\partial h_i^*}{\partial c_j} = \Delta_{i,2}, 
	\end{align}
	where $\Delta_{i,1} <0$, and 
	\begin{align}\label{half}
	%\Delta_{i,1} &<0,  
	%\qquad 
	\Delta_{i,2}>0 \quad &\Longleftrightarrow\quad \frac{h_i^*}{H^*}<\half. 
	\end{align} 
	%Explicit expressions for the derivatives are given in (\ref{derivC}), (\ref{derivGamma}), and (\ref{derivR}). 
	Furthermore, the sensitivities of miner $i$'s hash rate share to $c_i$ and $c_j$ satisfy 
	\begin{align}
	\frac{\partial }{\partial c_i}\frac{h_i^*}{H^*} = \widetilde\Delta_{i,1} + \widetilde\Delta_{i,2} < 0, 
	\qquad \frac{\partial }{\partial c_j}\frac{h_i^*}{H^*} = \widetilde\Delta_{i,2} > 0.
	\end{align} 
	where $\widetilde\Delta_{i,1}<0$ and $\widetilde\Delta_{i,2}>0$.
\end{prop}
\medskip 

%{We begin by discussing the sensitivity} of $h_i^*$ to the mining cost parameter $c_i$. 
In the derivative of $h_i^*$ with respect to $c_i$, the \emph{direct sensitivity} $\Delta_{i,1}$ %of $h_i^*$ to $c_i$ 
is negative, because the marginal cost of mining is increasing in $c_i$. The \emph{indirect sensitivity} $\Delta_{i,2}$ arises because the aggregate hash rate $H^*$ changes if $h_i^*$ changes, which in turn impacts the marginal gain of hashing. If $h_i^*/H^*<1/2$, 
the indirect sensitivity is positive %which partially offsets the direct sensitivity. % lowers the hash rate
which alleviates the hash rate reduction from the direct sensitivity when $c_i$ increases. %of the direct sensitivity. %lowers the reduction in hash rate when $c_i$ increases. 
If $h_i^*/H^*>1/2$, the indirect sensitivity is negative which causes a reduction in hash rate beyond that implied by the direct sensitivity. %, while the net effect is still negative. 
Regardless of miner $i$'s hash rate share, the net effect is that {$h_i^*$} gets smaller as miner $i$'s cost of hashing increases.

Observe that the indirect sensitivity $\Delta_{i,2}$ is exactly equal to the sensitivity of $h_i^*$ to the cost-per-hash $c_j$ of another miner. Intuitively, this is because both quantities capture the sensitivity of miner $i$'s hash rate to changes in the aggregate hash rate $H^*$. The only difference is that in the first case, the change in $H^*$ is caused by a change in $h_i^*$, but in the latter case it is caused by a change in $h_j^*$. 

Next, we discuss the condition which determines the sign of the indirect sensitivity in (\ref{half}), and provide intuition for how equilibrium hash rates are impacted by heterogeneity in mining costs. We begin by considering a system of $N=2$ miners, %, and highlight the intuition for how equilibrium hash rates are impacted by heterogeneity in mining costs. % the determinants of equilibrium hash rates 
%It follows from Proposition \ref{propH} that both miners are active, because the mining equilibrium includes \emph{at least} two active miners. This can be understood by observing that if no miner exerts a positive hash rate, then each miner has an incentive to marginally increase its hash rate and earn a positive profit. Similarly, if only a single miner applies a positive hash rate, then a marginal reduction in its hash rate will increase its profit. 
whose response functions are visualized in the left panel of Figure \ref{fig_2miners}. For miner $i$, the response function gives its profit-maximizing hash rate, given the hash rate of the other miner, i.e., 
\begin{align}\label{Ri}
\mathcal{R}_i(h_{-i}) := \argmax_{h_i\geq 0}\pi_i(h_i;h_{-i}).
\end{align}
%The unique equilibrium hash rates in Proposition \ref{propH} are determined by the point of intersection of the two curves. 
It is clear from the figure that if mining is more costly for the second miner, i.e., $c_2>c_1$, then this miner exerts a lower equilibrium hash rate relative to the one exerted if $c_1=c_2$. Surprisingly, the first miner also exert a lower hash rate if $c_2>c_1$.
%To understand these shifts in the miners' hash rates, we have the following proposition. 
%\medskip 
%\begin{prop}
%	The sensitivities of the hash rate $h_i$ to $c_i$ and $c_{-i}$ satisfy
%	\begin{align}
%	\frac{\partial h_i^*}{\partial c_{i}} > 0, \qquad 
%	\frac{\partial h_i^*}{\partial c_{-i}} > 0 \;\;\Longleftrightarrow\;\; \frac{h_i^*}{H^*} < \frac{1}{2}.
%	\end{align}
%\end{prop}
%\medskip 
%The reason for this is the negative externalities imposed by miners on each other through their mining activities.
%\note{Miner one is impacted by a change in the cost of miner two through the aggregate hash rate. This is the channel.}
%These shifts are determined by the externalities imposed by miners on each other through their hashing. 
%First note that without competition, a miner would exert minimal hash rate and earn the entire mining reward. This is because mining is costly and the mining reward is bounded and fixed. 
%%Miner one is impacted by a change in the cost of miner two through the aggregate hash rate. This is the channel.
The underlying reason %behind this equilibrium outcome is that, 
is that %in a system of two miners, 
the larger miner {(i.e., the one with a lower cost)} has an incentive to reduce its hash rate if the cost of the smaller miner increases, while the incentive of the smaller miner is to increase its hash rate if the cost of the larger miner increases. This pattern is reflected in the right panel of Figure \ref{fig_2miners}. %\note{Have we clarified the difference between small and large miner?}

%Because we can view each miner as competing against the cumulative hash rate of all other miners, we can generalize this reasoning to a system consisting of $n>2$ active miners: if the hash rate of miner $i$ satisfies $h_i^*/H^*>1/2$, then this miner responds to reduced activity of other miners by reducing its own hash rate. In particular, if $h_i^*/H^*$ is close to one, the value of $h_i^*$ is close to zero. 
%In particular, if $h_i^*/H^*$ is close to one, the value of $h_i^*$ is close to zero. 

Because we can view each miner as competing against the cumulative hash rate of all other miners, the above reasoning can be generalized to a system consisting of more than two active miners: if the hash rate of miner $i$ satisfies $h_i^*/H^*>1/2$, then this miner responds to cost increases of other miners by reducing its hash rate.
Put differently, this means that if a miner controls less than half of the aggregate hash rate, i.e., less than the hash rate of all other miners combined, then this miner behaves aggressively and attempts to gain hash rate share by filling in the void left by competing miners. If, instead, the miner already controls more than half of the system hash rate, then it becomes passive and manages its hash rate to fend off competition.

To understand this phase transition in the largest miner's behavior, observe that miner $i$ is impacted by a change in the cost of miner $j$ through the aggregate hash rate. %Specifically, 
Proposition \ref{propDeltaC} shows that, all else being equal, an increase in the cost of miner $j$ implies a decrease in its hash rate, which %, all else being equal, 
lowers the aggregate hash rate $H^*$. 
%To understand this phase transition in the largest miner's behavior, we observe that an increase in the cost of miner $j$ implies a decrease in its hash rate, which, all else being equal, lowers the aggregate hash rate $H^*$. %This follows directly from equation (\ref{Hstar}) for $H^*$, and c. All else being equal, 
A smaller $H^*$, in turn, impacts the marginal gain of miner $i$ in two different ways. 
First, %a smaller aggregate hash rate
it implies that each unit of hash in the system corresponds to a larger fraction of the mining reward, which {increases} the marginal probability of earning the reward for all miners. 
Second, a smaller $H^*$ %aggregate hash rate 
implies an increase in the aggregate hash rate {share} captured by miner $i$, which lowers its marginal probability of earning the reward. %\note{decreasing marginal probability of earning the reward}
%First, each unit of hashing applied by miner $i$ results in a larger share of the aggregate hash rate captured by $i$, which in turn increases the marginal probability of earning the reward.
%\add{Second, a rise in miner $i$'s hash rate share lowers the marginal probability of earning the reward.} 
These two effects offset each other if $h_i^*/H^*=1/2$, leaving the equilibrium hash rate of miner $i$ unchanged (i.e., $\Delta_{i,2}=0$). If $h_i^*/H^*<1/2$, the first effect is stronger, resulting in a higher marginal gain for miner $i$ and an increase in its hash rate $h_i^*$ (i.e., $\Delta_{i,2}>0$). 
The opposite happens if $h_i^*/H^*>1/2$, leading to a decrease in $h_i^*$ (i.e., $\Delta_{i,2}<0$). We refer to Appendix \ref{appOtherProofs} for the mathematical details.
%If, instead, $h_i^*/H^*>1/2$, the negative effect of an increasing hash share rate outweighs the positive effect of a lower mining difficulty, leading to a decrease in $h_i^*$. %Mathematically, $R/H^*(1-h_i^*/H^*)$ is ...	

%\add{so that the concavity of $h_i/H$ is lower,}\footnote{This means that while a larger value of $h_i$ translates into a larger probability of earning the reward, the gain gets smaller as $h_i$ becomes a larger proportion of $H$.} \add{and each additional hash contributes less the probability of winning.} The second effect gets stronger if $h_i/H$ gets larger, which reduces the increase in $h_i$. In the extreme case, if $h_i/H>1/2$, the second effect dominates and $h_i$ becomes smaller. %the marginal gain of miner $i$ becomes smaller.

%\item[(ii)]
It is also informative to consider the implications of very small or very large values of the cost parameter $c_2$.
If $c_2\to\infty$, the hash rate of the first miner vanishes. The intuition is that the second miner becomes so inefficient that its competitor can earn the entire mining reward by exerting minimal hash rate. If instead $c_2$ is very small, i.e., $c_2\to 0$, %i.e., if the second miner becomes extremely efficient, 
the hash rate of the first miner may not be pushed to zero (see the right panel of Figure \ref{fig_2miners}). The reason is that the capacity constraint prevents the second miner from exerting hash rates large enough to crowd out its competitor, even if its cost of hashing is very low.

\medskip 

\begin{rem} 
	In the context of blockchain-based cryptocurrency mining, the condition $h_i^*/H^*<1/2$ admits the following interpretation. If a mining organization were to control the majority of the system hash rate, it can effectively implement the so-called 51\% attacks. Such attacks include preventing transactions between users of the network, and reversing transactions in order to double-spend coins.
	\hfill\qed
	%Intuitively, in a system with more than two active miners, $h_i^*/H^*>1/2$ can only occur if miner $i$ is much more efficient than competing miners, in addition to the capacity constraint not being to restrictive. \hfill\qed
	%In the context of blockchain-based cryptocurrency mining, the condition $h_i^*/H^*<1/2$ is typically satisfied in equilibrium.\footnote{A single miner controlling at least half of the overall hash rate can be considered a failed state, e.g.\ due to 51\% attacks, in which case the blockchain ceases to exist.} 
	%This claim is supported by the sensitivity results in Table \ref{table0}, which show that the hash rate share of miner $i$ is decreasing in its own cost $c_i$, and increasing in the cost $c_j$ of any other miner $j\neq i$.  %\hfill\qed
\end{rem} 
%\note{This second part is a bit confusing and too technical. I would skip it.}
%Intuitively, an equilibrium with $h_i^*/H^*>1/2$ can only occur if the most efficient miner is significantly more efficient than the rest. For example, consider a fixed $c_1>0$. Then, for any $\alpha<1$, there exists a constant $K_{\alpha}$ such that if $c_2>K_{\alpha}$,\footnote{Since $c_2\leq c_3\leq \dots\leq c_N$, it follows that $c_k>K_{\alpha}$ for $2\leq k\leq N$.} the hash rate share of miner one satisfies $h_1^*/H^*>\alpha$. \hfill\qed
% if $c_1$ is small enough and for increasing values of $c_2$, the hash rate share of miner 1 can become arbitrarily large.%$c_1\to 0$ and $c_n\to\infty$ if needed.
%\end{itemize}

\begin{figure}[ht!]
	\centering
	%\hspace{-1 cm} %[width=20.0cm,height=6.0cm]
	\includegraphics[width=\textwidth]{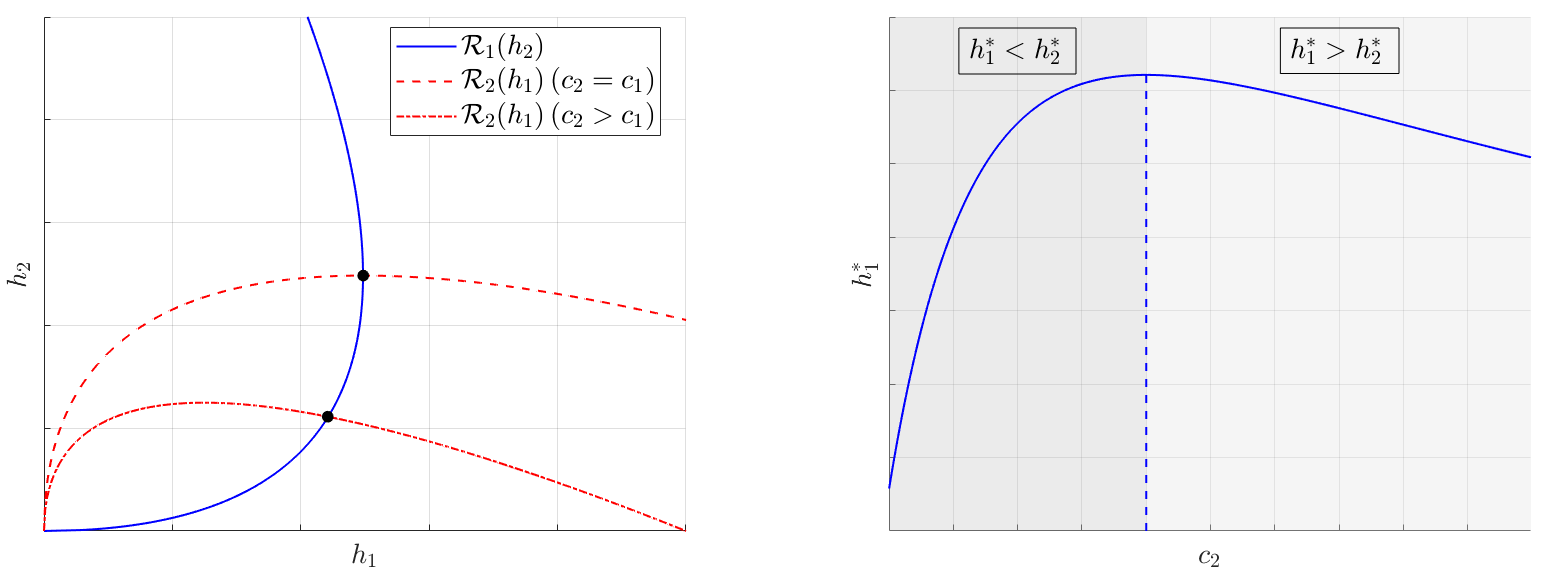}
	\caption{\small {Conceptual visualization of equilibrium hash rates in a system consisting of $N=2$ miners.} Left panel: Response functions. Solid blue curve is the response function of miner $1$ with cost $c_1>0$. Red curves are the response functions of miner $2$ for the symmetric case $c_2=c_1$ (dashed line), and for the case $c_2>c_1$ (dash-dotted line). {The equilibrium hash rates, characterized in Proposition} \ref{propH}, {are determined by the point of intersection of the two curves.} Right panel: Equilibrium hash rate $h_1^*$ of miner $1$ as a function of the cost $c_2$ of miner $2$. {The dashed line corresponds to the hash rate exerted at the point $c_2=c_1$, in which case $h_1^*=h_2^*$. The hash rate $h_1^*$ is decreasing in $c_2$ to the right of this point, but increasing in $c_2$ to the left of this point.} %\note{We should specify the parameter set used to produce these graphs, and to be consistent it would be good to use the same parameter set as the calibrated parameters.}
	} \label{fig_2miners}
\end{figure} 

The last part of Proposition \ref{propDeltaC} shows how a change in the cost-per-hash of a given miner imposes an externality on the hash rate shares of its competitors. In this case, the sign of the indirect sensitivity $\widetilde\Delta_{i,2}$ is always positive: if the cost of miner $j$ increases, the hash rate share of miner $i$ get larger, even if its hash rate goes down. In other words, if the cost-per-hash of miner $j$ increases, the share of the mining reward earned by all other miners increase at the expense of miner $j$.\footnote{In Proposition \ref{propPi}, we show that the same intuition extends to mining profits: If the cost-per-hash of miner $j$ increases, its profit decreases while all other miners benefit from miner $j$ becoming less competitive and see their profits increase.}

These findings have implications for mining centralization. First, we already know that a higher homogeneity among miners results in more of them being active in equilibrium, and thus increases mining decentralization (see Prop.\ \ref{propN}). Proposition \ref{propDeltaC} additionally shows that increased homogeneity contributes to decentralization, once the set of active miners is fixed. {For example, if miner $i$ has a relatively low mining cost, and thus a relatively large hash rate share,} then an increase in its cost transfers hash rate share from miner $i$ to other miners. {Similarly, if miner $i$ has a relatively high mining cost, and thus a relatively small hash rate share, then this miner would gain hash rate share from other miners if its cost were to decrease.

\subsubsection{Sensitivity of Individual Hash Rates to Mining Reward and Mining Capacity}\label{secDeltaRgamma}

In the previous section, we have argued that a miner's hash rate depends both directly and indirectly on mining costs. The same holds true for the dependence of hash rates on mining reward and mining capacity, where an indirect dependence again arises through the aggregate hash rate. These dependences are formulated in the following proposition; explicit expressions for all components are provided in equations (\ref{derivsh}) and (\ref{derivsHh}) of Appendix \ref{appProofs}.

\medskip 

\begin{prop}\label{propHdelta}
	The sensitivities of an active miner's hash rate $h_i^*$ to $\gamma$ and $R$ admit the following decompositions,
	\begin{align}
	\frac{\partial h_i^*}{\partial\gamma} &= \Delta_{i,1}^{(\gamma)} + \Delta_{i,2}^{(\gamma)}, 
	\mkern-36mu\mkern-36mu & \Delta_{i,1}^{(\gamma)} &<0, 
	\mkern-36mu\mkern-36mu & \Delta_{i,2}^{(\gamma)}>0 \;\;&\Longleftrightarrow\;\; \frac{h_i^*}{H^*}<\half, \\
	\frac{\partial h_i^*}{\partial R}     &= \Delta_{i,1}^{(R)} + \Delta_{i,2}^{(R)}>0, 
	\mkern-36mu\mkern-36mu & \Delta_{i,1}^{(R)} &>0, 
	\mkern-36mu\mkern-36mu & \Delta_{i,2}^{(R)}<0 \;\;&\Longleftrightarrow\;\; \frac{h_i^*}{H^*}<\half.
	\end{align} 
	%Explicit expressions for the derivatives are given in (\ref{derivC}), (\ref{derivGamma}), and (\ref{derivR}). 
	Furthermore, the  {derivatives of} $h_i^*/H^*$ {with respect to} $\gamma$ and $R$ are increasing in $i$, for $1\leq i\leq n$. %Explicit expressions are provided in (\ref{derivHh})-(\ref{derivPi}). 
\end{prop} 

\medskip 

%The impact of $\gamma$ on a miner's hash rate depends on the miner's size. A higher value of $\gamma$ reduces the hash rates of miners that {exert large hash rates} and are thus capacity constrained. In turn, this allows miners who exert lower hash rates and are thus not capacity constrained to expand. For these miners, the indirect sensitivity of $h_i^*$ to $\gamma$ is positive and large enough to outweigh the direct sensitivity.\footnote{The condition on $h_i^*/H^*$ determining the signs of $\Delta_{i,2}^{(\gamma)}$ and $\Delta_{i,2}^{(\gamma)}$ can be interpreted in the same way as the condition determining the sign of $\Delta_{i,2}$ in Proposition \ref{propDeltaC}.} {Intuitively, this is because the marginal gain of hashing increases as miners who are capacity constrained lower their hash rates.} Observe {from Table} \ref{table0} that the net effect of an increase in $\gamma$ on the aggregate hash rate is negative. That is, increased hashing of smaller miners does not fully compensate for the reduction in hash rates by larger miners.

The impact of $\gamma$ on a miner's hash rate depends on the miner's size. First, the direct sensitivity $\Delta_{i,1}^{(\gamma)}$ is negative for all miners, because a higher value of $\gamma$ increases the marginal cost of hashing. However, the explicit expression provided in Equation (\ref{derivsh}) {for $\Delta_{i,1}^{(\gamma)}$ and $\Delta_{i,2}^{(\gamma)}$ shows that for small enough miners,} the indirect sensitivity is positive and large enough to outweigh the direct sensitivity.\footnote{The condition on $h_i^*/H^*$, which determines the signs of $\Delta_{i,2}^{(\gamma)}$ and $\Delta_{i,2}^{(R)}$, can be interpreted in the same way as the condition which determines the sign of $\Delta_{i,2}$ in Proposition \ref{propDeltaC}.} Intuitively, this is because miners who exert large hash rates, and are thus capacity constrained, are forced to reduce their hash rates. This, in turn, increases the marginal gain of mining, and allows smaller miners who are not capacity constrained to expand. Observe from Table \ref{table0} that the net effect of an increase in $\gamma$ on the aggregate hash rate is negative. That is, increased hashing of smaller miners does not fully compensate for the reduction in hash rates by larger miners.

From the above analysis it follows that a tighter capacity {constraint (i.e., higher $\gamma$)} increases decentralization - smaller miners gain hash rate share at the expense of larger miners. {The proposition shows that a larger mining reward has the same effect}, even though the hash rate of each miner increases. 
This can again be understood in terms of the capacity constraint{. Specifically, a larger mining reward increases the marginal gain of hashing, leading to larger hash rates, but smaller miners are able to raise their hash rates more than larger miners.}
%\remove{This effect can again be understood in terms of the capacity constraint, as smaller miners are able to raise their hash rates more than larger miners, and thus increase their hash rate shares.}\note{This sentence seems a repetition of the previous sentence. There is nothing specific about mining reward in that sentence. We should add some specificity for mining reward.} 
Hence, a reduced mining capacity and an increased mining reward 
{not only improves decentralization by increasing the number of active miners (as shown in Section} \ref{secNumMiners}), but also creates a second decentralization effect by reducing the hash rate of large miners and increasing the hash rate of small miners.
This second decentralization effect is material, considering that a fair amount of time and planning is needed to establish new mining operations, so the total number of active miners stays unchanged over short periods of time (see also the discussion in Section \ref{secEDA}).

	\section{Miners' Hardware Investment}\label{secMiningII}

In this section, we study the {investment in hardware made by miners} before {participating} in the mining {competition}. We characterize the {equilibrium investment profile} in Section \ref{secEqBeta}. We study how investment affects the mining equilibrium in Section \ref{secEffectBeta}.

\subsection{Equilibrium Investment Profile}\label{secEqBeta}

We begin by showing the existence of an equilibrium investment $\beta^*$, i.e, {such that $(\beta^*,h^*(\beta^*))$} satisfies equations~(\ref{eq1})-(\ref{eq2}), 
where $h^*(\beta^*)$ is the unique equilibrium hash rate profile {corresponding to investment $\beta^*$, }characterized in Proposition \ref{propH}.

\medskip 

\begin{prop}\label{propBeta}
	There exists a {unique} equilibrium investment $\beta^*$ given by 
	\begin{align}\label{opt_beta}
	\beta_i^* = \left\{\begin{array}{ll} \min\big\{\frac{1}{\eta},1\big\},&\; 1\leq i\leq n, \\
	%\min\Big\{\frac{u_i}{\eta_i},1\Big\},&\; 1\leq i\leq n(\beta_i^*), \\
	0,&\; n<i\leq N,
	\end{array}\right. 
	\end{align}
	for some $2\leq n\leq N$. The equilibrium set of active miners defined in Eq.~(\ref{A}) is such that $A(0)\subseteq A(\beta_i^*)$. Furthermore, the {number $n$ of active miners} is decreasing in the fixed cost $K$ incurred by miners. % (\ref{pi_i}). 
\end{prop}

\medskip 

{Observe that the equilibrium number $n$ of active miners depends on the investment profile $\beta^*$, and is thus an endogenous quantity. However, this number can be characterized in terms of model primitives as the largest $i\geq |A(0)|$ such that miner $i$ {is active} in the mining equilibrium corresponding to the investment profile $\beta^{(i)}$ given by $\beta_j^{(i)}=\min\{1/\eta,1\}$ for $1\leq j\leq i$, and $\beta_j^{(i)}=0$ for $i<j\leq N$.}

With investment, the number of active miners either increases or stays the same as without investment. Because of gains in efficiency, miners who already were active without investment remain active after investment, although their shares of the aggregate hash rate may change. Inactive miners, instead, may not be able to benefit sufficiently from investment to enter the mining competition, either because of too high initial costs, or because of the fixed cost incurred by entry (see Eq.~\eqref{pi_i}).

An active miner $i$ chooses a level of investment $\beta_i^*$ which minimizes its cost-per-hash $c_i(\beta_i)$. %\note{Is this shown in the proof or it is evident from the statement} 
This follows %given the cost of other miners, 
from the fact that the profit of miner $i$ is decreasing in its cost of mining (see Prop.\ \ref{propPi}). As a result, the investment of miner $j\neq i$ does not impose an externality on the investment of miner $i$.
%\remove{That is, given the investment $\beta_{-i}$ of other miners, the best response $\beta_i$ of miner $i$ is independent of $\beta_{-i}$ and obtained by minimizing $c_i(\beta_i)$.} 

The absence of these investment externalities is consistent with existing practices, where each miner aims to maximize its own efficiency to remain competitive. Namely, while the overall system hash rate is readily observed, each miner does not observe the hash rates and mining costs of its competitors, or even the number of active miners. This is reflected in the objective function of miner $i$, where the {only quantity that depends on the actions of other miners is the aggregate hash rate}. 

It follows directly from equation (\ref{c_beta}) that the reduction in cost-per-hash of an active miner $i$ because of investment is
\begin{align}\label{Ieq}
I_i := c_i(0) - c_i(\beta_i^*) = \left\{\begin{array}{ll}
\frac{1}{2\eta}(\tilde c_i-\tilde c_0), &\; \eta>1, \\
\big(1-\frac{\eta}{2}\big)(\tilde c_i-\tilde c_0), &\; \eta\leq 1.
\end{array}\right.
%\qquad \bar I :=  \sum_{i=1}^nI_i,
\end{align}
The above formula highlights how initial mining efficiencies and adjustment costs impact the efficiency gains from investment. First, the cost reduction is greater if the miner is less efficient to begin with. Second, the cost reduction is decreasing in %The numerator $u_i$ is the reduction in cost-per-hash that miner $i$ can get from investment, and the denominator is 
the adjustment cost parameter $\eta$, which reflects frictions faced by miners when it comes to investing in new hardware. 
%It follows that for a constant value of $\eta_i$, i.e., if $\eta_i\equiv\eta$, less efficient miners have a higher level of investment.
%\footnote{The initial cost-per-hash of miner $i$ may be decomposed as $\tilde c_i=\tilde c_{i,1}\tilde c_{i,2}$, where $\tilde c_{i,1}$ is the number of units of energy required per hash (i.e., efficiency of hardware), and $\tilde c_{i,2}$ is the price per unit of energy. It follows that the reduction in cost from investment is $\beta_i(\tilde c_{i,1}-\tilde c_0)\tilde c_{i,2}$, where $\tilde c_0$ is the efficiency of the latest hardware. Hence, the cost reduction is increasing in both of the miner's cost factors, $\tilde c_{i,1}$ and $\tilde c_{i,2}$. In our model, we combine the two factors into one, i.e., work directly with $\tilde c_i$, as the implications of investment are unchanged. That is, miners with a high initial value of $\tilde c_i$ have the largest cost reduction.} 
Overall, the intuition is that for miners with a low initial cost, the cost reduction from investing in new hardware is limited, as they are already highly efficient. Hence, investment reduces the technological gap between miners.

\subsection{Investment and Mining Equilibrium}\label{secEffectBeta}

In this section, we study how the mining equilibrium changes with low levels of investment. Miners make small investments if the adjustment cost is sufficiently convex, i.e., if $\eta$ is large. % (see (\ref{opt_beta})).
Specifically, it can be seen from (\ref{opt_beta}) {that $\beta^*$ is decreasing in $\eta$. 
{It follows that} if $\eta$ is sufficiently large, the set of active miners remains the same after investment, i.e., $A(\beta^*)=A(0)$. 

{We use $\bar I:=\sum_{i=1}^nI_i$ to denote the aggregate cost reduction for all active miners, where $I_i$ is given by} (\ref{Ieq}), {and
	we denote by $I_{-i}:=\bar I-I_i$ the cost reduction of all miners except $i$. } %Also recall from (\ref{pi_i}) and (\ref{cn}) that $\tilde c_i$ is the initial cost of miner $i$, and 
{We also recall that $c^{(n)}(0)$ is the cumulative cost-per-hash of all active miners without any investment.}

\medskip 

The following propositions provide approximate formulas for how hash rate levels, hash rate shares, and profits change with investment. {The accuracy of these} approximations is theoretically guaranteed when the adjustment costs are large. In Appendix~\ref{secNumInvest}, we verify numerically that they remain accurate for small adjustment costs.

%In Section \ref{secNumInvest}, we visualize the effect of investment on the mining equilibrium, and test the accuracy of the approximation results in Section \ref{secEffectBeta}.

%We begin by analyzing how the equilibrium aggregate hash rate $H^*$ changes with investment $\beta$. 

\medskip 

\begin{prop}\label{propDeltaH} 
	The aggregate hash rate in equilibrium, with and without investment, satisfies\footnote{We use $O(x)$ to denote a function $h(x)$ such that $\limsup_{x\to\infty}|h(x)|/x<\infty$. It follows from (\ref{opt_beta}) that the error term can be equivalently written as $O({1}/{\eta^2})$.}
	\begin{align}
	H^*(\beta^*) &=  H^*(0) %+ \frac{H^*(0)}{c^{(n)}(0) + 2\gamma H^*(0)}\bar I 
	+ a H^*(0)\bar I + O(\bar I^2).
	\end{align} 
	The {coefficient} $a>0$ admits an explicit expression given in (\ref{Hstar}). Furthermore, $a$ is increasing in $c^{(n)}(0)$, and decreasing in $\gamma$. %, and admits an explicit expression given in (\ref{Hstar}). 
\end{prop} 

\medskip 

The proposition shows that the investment of each miner increases the aggregate hash rate, and that the relative increase $H^*(\beta^*)/H^*(0)$ is larger if miners are collectively less efficient to begin with, i.e., if $c^{(n)}(0)$ is large.
The reason is that the reduction in mining costs is increasing in the miners' initial costs; thus, investment has a stronger impact if the initial costs are large. {Moreover, investment has a greater impact {on the aggregate hash rate} if $\gamma$ is small, because less capacity constrained miners are able to increase their hash rates more}.

\medskip 

\begin{prop}\label{propDeltah}
	The relation between the hash rate of miner $i$, with and without investment, is as follows,
	\begin{align} 
	h_i^*(\beta^*) &= h_{i}^*(0) + a_{i}I_i + a_{-i}\bar I_{-i} + O(\bar I^2),
	\end{align}
	where the {coefficients} $a_i$ and $a_{-i}$ are given explicitly in (\ref{a_eq}), and satisfy
	\begin{align}
	a_{i} > 0, \qquad 
	a_{-i} < 0 \;\;\Longleftrightarrow\;\; \frac{h_{i}^*(0)}{H^*(0)} < \half.
	\end{align}
	The equilibrium hash rate share of an active miner $i$, with and without investment, satisfies	
	\begin{align}\label{deltahH}
	\frac{h_i^*(\beta^*)}{ H^*(\beta^*)} &= \frac{h_{i}^*(0)}{H^*(0)} + \alpha_0\big((1-\alpha_i)I_i - \alpha_i \bar I_{-i}\big)  + O(\bar I^2),
	\end{align}
	where the {coefficients} $\alpha_0>0$ and $\alpha_i\in(0,1)$ are given explicitly in (\ref{alpha_eq}). Furthermore, $(1-\alpha_i)I_i - \alpha_i$ is increasing in the initial cost $\tilde c_i$ of miner $i$.
\end{prop} 
\medskip 

The investment of miner $i$ leads to an increase of its own hash rate. However, there is an externality equal to $a_{-i}\bar I_{-i}$ imposed by the investment of other miners {on the hash rate of miner $i$} (see Section \ref{secDeltaC} for a discussion of the condition on $h_{i}^*(0)/H^*(0)$). Since other miners also invest to become more competitive, miner $i$ is unable to capture the full benefit of his investment.
%It is also observed that the sensitivity of miner $i$ is the same to the investment of all other miners. This is because each miner competes against all other miners, and is thus affected in the same way by competing hash rate, irrespective of which miner it comes from.

{In accordance with intuition, the hash rate share of each miner does not change with investment if miners are initially equally efficient.}\footnote{In this case, $\alpha_i=1/n$, and $I_i$ is independent of $i$, so $(1-\alpha_i)I_i - \alpha_i \bar I_{-i}=0$.
	%\begin{align}
	%(1-\alpha_i)I_i - \alpha_i \bar I_{-i}
	%&= \Big(1-\frac{1}{n}\Big)I_i - \frac{1}{n}(n-1) I_{i} = %0.
	%\end{align}
} However, this is no longer the case if there is heterogeneity in the miners' initial costs.
 %\remove{Since the total hash rate share is equal to one,
	%\footnote{The proof of Proposition \ref{propDeltaH} shows that in the first-order approximation (\ref{deltahH}), the hash rate shares sum to one.} 
	%an increase in the hash rate share of a miner implies a decrease in the share of other miners. }
%\remove{How this plays out depends on the initial costs of the miners, and we analyze this further for the important case where the adjustment cost parameter $\eta_i$ is given in} (\ref{eta}).
{It then follows from the proposition that the change in hash rate share is larger for miners with higher initial costs.} Hence, \emph{investment leads to greater decentralization}, i.e., the hash rate share of smaller miners increases while that of larger miners decrease. {The mining capacity constraint plays a key role in producing this effect.} Because smaller miners are less capacity constrained to begin with, the efficiency gains from their investment allows them to overcome the negative externalities imposed by the investment of other miners. As the capacity constraint becomes less binding, i.e., as $\gamma$ gets smaller, this advantage is reduced and minimized in the limiting case of $\gamma=0$, i.e., when all miners have unbounded capacity.
%, and in the limiting case $\gamma=0$ the hash rate shares of miners are unaffected by investment. In other words, even though investment leads to a larger increase in the efficiency of smaller miners (see (\ref{opt_beta})), they are not able to capitalize on it by obtaining a larger hash rate share.{\Red This is not true.}

\medskip 

Next, we analyze the impact of investment on mining profits. We denote by $\pi_i^*(\beta)$ the profit of miner $i$ in the mining equilibrium corresponding to the investment profile $\beta$.

\medskip 

\begin{prop}\label{propDeltaPi}
	{The relation between the profit of miner $i$, with and without investment, is as follows,}
	\begin{align}\label{pii}
	%\frac{\pi^*_i(\beta^*) - \pi^*_{i}(0)}{h^*_{i}(0)}
	{\pi^*_i(\beta^*)}
	&= \pi^*_{i}(0) + h_i^*(0)\big(b_i I_i + b_{-i} \bar I_{-i}\big) + O(\bar I^2),
	\end{align}
	where the {coefficients} {$b_i>0$} and $b_{-i}<0$ are given {explicitly} in (\ref{b_eq}). Furthermore, $b_i I_i + b_{-i} \bar I_{-i}$ is increasing in the initial cost $\tilde c_i$ of miner $i$, and negative for small enough values of $\tilde c_i$. 
\end{prop}

\medskip 

%\remove{While miner $i$ benefits from his own investment, he suffers from negative externalities imposed by the investment of other miners. Those miners are able to extract a larger share of the mining reward by increasing their mining efficiency, which cuts into the profit of miner $i$.}
Investment impacts the profit of a miner similarly to how it affects its hash rates. In Proposition \ref{propDeltah}, we argue that smaller miners are able to gain hash rate share at the expense of larger miners. Proposition~\ref{propDeltaPi} {additionally states that smaller miners} are those who increase their profits the most, {\it while the profits of larger miners can even decrease}.
These theoretical findings are supported by the global cryptocurrency studies discussed in Remark \ref{remEta}, where it is reported that the primary concern of large miners is earning lower profits due to increased competition. % (see Remark \ref{remEta}).

Next, we study welfare in the mining ecosystem, defined as the sum of all miners' profits. We denote by $\Pi^*(\beta):=\sum_{i=1}^n\pi_i^*(\beta)$ the aggregate profit in the mining equilibrium corresponding to the investment profile $\beta$. As we show in the next proposition, welfare increases with investment in a homogeneous system. 

\medskip 

\begin{prop}\label{cor1}
	If miners are homogeneous in their costs, then the relation between the aggregate profit of all miners, with and without investment, is as follows,
	%\begin{align}\label{deltaPi}
	%\pi^*_i(\beta^*) = \pi^*_{i}(0) +  {h^*_{i}(0)}\Big(1- \frac{c^{(n)}(0) + \gamma H^*(0)}{c^{(n)}(0)+2\gamma H^*(0)}\Big)\bar I + O(\bar I^2) > 0.
	%\end{align}
	\begin{align}\label{deltaPi}
	\Pi^*(\beta^*) = \Pi^*(0) + bH^*(0)\bar I + O(\bar I^2), 
	%\pi^*_i(\beta^*) = \pi^*_{i}(0) +  b\,{h^*_{i}(0)}\bar I + O(\bar I^2),
	\end{align}
	where the {coefficient} $b>0$ is given explicitly in (\ref{b}). Furthermore, $b$ converges to zero as $\gamma\to 0$.
	%$b$ is increasing in $\gamma$, decreasing in $n$, and converges to zero as $\gamma\to 0$ or $n\to\infty$.
\end{prop}

\medskip 

The above proposition implies that more efficient mining allows capacity constrained miners to reap the benefit of their investment. However, as mining capacity grows unbounded, i.e., as $\gamma\to 0$, the profit becomes independent of investment. That is, with unbounded capacity, the benefits of a higher mining efficiency are exactly offset by the costs of a higher aggregate hash rate. % each miner applying a higher hash rate. %a higher hash rate applied by all miners.

%We also observe that the marginal profit becomes smaller\note{Why the marginal profit?} as the number of active miners increases. This is consistent with the classical rent-seeking model of \cite{Tullock}, where the profit of each miner is driven to zero by increased competition (see also the discussion at end of Section \ref{secH}).

If miners are heterogeneous in their initial hashing costs, the aggregate profit may either increase or decrease with investment. 
If there is little heterogeneity, then the profit increases, as in the fully homogeneous case discussed above. However, if there is significant heterogeneity, the profit may decrease. The intuition is that while smaller miners are able to increase their profits, the scale of their operations is smaller. Hence, the increase in their profits does not fully compensate for the profit reduction of larger miners - such miners do not benefit as much from investment, either because they are already very efficient in their mining, or because they are too capacity constrained. Formally, this can be seen from the profit formula (\ref{pii}), where small and negative values of the term $b_i I_i + b_{-i} \bar I_{-i}$ are associated with smaller values of $\tilde c_i$, and thus multiplied by larger values of $h_i^*(0)$.

\section{Mining Centralization and Network Security}\label{secParams}%\label{secNumerical}

{In this section, we construct measures of mining centralization and network security. We then evaluate these measures at the mining equilibrium, and discuss how they depend on investment, mining reward, and mining capacity.}

We begin by estimating the model parameters based on statistics from the Bitcoin network and using the model's equilibrium solution.

\paragraph{(i) Mining reward:} We set $R=\$20\times 10^6$, which is the average daily Bitcoin mining reward in the period from January 1st, 2020, to April 30th, 2021 (see Fig.\ \ref{fig_reward_price}). %01/01/20 and 04/30/21 (see Fig.\ \ref{fig_reward_price}).

\paragraph{(ii) System hash rate:} We estimate $H$ by taking the average Bitcoin network hash rate in the period from January 1st, 2020, to April 30th, 2021. This yields $H=120$ million TH/s (see Fig.\ \ref{fig_reward_price}).
%\note{similarly in what sense? As an average? Please check if my fix is correct.}

%which is based on the number of terahashes per second performed by the Bitcoin network per day in the past year.\footnote{See: \url{https://www.blockchain.com/charts/hash-rate}} ``Tera'' denotes multiplication by one trillion, or $10^{12}$.

\paragraph{(iii) Number of miners:} We base our estimate on the Global Cryptocurrency Benchmarking Study (\cite{HilemanRauchs}), where 11 of the participants are designated as large mining organizations, and estimated to cover over 50\% of the total professional mining sector in terms of hash rate. Hence, we approximate the number of miners to $N=20$.\footnote{{In practice, the number of miners is not directly observable. In the 2nd Global Cryptoasset Benchmarking Study} (\cite{Rauchs}), {it is mentioned that while there are at least several dozen operators of facilities around the globe, a small number of large hashers have a dominant position. We report our results for $N=20$, but also carried out the analysis for a range of values, including $N=10$ and $N=50$, verifying that our results are not sensitive to the exact value of $N$.}}

\paragraph{(iv) Mining costs:} We measure the cost-per-hash of a single miner as the cost of generating one million TH/s for one day (24 hours).  %producing 110 TH/s while consuming 3.25 kW,
	We base our estimates on the power consumption of AntMiner S19 Pro, which is Bitmain's latest model, released in May 2020. As of April 2021, it is the most energy efficient commercially available mining hardware, together with MicroBT's WhatsMiner M30S$++$. AntMiner S19 Pro has an energy efficiency of 29.5 J/TH,
	%Since 1 watt is equal to 1 joule per second, this means that the energy efficiency is 3250/110 = 29.5 J/Th.
	and {we assume} an electricity cost of \$0.05 per kWh. The latter is a standard estimate of the average electricity cost incurred by large miners.\footnote{For example, the Cambridge Bitcoin Electricity Consumption Index uses this cost estimate ``based on in-depth conversations with miners worldwide and to be consistent with estimates used in previous research'' (see: \url{https://cbeci.org/cbeci/methodology}).} % (one kW for one hour), 
	Altogether, this yields a cost of %$3.25/110\times 0.05\times 24=\$0.0355$
	$0.0295\times 0.05\times 24=\$0.0355$ per TH/s for one day. We set the cost of the most efficient miner to this value, i.e., $c_1 = 0.0355$. 
	
	Recall from Section \ref{secNumMiners} that miner $i$ exerts a positive hash rate in equilibrium if its cost $c_i$ is lower than the break-even cost $R/H^*$. 
%Recall that the reward-per-hash in equilibrium is $R/H^*$. Miner $i$ exerts a positive hash rate if $c_i<R/H^*$, and it exerts zero hash rate if $c_i=R/H^*$.
We set the mining costs $(c_i)_{1\leq i\leq N}$ as $N$ evenly spaced points between $c_1$, as estimated above, and $R/H$, where $H$ is estimated as in (ii). 

%AntMiner S9 SE was a previously dominant hardware, released in July 2019. It's efficiency is 80 J/TH, yielding cost-per-hash $c_i=\$0.096\time 10^6$. 
%$c_i=0.08\times 0.05\times 24=\$0.096$. 

%AntMiner S19 Pro: 110 TH/s, 3.25 kW, 29.5 J/Th
%AntMiner S19    :  95 TH/s, 3.25 kW, 34.5 J/Th
%AntMiner S19j   :  90 TH/s, 3.10 kW, 34.5 J/Th (3.25?)
%AntMiner T10    :  84 TH/s, 3.15 kW, 37.5 J/Th (3.15?)

%AntMiner S9 SE  :  16 TH/s, 1.28 kW, 80.0 J/Th 
%(dominant machine in 2018 according to Humoud)

%\footnote{See: \url{https://shop.bitmain.com/}. Since 1 watt is equal to 1 joule per second, this means that the energy efficiency is 3250/110 = 29.5 J/Th. Antminer S19 Pro is the latest model and costs \$3769. Corresponding numbers for Antminer S19, which costs \$2767, are 95 TH/s and 3.25 kW, yielding 34.5 J/Th.} 

\paragraph{(v) Capacity constraint:} 
We invert the formula for the equilibrium hash rate $H^*$, given in equation (\ref{optH}), and imply the parameter value
\begin{align}
\gamma = \frac{1}{H^*}\Big(\frac{(n-1)R}{H^*}-c^{(n)}\Big).
\end{align}
We set $H^*$ to the value estimated in (ii). If all miners are active, i.e., $n=N=20$, then %$\gamma=0.0041\times 10^6$ and 
$\gamma=0.0095\times 10^6$. %, respectively.
%By definition, this value of $\gamma$ results in equilibrium hash rate $H^*=H$, %and, 
%\remove{Together with the mining costs specified in (iv), this value of $\gamma$ implies that all miners are active in equilibrium, and the marginal miner exerts zero hash rate.}

\paragraph{(vi) Adjustment costs:} 

Recall that the adjustment costs are of the form (\ref{eta}), and $\eta\geq 0$ governs how much miners are able to increase their efficiency. It follows from formula (\ref{Ieq}) that %for miner $i$, 
\begin{align}\label{eta1}
c_i(\beta^*_i)= \tilde c_{i} - \frac{\tilde c_{i}-\tilde c_0}{2} \quad\Longleftrightarrow\quad \eta=1.
\end{align}
That is, less than half (more than half) of the efficiency gap $\tilde c_{i}-\tilde c_0$ is bridged by investment if $\eta>1$ ($\eta<1$). A reasonable lower bound for the value of $\eta$ is therefore given by $\eta=1$, as in that case $\beta_i^*=1$ and miner $i$ upgrades its entire stock of hardware. 
%Furthermore, we show that the order of the miners in terms of cost-per-hash is unchanged with investment, 
%\begin{align}\label{eta2}
%c_i(\beta_i^*) \leq c_i(\beta_i^*).
%\end{align} 

We next develop and analyze the measure of mining centralization and network security. 
%\subsection{Mining Centralization and Network Security\note{(i) A lot of changes (ii) Need to change this title or the title of Section 5}}\label{secmincentr} 
%\note{To be done: Please, remove the approximation line from the plots in this section. It is enough to have the section in the Appendix dedicated to evaluate the accuracy of this approximations. Here just report the exact formulas. Also, please add the part on network security currently in the note environment. Also, discuss the practical implications of this network security measure in more detail by refering to the MIT report and empirical facts.}
For Bitcoin and other proof-of-work cryptocurrencies, the security of the network depends on the \emph{distribution of hashrate} among miners, i.e., the level of decentralization. To quantify the implications of investment on mining centralization, we defining the following function, 
	\begin{align}\label{CDF}
	F(k) := \left\{\begin{array}{ll} 0, &\; k=0, \\
	\sum_{i=1}^k \frac{h_i^*}{H^*}, &\; k=1,2\ldots,n, \\
	1, &\; k=n. \end{array} \right. 
	\end{align}
	It follows directly from the definition that $F(k)$ is the total hash rate {share} of the $k$ largest miners. 
	%Then, for a given $0<\alpha<1$,
	%\begin{align}
	%n^{(\alpha)} := \min\big\{1\leq k\leq n:f(k)>\alpha\big\}
	%\end{align}
	%is the smallest number of miners whose combined hash rate share exceeds $100\times\alpha\%$ of the aggregate hash rate. {For example, if $\alpha=0.5$,  the largest $n^{(\alpha)}$ miners control more than half of the overall hash rate.}
	%\note{$n^{(\alpha)}$ is introduced but never used. Either we remove it, or we use it. A possibility would be to present a table with $\alpha=10\%,\alpha=20\%, \alpha=30\%, \alpha=40\%$, and $\alpha=50\%$, and for each value of $\alpha$ report the corresponding $n^{(\alpha)}$.} 
	The domain of the function $F$ can be extended to non-integer values via linear interpolation. The resulting function is increasing and concave, and degenerates to a straight line if and only if miners are homogeneous and thus all exert the same hash rate. The steeper the function $F$ is for small values of $k$, the greater the extent of mining centralization. 
	
	It is evident from the left panel of Figure \ref{fig_CDF} that $F$ is closer to a straight line if the equilibrium hash rates account for investment, and more so if adjustment costs are smaller. This confirms {the implications of} our model {highlighted in Section} \ref{secEffectBeta}, i.e., that investment steers the mining system towards decentralization - a larger number of miners controls any given fraction of the overall hash rate. 
	
	The right panel of Figure \ref{fig_CDF} shows that a higher mining reward {also} leads to greater decentralization, consistent with our analysis in Section \ref{secDeltaRgamma}. This implies that if the market capitalization of a coin increases, mining {would} not become increasingly centralized, i.e., exhibit ``the rich getting richer'' phenomenon. The rationale offered by our model is that existing mining facilities are not able to increase their hash rate indefinitely, which allows smaller miners to grow and new miners to enter. This is consistent with the fact that Bitcoin mining seems to be less concentrated, both geographically and in terms of hash rate ownership, than commonly indicated by public discourse (see \cite{Rauchs}). 
	 
	\medskip 
	\begin{rem}\label{ghash}
		As the Bitcoin network has grown larger, the cost of getting anywhere close to the 51\% attack threshold %the cost for coming anywhere close to the 51\% attack threshold %and holding it persistently 
		has become out of reach for most entities. 
		In addition to the network consuming as much electricity as a small country, % the size of Malaysia or Sweden, 
		the sheer amount of mining hardware required, which is in short supply, is all but impossible to obtain. This is consistent with our model, where a miner with \emph{zero cost-per-hash} still does not manage to dominate the mining competition, because of limited capacity. 
		
		In practice, a small number of large mining pools commonly accounts for over 50\% of the Bitcoin network hash rate\footnote{The market share of the most popular Bitcoin mining pools can be viewed at \url{https://www.blockchain.com/charts/pools}.}, but this is unlikely to lead to an attack. 
		Even if a mining pool operator with malicious intent were to successfully implement a 51\% attack, such an event would be discovered quickly, and be self-destructive for the mining pool because miners, whose business model is tied to the success of Bitcoin, would quickly switch to other mining pools.\footnote{{Additionally, the ability of mining pools to control the hash rate directed to them has been significantly reduced by developments such as \emph{Stratum V2}, which enables miners to choose which transactions to include in blocks, rather than mining a block proposed by the pool.}}   In fact, the mining pool Ghash.io came close to the 51\% threshold in 2014, seemingly with no intent to attack, but this still prompted miners to voice their concerns and leave the mining pool.  \hfill\qed 
	\end{rem} 
	\medskip

	The security of proof-of-work cryptocurrencies relies not only on the network hash rate being distributed among multiple miners, but also on how expensive it is to gain control of a large hash rate share. 
	In fact, the main benefit of a \emph{high network hash rate} is that it makes the network more secure from attacks. We have seen that technological advancements and investment lead to a higher network hash rate (see Table \ref{table0}). However, more efficient hardware also leads to lower cost of hashing. To study the net effect of investment on the cost of attacks, observe that $C_k:=\sum_{i=1}^k(c_ih_i^*+\frac{\gamma}{2}(h_i^*)^2)$ is the cost of capturing a fraction $p_k := \sum_{i=1}^kh^*_i/{H^*}$ of the network hash rate. {We then define} the {cost-of-attack} function $TC:[0,1]\mapsto[0,\infty)$ by 
	\begin{align}\label{TC}
		%TC(p) = \sum_{i=1}^k\Big(c_ih_i^*+\frac{\gamma}{2}(h_i^*)^2\Big) \quad\textrm{if}\quad  p=\frac{\sum_{i=1}^kh^*_i}{H^*},
		{TC(p_k) = C_k, \quad 1\leq k\leq n,}
	\end{align}
	%for $k=1,\dots,n$, 
	and by linear interpolation for $p\notin\{p_k,\; 1\leq k\leq n\}$.
 The left panel of Figure \ref{fig7} shows that investment has a small impact on this function. However, the right panel shows that the value of the mining reward has a significant impact. Intuitively, this is because a higher mining reward leads to a higher hash rate, just like investment does, but the cost of mining stays the same, resulting in larger values of the function $TC$. 

	Since a smaller mining reward goes hand in hand with a smaller hash rate, this supports the common belief that coins with low hash rates are susceptible to cheap 51\% attacks - only a small number of miners from larger coins need to switch to a smaller coin in order to control 51\% of the smaller coin's network hash rate (see also Remark \ref{remCloud}).
	In particular, this means that emerging coins with low market capitalization are less secure. {It is also worth noting that concentration of hash rate below the 51\% barrier also presents danger to the network, for example due to selfish mining attacks} (see, e.g., \cite{Eyal}).
	
	The effect of mining reward on the cost of attacks also has implications for more mature cryptocurrencies like Bitcoin.
	%Currently, the vast majority of the cost per transaction is in the form of block rewards, which is a form of inflation that doesn't affect the user directly, and instead affects the whole network.
	To see that, first note that transaction fees and block rewards represent the \emph{security spending} of the network, which is paid for by users of the network, through transaction fees, and by holders of coins, through inflationary block rewards. %and processing budget for the store of value and payment settlement network. 
	%Figure \ref{fig8} shows that %he reward paid to miners as a fraction of the market capitalization of Bitcoin. 
	%as Bitcoin's market capitalization has grown, the security spending has grown as well, but the percentage of the market capitalization spent on security has declined. Observe that there is a noticeable drop in this percentage around the Bitcoin halving events in November 2012, July 2016, and May 2020, when the reward for mining a block was halved. 
	In principle, the {security spending} of the network should be large enough in absolute terms to deter attacks, i.e., make them expensive, but also large enough as a percentage of the market capitalization
	%The traditional view is that security spending needs to remain proportional to the market capitalization of the network, %asset being stored, 
	{because the damage of a successful attack varies with the market value of Bitcoin.} In coming years, Bitcoin will continue its gradual shift from paying miners primarily through block reward to paying miners primarily through transaction fees. %Hence, %with diminishing block rewards, 
	%while not so large as to make normal settlement transactions uneconomic due to needlessly high fees. 
	%In other words, 
	%With diminishing block rewards, 
	%Bitcoin needs to %sustain a sufficient market capitalization, and 
	%develop a persistent fee market structure to maintain security.
	%In coming years, Bitcoin will continue its gradual shift from paying miners primarily through block reward to paying miners primarily through transaction fees. 
	%while not so large as to make normal settlement transactions uneconomic due to needlessly high fees. 
	%In other words, 
	%Figure \ref{fig8} shows that 
	With diminishing block rewards, Bitcoin therefore needs to %sustain a sufficient market capitalization, and 
	develop a persistent fee market structure to sustain the mining reward and maintain security (see Figure \ref{fig8}). Whether that will happen % largely depends on Bitcoin's eventual level of adoption. That is, 
	%\remove{i.e., the economic feasibility of network security through fees,} 
	largely depends on the eventual level of Bitcoin adoption.  %is based on sufficient
	%sufficient long-term adoption of Bitcoin. 
	If fees are not sufficient to sustain the mining reward, our results indicate that the network becomes vulnerable to attacks. This rhymes with one of Nakamoto's original propositions regarding the future of Bitcoin: 
	%Ultimately, since security through fees is economically workable based on achieving sufficient long-term adoption, it boils down to one of Satoshi's original propositions that Bitcoin will have a rather binary outcome: 
	``When the reward gets too small, the transaction fee will become the main compensation for nodes. I'm sure that in 20 years there will either be very large transaction volume or no volume''.\footnote{Posted by ``satoshi'' on the Bitcointalk forum on February 14, 2010 (see: \url{https://bitcointalk.org/index.php?topic=48.msg329#msg329}).}
	%transition to a fee-based reward system
	%This shows how transaction fees need to adapt in order to maintain this proportion. Assuming that market capitalization does not die down, this shows how fees need to grow. 
	{An alternative view is that security spending should be proportional to transactional volume, i.e., a \emph{flow} variable rather than a \emph{stock} variable} (\cite{Budish}). {This is based on the notion that 51\% attacks can only endanger the last few blocks appended to the blockchain, so the reward from such an attack is proportional to the value of recent transactions. 
	In this case, the same principle applies, i.e., that transaction fees have to grow to sustain the mining reward, as the network transitions to a fee-based reward system.}
	
	%In principle, the security spending of the network should be large enough in absolute terms to deter attacks, i.e., make attacks expensive, but also large enough as a fraction of the market capitalization.
	%make attacks uneconomical - if mining revenue is too small relative to the market capitalization, double-spending attacks become more viable.\note{Let us discuss this last sentence} 
	%Figure \ref{fig8} shows that security spending was large in the early days relative to market capitalization, which is expected for emerging coins. Since then it has tapered off and was largely constant between the 2016 and 2020 halving events. %, to be a large percentage of market capitalization in the early days of a coin, and taper off with growing adoption and increased market capitalization.

		\medskip 
	\begin{rem}\label{remCloud}
		The function $TC$ in (\ref{TC}) {gives the cost, for existing miners,  of generating a given share of the network hash rate. In addition to this cost,} rational self-interest can be considered a backup defense for attacks. Miners invest large amounts of capital into mining facilities whose profitability is tied to the success of the network. %their rigs and generally own a lot of coins; 
		Hence, even if {miners} were to achieve a successful attack, % on bitcoin and threaten the security of the system, 
		it would likely damage the market capitalization of the network, and thus their own income and net worth. %resulting in a reduction in their income and net worth, even if they were able to steal some coins in the attack.
		
		However, attacks from entities outside of the network have become of increased concern with the emergence of cloud mining services that offer customers the possibility to participate in mining without having to own hardware themselves. That is, the attacker is not exposed to the opportunity cost of future revenue generated by the hardware. 
		For many smaller coins, there is hash rate available to rent that is orders of magnitude larger than their network hash rates. %\footnote{Cloud mining also makes attacks more viable because they are likely to hurt the market capitalization of the coin, but the attacker is not concerned with the opportunity cost of future revenue generated by the underlying hardware.}
		As part of the \emph{Digital Currency Initiative}, launched by the MIT Media Lab, a system was constructed to actively monitor proof-of-work cryptocurrencies with the goal of detecting 51\% attacks. Since launching the system in June 2019, over 40 successful attacks have been detected, with evidence that hash rate rental markets have been used to perform some of them.\footnote{We refer to \url{https://dci.mit.edu/51-attacks} for further details.} \hfill\qed
		%This is of increased concern with the emergence of cloud mining services that offer the possibility to participate in mining without having to own hardware themselves. Namely, besides the consensus network protocol, rational self-interest is the backup defense for a 51\% attack - miners invest large amounts of capital in mining facilities and generally own coins themselves. Hence, even a successful double-spending attack would likely be detrimental to the attacker because of damage to the market capitalization of the network. %, and likely reduce their income and net worth. 
		%Miners invest a ton of capital into their rigs and generally own a lot of coins; if they were to achieve a successful 51\% attack on bitcoin and threaten the security of the system, it would likely damage the market capitalization of the network, resulting in a reduction in their income and net worth, even if they were able to steal some coins in the attack.
		%However, cloud mining essentially reduces this cost for an outside attacker to zero, as it %only needs to rent hash rate for the duration of the attack, without the 
		%is not concerned with the opportunity cost of future revenue generated by the underlying hardware. %A large number of Proof-of-Work altcoins have many multiples of their network hashrate available to rent, leading to a number of high-value attacks in the wild.
		%This is known to have resulted in a large number of high-value attacks. For example, 
	\end{rem} 
	\medskip

	\begin{figure}[ht!]
	\centering
	%\hspace{-1 cm} %[width=20.0cm,height=6.0cm]
	\includegraphics[width=\textwidth]{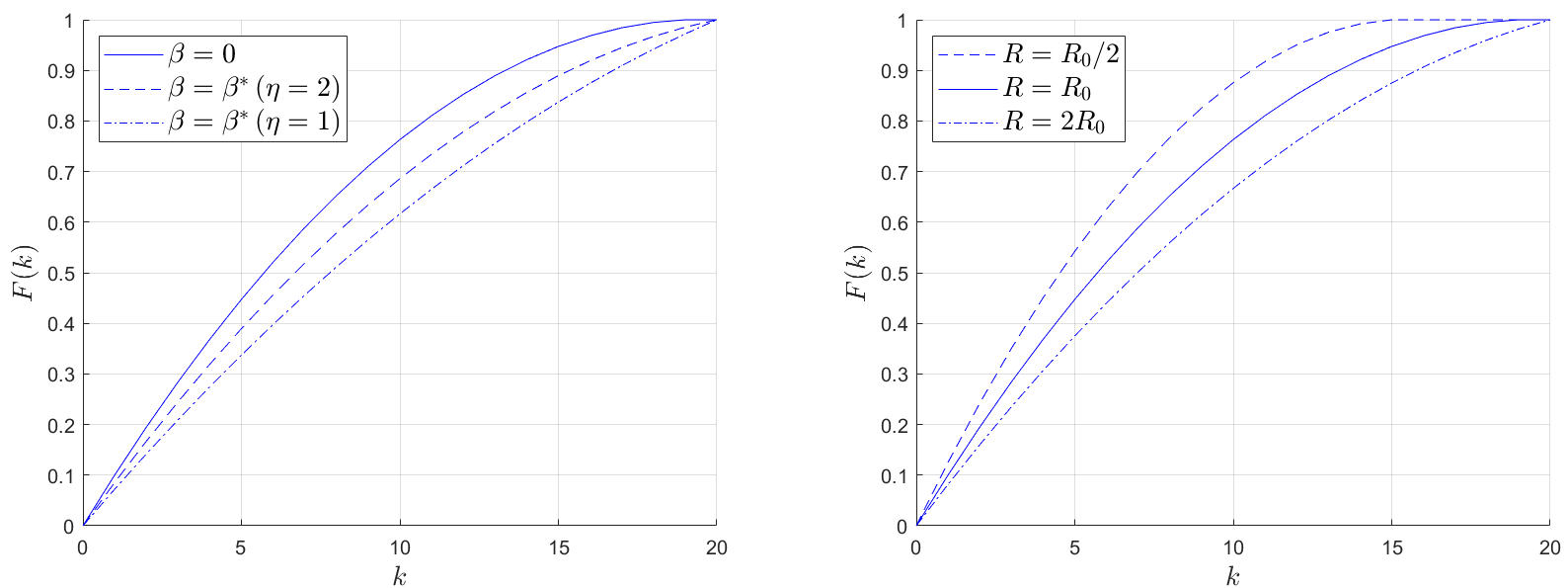}
	\caption{\small Plot of the hash rate function in (\ref{CDF}). Left panel: {Effect of investment on mining centralization.} The solid line correspond to the mining equilibrium without investment ($\beta=0$), and the dashed and dash-dotted lines correspond to mining equilibria with investment ($\beta=\beta^*$). % and $\eta=2$ and $\eta=1$, respectively. %and dashed blue lines correspond, respectively, to mining equilibria with investment ($\beta=\beta^*$) and without investment ($\beta=0$). %The red dashed line corresponds to the approximation from Proposition \ref{propDeltah}. 
		We set the model parameters as described in Section \ref{secParams}. %We set $\eta=2$ in the left panel, and $\eta=1$ in the right panel.
		Right panel: {Effect of mining reward on mining centralization.} The model parameters are chosen as described in Section \ref{secParams}, with $R_0=20\times 10^6$ as in part (i) therein.
	}\label{fig_CDF}
\end{figure} 

\begin{figure}[ht!]
	\centering
	%\hspace{-1 cm} %[width=20.0cm,height=6.0cm]
	\includegraphics[width=\textwidth]{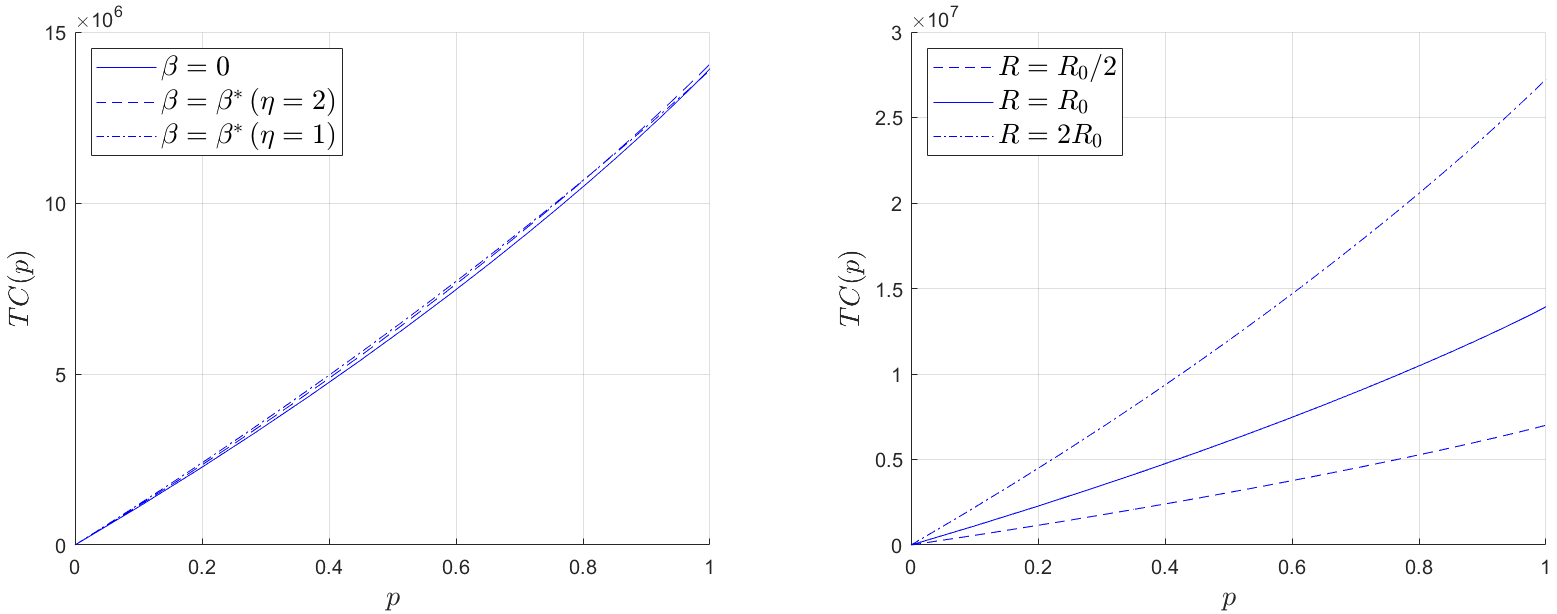}
	\caption{\small Plot of the total cost function in (\ref{TC}). Left panel: {Effect of investment on the cost of attacks}. Right panel: {Effect of mining reward on the cost of attacks}. The model parameters are chosen as in Figure \ref{fig_CDF}.  %The function $TC(p)$ in (\ref{TC}), with investment ($\beta=\beta^*$) and without investment ($\beta=0$). We set the parameter values as  described in Section \ref{secParams}, with $\eta=2$. Right: The function $TC(p)$ for different values of the mining reward. We set the parameter values, including the value of $R_0$, as described in Section \ref{secParams}.
	}\label{fig7}
\end{figure}

	\section{Empirical Analysis}\label{secEmpirical}
	
	In this section, we provide empirical support for the most significant equilibrium implications of our model. In Section \ref{secEDA}, we present empirical patterns of the Bitcoin mining network. In Section \ref{secEmpRH}, we provide statistical evidence for the predicted relationship between mining reward and the aggregate hash rate. All data used in this section is obtained from \url{https://www.blockchain.com/}.
	
	\subsection{Empirical Patterns of Cryptocurrency Mining}\label{secEDA}
	
	The mining reward, i.e., the revenue from mining, consists of coinbase block rewards plus transaction fees paid to miners. These fees are voluntarily attached by users to increase the probability that their transactions are included in mined blocks appended to the ledger. It is evident from Figure \ref{fig_reward_price} that transaction fees are a small component of the mining reward, and that there is a strong positive co-movement between the mining reward and the price of Bitcoin. We also observe a positive relation between the value of transaction fees and the price of Bitcoin. Intuitively, a higher price is the result of increasing demand for Bitcoin, which in turn implies a larger number of transactions. Hence, higher fees are required to incentivize miners to include a given transaction in a block. 

The evidence provided above indicates that transaction fees have so far had a limited impact on the incentives of miners, and that both coinbase block rewards and fees are primarily driven by a common factor - the price of Bitcoin. 
%indicates that there is not a need to consider separately the effect of coinbase block rewards and transaction fees on the incentives of miners, as the latter is a significantly smaller component, and both components are primarily driven by a common factor - the price of Bitcoin. 
 A notable exception is the dramatic increase in transaction fees observed in December 2017.\footnote{This surge was in part due to an increased demand for Bitcoin. However, during this period the Bitcoin network was also flooded with spam transactions, widely believed to be the result of a coordinated attack.} % (i.e., not real-world transactions). 
%To this day, the source of this event is unclear.} %https://www.ft.com/content/abd7ba0b-40ef-3876-857c-cd5d49ec557e
The subsequent decline in fees is largely attributed to increased adoption of the so-called SegWit update to the Bitcoin protocol, % called Segregated Witness (SegWit), 
which effectively acted as a block-size increase and thus allowed more transactions to be processed per block. %\add{As discussed in Section} \ref{secParams}, \add{the proof-of-work protocol will eventually rely on transaction fees as reward from mining. 
While Bitcoin has been in existence for over a decade, and around 90\% of its 21 million coins have already been mined, transaction fees still constitute a small component of mining revenues. As pointed out in} \cite{Easley}, the {growing market capitalization of a coin can delay the point where transaction fees take over, and this has been the case for Bitcoin.}

%\note{Our model does not say anything about a decomposition between mining reward and transaction fees. Why are we decomposing into the two? Can this lead to questions?}
%https://arstechnica.com/tech-policy/2018/02/bitcoins-transaction-fee-crisis-is-over-for-now/?comments=1
%The problem is that users had to modify their bitcoin software to use a new, more efficient transaction format.

%\footnote{\cite{Easley} model the behavior of miners and Bitcoin users and show that transaction fees can only play a secondary role in the willingness of miners to participate, but play a bigger role in affecting the participation of users. Indeed, they demonstrate how equilibrium transaction fees evolve in the Bitcoin ecosystem, with the arrival rate of potential transactions being the key determinant. If the arrival rate of potential transactions is low, transactions without fees attached are written to the blockchain but, as the arrival rate of potential transactions increases, the equilibrium shifts and only transactions with fees attached are posted to the blockchain.}

%Transaction fees and block rewards represent the \emph{security spending} of the network, which is paid for by users of the network, through transaction fees, and by holders of coins, through inflationary block rewards. 
Figure \ref{fig8} shows that as Bitcoin's market capitalization has grown, the security spending (see Section \ref{secParams}) has grown as well, but the percentage of the market capitalization spent on security has declined. Observe that there is a noticeable drop in this percentage around the Bitcoin halving events in November 2012, July 2016, and May 2020, when the reward for mining a block was halved. Figure \ref{fig8} shows that security spending was large in the early days relative to market capitalization, which is expected for emerging coins. Since then it has tapered off and was largely constant between the 2016 and 2020 halving events.

%traditional notion that The security expenditure should vary with value of the asset being stored
%the value of a successful attack varies with the market value of Bitcoin itself
%Rewardsfrom51\% attacks (which are a function of txnvalue) must be offset by high fees to honest miners
%Flow(Budish1): fees must be large relative to transactional volume

\begin{figure}[ht!]
	\centering
	%\hspace{-1 cm} %[width=20.0cm,height=6.0cm]
	\includegraphics[width=\textwidth]{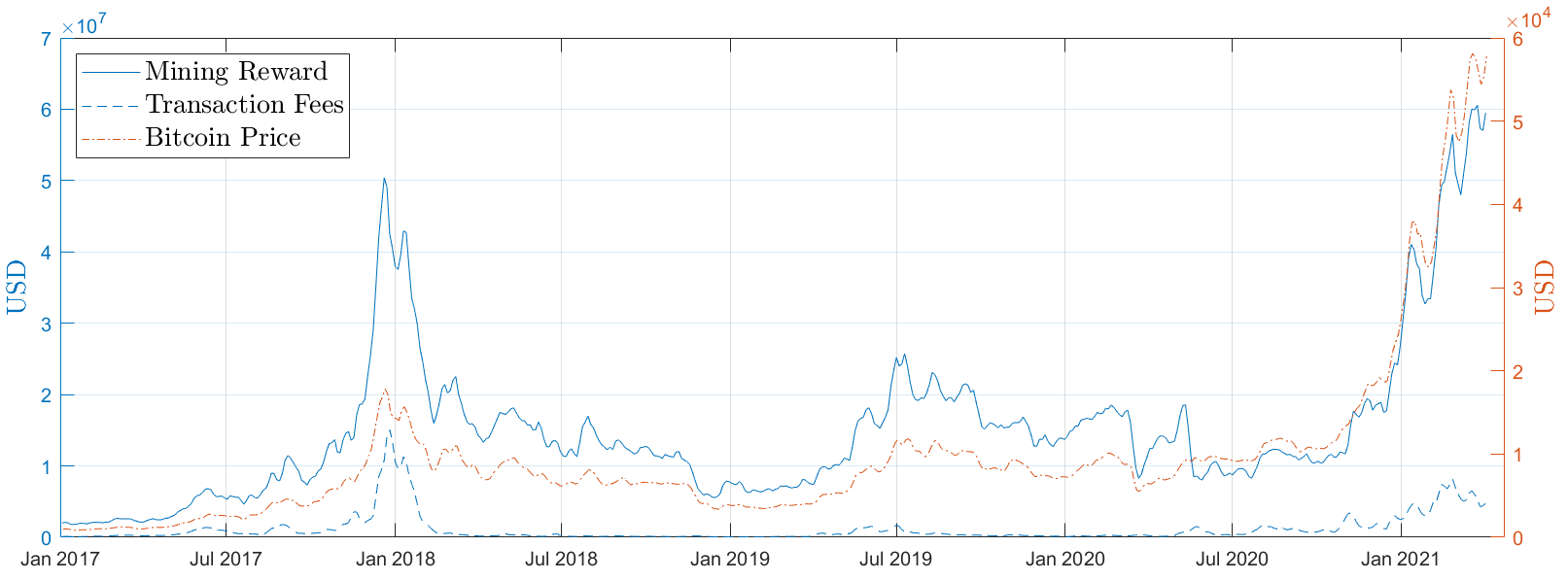}
	\caption{\small Left y-axis: the total reward earned by miners (coinbase block rewards and transaction fees), and the component of total reward due to transaction fees. Right y-axis: price of Bitcoin. The plot shows seven-day averages, computed every three days between January 2017 and April 2021.} \label{fig_reward_price}
\end{figure} 

Figure \ref{fig_hash_price} shows that there exists an overall positive relationship between the system hash rate and the Bitcoin price, but with a few notable discrepancies. For example, in the first half of 2018, there was a significant drop in the price of Bitcoin,\footnote{This drop was due to a large wave of cryptocurrency sell-off, known as the 2018 Cryptocurrency Crash. By the end of November, Bitcoin had fallen by over 80\% from its historical peak.} but the hash rate kept climbing until finally dipping at the end of that year. Importantly, this period was preceded by an unprecedented Bitcoin boom. Such periods of Bitcoin mania are likely to trigger significant investments in mining facilities, which then become operational with a time lag. In this specific case, the effect was to raise the system hash rate during the market drop in 2018. % such investments became operational during the market drop in 2018, and raised the system hash rate.
The same pattern is observed in the second half of 2019, where the Bitcoin price trended downwards while the hash rate kept climbing. Again, this occurred following a rally in the first half of 2019, leading to investments that contributed to {increasing} the hash rate during a time when the Bitcoin price was dropping. The third, and perhaps  best representation of the above pattern, is the extreme surge in the price of Bitcoin during the second half of 2020, which has not yet been associated with an equally large increase in the system hash rate. {This can be explained by the inability of miners} to expand their operations immediately,\footnote{An important limiting factor is the severe shortage of mining hardware. {See footnote} \ref{f1}. } and suggests that the hash rate may keep rising in the second half of 2021, even if the price of Bitcoin starts falling. {The above argument is supported by the 2nd Global Cryptoasset Benchmarking Study} (\cite{Rauchs}). {Based on proprietary data {covering six major proof-of-work cryptocurrencies, including Bitcoin and Ethereum,} it is reported  that ``hashers often cannot immediately increase production when running at full capacity'', resulting in hash rate growth lagging behind market price growth.}

%\note{This is good, but as written the section appears disconnected from the theoretical results and model assumptions in the earlier sections. Can you link the theoretical results to the empirical regularities presented in this section. Also would it makes sense to compute a correlation coefficient to statistically quantfy the correlation between system hash rate and mining reward, and system hash rate and Bitcoin price? Or not?} 

%Note that 2018 cryptocurrency crashhe was a sell-off of cryptocurrencies from Jan 2018. After an unprecedented boom in 2017, the price of Bitcoin fell by about 65 percent during the month from 6 Jan to 6 Feb 2018. Subsequently, nearly all other cryptocurrencies which had also peaked from Dec 2017 through Jan 2018, then followed Bitcoin's crash. By Sep 2018, cryptocurrencies collapsed 80% from their peak in Jan 2018, making the 2018 cryptocurrency crash worse than the Dot-com bubble's 78% collapse. By 26 Nov, Bitcoin also fell by over 80% from its peak, having lost almost one-third of its value in the previous week.

\begin{figure}[ht!]
	\centering
	%\hspace{-1 cm} %[width=20.0cm,height=6.0cm]
	\includegraphics[width=\textwidth]{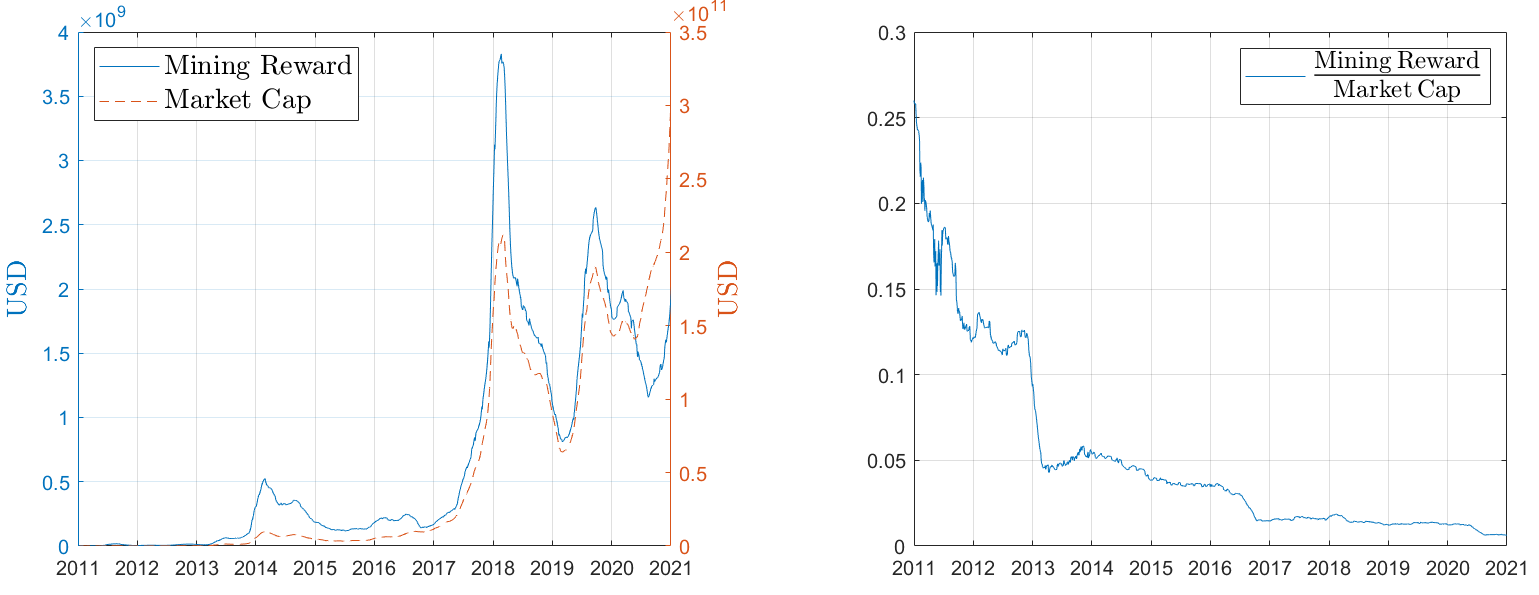}
	\caption{\small Left panel: Solid line plots total mining reward earned by miners, computed as the annualized sum of mining reward over a running three-month window. Dashed line shows the Bitcoin market capitalization, computed as the average market capitalization over a running three-month window. Right panel: Mining reward as a proportion of Bitcoin market capitalization, with both quantities computed as in the left panel.}\label{fig8}
\end{figure} 

\begin{figure}[ht!]
	\centering
	%\hspace{-1 cm} %[width=20.0cm,height=6.0cm]
	\includegraphics[width=\textwidth]{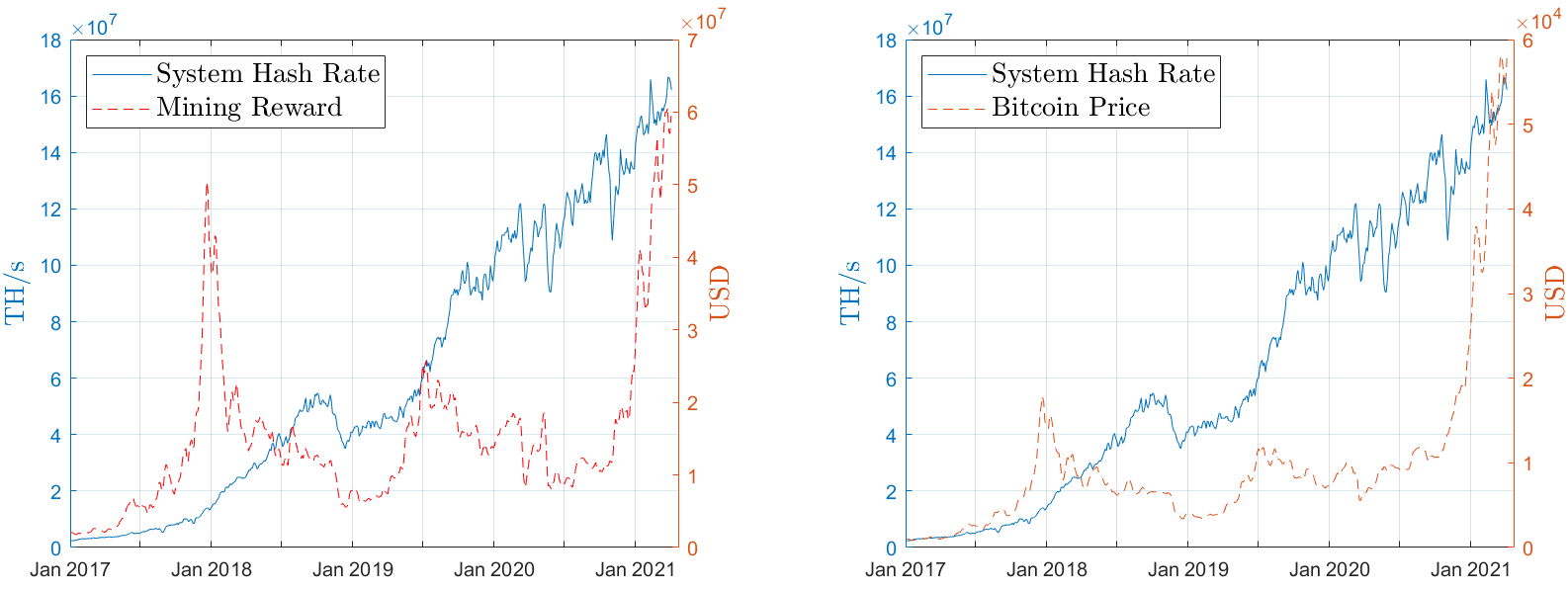}
	\caption{\small Left panel: System hash rate (left y-axis) and total reward earned by miners per day (right y-axis). Right panel: System hash rate (left y-axis) and price of Bitcoin (right y-axis). The plots show seven-day averages, computed every three days between January 2017 and April 2021.}\label{fig_hash_price}
\end{figure}

\subsection{Hash Rate vs.\ Mining Reward: Statistical Evidence}\label{secEmpRH}

In Proposition \ref{propH}, we have shown that if $\gamma>0$, the equilibrium hash rate $H^*$ increases %\remove{approximately} 
like the square root of the mining reward $R$ (see also discussion in Section \ref{secDeltaH}). It follows that 
\begin{align}
1+r_{H^*}(t) %= \frac{\sqrt{R_{t+1}} - \sqrt{R_t}}{\sqrt{R_t}} 
\approx \sqrt{1+r_R(t)},
\end{align}
where $r_{H^*}(t)$ and $r_R(t)$ are equilibrium hash rate and mining reward returns between times $t-1$ and $t$. 
%\add{Specifically, if $H^*\approx\alpha \sqrt{R}$, for some $\alpha>0$, then} 
%	\begin{align}
%	1 + r_{H^*} := 1 + \frac{H_{t+1}^*-H_t^*}{H_t^*} %= \frac{\sqrt{R_{t+1}} - \sqrt{R_t}}{\sqrt{R_t}} 
%	\approx \sqrt{1+\frac{R_{t+1}-R_t}{R_t}} =:\sqrt{1+r_R}.
%	\end{align}
More generally, if $H^*$ increases  like a power $\beta$ of $R$, we have
\begin{align}\label{reg}
%H^* \approx \alpha R^{\beta}
%\quad\Longrightarrow\quad \frac{H_{t+1}^*-H_t^*}{H_t^*} %= \frac{R_{t+1}^{\beta} - R_t^{\beta}}{R_t^{\beta}} 
%\approx \Big(1+\frac{R_{t+1}-R_t}{R_t}\Big)^{\beta}-1
1+r_{H^*}(t) \approx (1+r_R(t))^{\beta}.
\end{align}
To {statistically test} this relationship, we consider the linear regression model 
\begin{align}\label{reg2}
\log\big(1+r_{H}(t)\big) = \alpha + \beta\log\big(1+r_R(t-1)\big) + \epsilon(t),
\end{align}
where %$r_H(t):=(H(t)-H(t-1))/H(t)$ and $r_R(t):=(R(t)-R(t-1))/R(t)$ 
%s, and 
$\epsilon$ is a vector of idiosyncratic errors and $r_H(t)$ and $r_R(t)$ are, respectively, historical three-month hash rate and mining reward returns. We compute these returns using monthly averages of hash rates and mining rewards. For example, if $t$ is equal to July 15, then $r_H(t)$ is the simple return between the average hash rate in April and the average hash rate in July, and $r_R(t-1)$ is the simple return between the average mining reward in January and the average mining reward in April. We use lagged values for mining reward, because hash rates typically respond with a lag to changes in the mining reward, as documented in Section \ref{secEDA}. 

It is evident from Figure \ref{fig_5} that there exists a sublinear relationship between $r_H(t)$ and $r_R(t-1)$. This pattern is statistically confirmed by our regression analysis, which yields the estimate $\hat\beta=0.34$ (see Table \ref{table1}). {These findings are consistent with the equilibrium hash rates when $\gamma>0$,} given in Proposition \ref{propH}, %\remove{while the case $\gamma=0$ would be consistent with the estimate $\hat\beta=1$.} 
and also reflect economic intuition. {That is, we expect capacity constraints to prevent miners from increasing their hash rates proportionally to increases in the mining reward, as the marginal cost of mining becomes higher as the exerted hash rate increases.}
%\remove{the marginal cost of mining to be increasing in the exerted hash rate, which is the case if $\gamma>0$, and prevents miners from increasing their hash rates proportionally to increases in the mining reward (see discussion following Prop.\ }\ref{propR}). %\footnote{If the marginal cost of mining were constant, i.e., $\gamma=0$, we would expect $\hat\beta\approx 1$.}. 

%The right panel of Figure \ref{fig_5} and the second row of Table \ref{table1} show the results of the same regression analysis, but with the returns $r_R(t)$ computed using Bitcoin prices than mining rewards. The results are similar, which is expected as the Bitcoin price is a good proxy for the mining reward (see Fig.\ \ref{fig_reward_price}).

%{Figure \ref{fig_HvsR2} shows nonlinear least squares estimates for the slightly more general model  
%\begin{align}\label{reg2}
%1+r_{H^*} \approx \alpha(1+r_R)^{\beta},
%\end{align}}
%\note{Can we please provide a table with all regression outputs, $R^2$, statistical significance, etc...? See, for instance, Table 5 on pag. 33 of} \url{http://www.columbia.edu/~ac3827/assets/files/CollateralRuleEmpirics_JME.pdf}. \note{We can also control for time fixed effects.}

\medskip 

\begin{rem} 
	The objective function (\ref{pi_i}) can be generalized to 
	\begin{align}\label{Happrox}
	\pi_i = \frac{R}{H}h_i-c_ih_i-\frac{\gamma}{1+\delta}h_i^{1+\delta},
	\end{align}
	where $\delta> 0$, and the expression in (\ref{pi_i}) is then recovered if $\delta=1$. For $\delta\neq 1$, the aggregate hash rate $H^*$ does not admit a closed-form representation. However, we can mimic the steps in the proof of Proposition \ref{propH}, and show that if miner heterogeneity is low, then $H^*$ behaves approximately like a power $1/(1+\delta)$ of the mining reward $R$. %satisfies 
	%\begin{align}
	%H^* \approx K R^{\frac{1}{1+\delta}},
	%\end{align} 
	%where $K>0$ is a constant.
	Hence, it follows from (\ref{reg})-(\ref{reg2}) that the regression estimate $\hat\beta\approx 1/3$ in Table \ref{table1} is consistent with {a coefficient} $\delta=2$. This corresponds to a convex capacity constraint which is cubic rather than quadratic. Using the objective function~\eqref{Happrox} would produce results qualitatively similar to those obtained with a quadratic objective function~(\ref{pi_i}), the key point being that the marginal cost of mining is increasing in both cases. \hfill\qed
\end{rem}

\begin{comment}
\begin{figure}[ht!]
\centering
\includegraphics[width=\textwidth]{fig3_app3_MSE_noline.png}
\caption{\small Left panel: Scatter plot of three-month mining reward returns (x-axis) and three-month returns of the system hash rate (y-axis). Returns are computed biweekly between January 2017 and April 2021. At a given time $t$, the return $r_{H^*}$ is the simple return between $\bar H^*_{t-4m,t-3m}$ and $\bar H^*_{t-1m,t}$, where $\bar H^*_{t_1,t_2}$ denotes the average daily hash rate in the time interval $(t_1,t_2)$. The return $r_R$ is the simple return between $\bar R_{t-7m,t-6m}$ and $\bar R_{t-4m,t-3m}$, where $\bar R_{t_1,t_2}$ denotes the average daily mining reward in the time interval $(t_1,t_2)$. Right panel: Root-Mean-Squared-Error (RMSE) for the model in (\ref{reg}) with $\beta$ between 0 and 1.\note{I do not think it is necessary to plot the root means square error. We can just mention in the main text the value of $\beta$ which we find to minimize the RMSE.}
}\label{fig_HvsR}
\end{figure} 
\end{comment}

\begin{figure}[ht!]
	\centering
	\includegraphics[width=\textwidth]{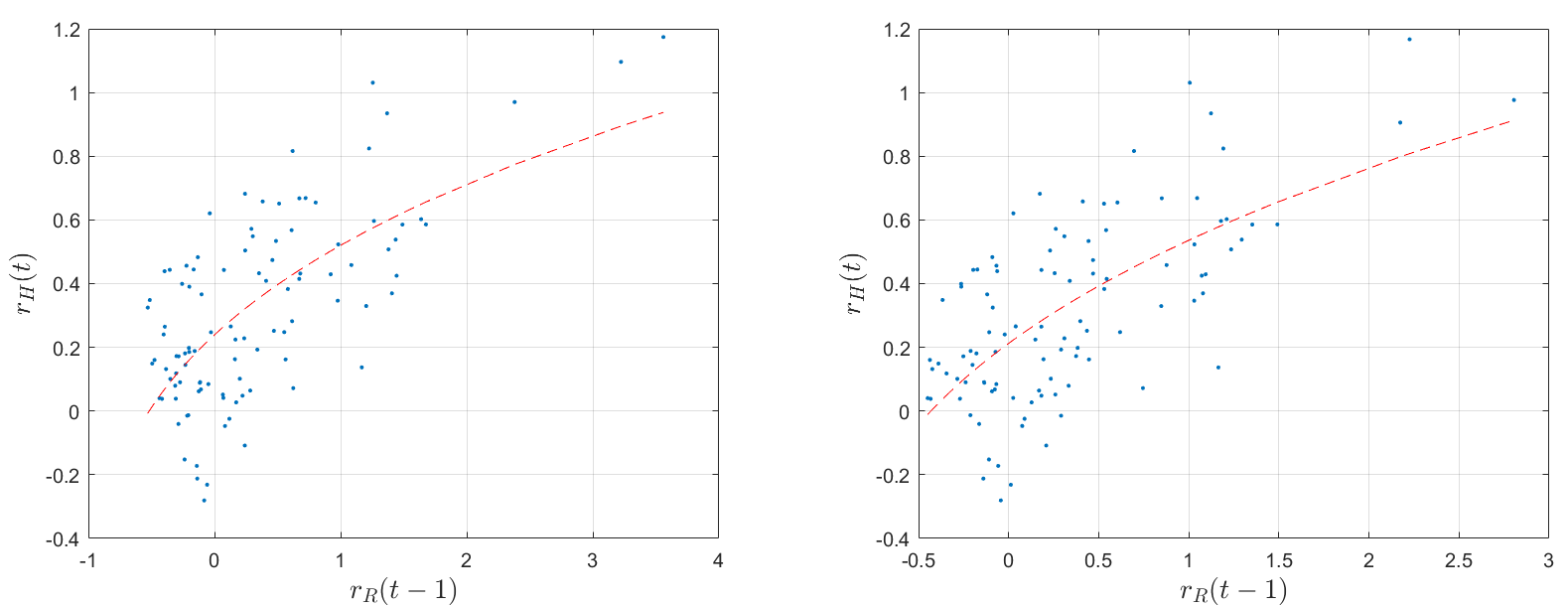}
	\caption{\small Left panel: Scatter plot of three-month system hash rate returns (y-axis) against lagged three-month mining reward returns (x-axis), computed for biweekly time points between January 2017 and April 2021. 
		%At a given time $t$, the return $r_{H}(t)$ is the simple return between $\bar H_{t-4m,t-3m}$ and $\bar H_{t-1m,t}$, where $\bar H_{t_1,t_2}$ denotes the average daily hash rate in the time interval $(t_1,t_2)$. The return $r_R(t-1)$ is the simple return between $\bar R_{t-6m,t-5m}$ and $\bar R_{t-4m,t-3m}$, where $\bar R_{t_1,t_2}$ denotes the average daily mining reward in the time interval $(t_1,t_2)$.
		The dotted line corresponds to the best fit of the linear model (\ref{reg2}) - the regression statistics are reported in Table \ref{table1}. 
		Right panel: Same plot as in the left panel, but with the returns $r_R(t-1)$ computed using Bitcoin prices rather than mining rewards.
		The patterns in both panels are similar, consistently with the fact that the Bitcoin price is a good proxy for the mining reward (see also Fig.\ \ref{fig_reward_price}).
	}
	\label{fig_5}
\end{figure} 

%The returns $r_H(t)$ and $r_R(t-1)$ are three-month returns between monthly averages of hash rates and mining rewards, computed using nonoverlapping time intervals.  
%For example, for $t$ equal to July 15, $r_H(t)$ is the return between the average hash rates in April and July, and $r_R(t-1)$ the return between the average mining rewards in January and April.
%Returns are computed biweekly between January 2017 and April 2021, resulting in $102$ data points.

%At any time $t$, $\bar H(t)$ is the 30-day average daily hash rate in $(t-m/2,t+m/2)$, and $r_H(t)$ is the simple return between $\bar H(t-3m)$ and $\bar H(t)$. Similarly, $r_R(t-1)$ is the simple return between $\bar R(t-6m)$ and $\bar R(t-3m)$. For example, for $t$ equal to April 15, $r_H(t)$ is the return between the average hash rates in January and April, and $r_R(t)$ the return between the average hash rates in October and January. 

\renewcommand*{\arraystretch}{1.2}
\begin{table}[h!]
	\begin{center}
		%\begin{tabular}{|c|c|c|c|c|} \hline % <-- Alignments: 1st column left, 2nd middle and 3rd right, with vertical lines in between
		\begin{tabular}{|>{\centering}p{0.05\textwidth}|>{\centering}p{0.1\textwidth}|>{\centering}p{0.1\textwidth}|p{0.1\textwidth}<{\centering}|} \hline
			& $\hat\alpha$ & $\hat\beta$ & $R^2$   \\\hline
			$R$   & $0.22$  &  0.29  &  0.44      \\
			\hline
			$P$   & 0.19  &  0.34 &   0.43    \\ \hline
		\end{tabular}
	\end{center}
	\vspace{-10pt}
	\caption{\small Regression estimates for the linear model in Equation (\ref{reg2}). In the first row of the table ($R$), $r_R(t-1)$ is computed using mining rewards. In the second row of the table ($P$), $r_R(t-1)$ is computed using Bitcoin prices. The number of observations is $102$, and all parameter estimates are statistically significant with $p$-values smaller than 0.001. Because the Bitcoin price is a good proxy for the mining reward (see Fig.\  \ref{fig_reward_price}), the regression estimates in the two rows of the table are similar.
	}\label{table1}
\end{table}

\section{Conclusions}\label{secConclusions}

%Cryptocurrencies such as Bitcoin remove the need of a trusted third-party who manages the centralized ledger for a fee. Instead, they rely on a decentralized network of validators to maintain and update the ledger. This decentralization brings forth many benefits over traditional payment systems such as speed of transactions, low transaction costs and anonymity. However, this decentralization does not come without a cost. To ensure consensus, validators compete for the right to update the ledger by solving a computationally costly problem, a process known as mining. Recently, concerns have grown about the large consumption of energy from these mining activities.
%and its social efficiency of this race. %We consider a two-stage game where firms first invest in R\&{D} to subsequently compete in a Bitcoin mining game at reduced costs.

%We study the key factors contributing to the arms race in mining activities. We show that the ability of firms to compete depends crucially on the cost effectiveness of their mining operations, highlighting the importance of R\&{D} in the Bitcoin mining industry.

We develop a model of cryptocurrency mining where miners invest in new hardware to improve the efficiency of their operations in a subsequent proof-of-work mining competition. %Our model embeds two critical properties of the proof-of-work protocol. First, the expected proportion of the mining reward attained by a miner is equal to the fraction of the network hash rate it contributes. Second, the mining reward is independent of the network hash rate. %in a given period is fixed, regardless of the total hashing power exerted. 
%\remove{These properties reflect intrinsic characteristics of the hashing functions used in the proof-of-work protocol, and the adaptive difficulty level of the hashing problem.} 
We demonstrate that the nature of the mining competition excludes inefficient miners from participating, in contrast to the original vision of Nakamoto. We argue that mining centralization is lower than indicated by earlier research due to capacity constraints faced by miners. Additionally, we show that {the ability of miners to invest in new technology leads to greater decentralization in mining}. Hence, while the emergence of specialized mining hardware has resulted in  individual miners being replaced by relatively large mining operations, {larger miners do not inherently increase their advantage over smaller ones}. As advancements in ASIC mining hardware slow down, our model suggests that mining will become a race for access to low-cost electricity.

Our model indicates that investment in new hardware may not have a large impact on the cost of attacking a cryptocurrency network, even though it increases the aggregate hash rate in the network. In contrast, a higher mining reward {significantly increases} both mining decentralization and network security, which confirms the common wisdom that smaller coins are more vulnerable to attacks. These findings imply that the majority {of coins} in existence may not be long-lived. They also raise concerns regarding whether transaction fees will be sufficient to sustain mining reward, and thus network security, for maturing cryptocurrencies.

Our study focuses on the behavior of miners and takes other components of the mining value chain as given. We do not directly model the activities of hardware manufacturers, but rather assume the existence of a monopolistic manufacturer that supplies miners with new hardware. % produces the hardware that miners can purchase.
%take the existence of cutting-edge hardware that miners can invest in as given. \add{The assumption of an exogenous manufacturing sector is consistent with the fact that there is little overlap between the activities of miners and manufacturers.} %, and since the emergence of ASIC mining hardware, there has been a consistent increase in its efficiency.} 
%\remove{Assuming} %that there is only a single 
From the miners' point of view, the ASIC hardware industry has %to a very large extent been 
been dominated by a single firm, Bitmain, and %, at any given time, 
the hardware of other prominent manufacturers can safely be treated as a substitute good. {We also remark that our implicit assumption of a monopolistic manufacturer is consistent with} \cite{Ferreira}, {who argue that a single manufacturer will dominate in the blockchain ecosystem.}
%\change{, and,}{.} \change{a}{A}t any given time, the \remove{ASIC} hardware of other prominent manufacturers \add{(e.g., MicroBT)} %(such as MicroBT in the current landscape) 
%can safely be considered as a substitute good. %\note{Are we saying that an exogenous manufacturing sector is the same as a homogenous manufacturing sector? {\Red Modified. Saying that from the miners' point of view, there simply exists an exogenous hardware sector, and it can also be considered to be a homogeneous sector (the products of established manufacturers are the same for all intents and purposes - miners just buy the latest release)}}

We also do not consider mining pools in our analysis. Empirical evidence indicates that the size of mining pools is mean-reverting, with honest miners leaving a mining pool that becomes too large (see also \cite{Cong}). Additionally, significant constraints have been placed on the ability of pool managers to act in bad faith (see Remark \ref{ghash}).
%\cite{Cong} argue that the size of mining pools mean-revert endogenously because of the ability of larger pools to charge higher fees. However, they assume infinitesimal miners and are thus not concerned with distribution of mining power among {actual} miners. %Not accounting for bad intentions of pool managers\note{new protocols, see also remark}, 
%Not accounting for the intentions of pool managers,
When it comes to actual mining, the availability of mining pools is likely to contribute to decentralization.
%If anything, as a financial innovation, mining pools are likely to contribute to decentralization. 
This is because their absence would be more detrimental to smaller miners, who are likely to be worse equipped to face a high variance in their mining revenue, and for which the average time between mined blocks is larger. %In their absence, mining would be completely inaccessible for small miners, while larger miner are better equipped to tolerate fluctuations in mining revenue. 

%solo miners are expected to face a high variance of payouts, depending on their share of the overall mining power. Consequently, miners collude to form so-called mining pools, where participants work together towards finding the next block and share rewards based on each miner’s contribution and according to some reward distribution scheme. As such, since the appearance of the first mining pools in Bitcoin, the fraction of blocks generated by solo miners has steadily declined and is negligible today

%have empirically tested mining centralization by analyzing the structure of mining pools. They assume infinitesimal miners. There is also indication of miners leaving mining pools that are too large - ghash. 
%Mining pools are in fact more likely to have a decentralization effect and to be integral to the survival of a large network. Without mining pools, mining would be completely inaccessible to smaller miners. Of course one can think about the mining pool manager having bad intentions, but there are protocols meant to stop that. 

Alternatives to the proof-of-work protocol have been proposed, most prominently the \emph{proof-of-stake} protocol, where miners do not use computational resources to compete for the right to validate transactions, but rather are chosen in proportion to their ownership of the underlying coin. This implies a clear path to centralization in mining: a miner who starts with a larger number of coins has a higher probability of further increase its advantage than a smaller miner has of reducing the initial gap. {Our model can be extended to such a proof-of-stake setting where the miners' initial probabilities of mining a block are determined by their initial coin holdings, and the hardware investment stage is replaced by a stage where miners invest in the underlying coin.}

%Due to the fact PoS doesn't require you to fight with hashing power, the energy cost and hardware costs don't rise with odds of being chose to mine the next block (the mining lottery).

%This leads to centralization of mining and the rich get richer, faster.

%To be online, it requires resources (hardware, electricity, and internet connection)

While the size of the Bitcoin network currently dwarfs the size of other networks, there are by now multiple established cryptocurrencies that miners can move their hash rate between. %Considering that miners are able to move their hash rate between cryptocurrencies, 
{Arguably, for a given level of mining power, only a finite number of networks of significant size can be secured. However, most existing research focuses on a single network, and it remains an open question to understand how allocation of mining power between different networks will impact centralization in individual networks. } %cryptocurrency mining. } impact centralization in individual networks. 
 %Will fragmentation of mining power among competing cryptocurrencies be beneficial or detrimental? 
%whether mining will be sufficient to support multiple cryptocurrency networks, or whether fragmentation of mining power among competing cryptocurrencies will be too great. It is clear that for a given mining power, only a finite number of networks can be secured. 

{In a dynamic extension of our model, we expect the equilibrium hash rate shares of miners to converge over time. A dynamic model would provide a deeper understanding of centralization forces in cryptocurrency mining. For example, the effect of investment is to some extent permanent, and its benefit is path-dependent. Investment thus effectively allows miners to remain competitive over time and protected against a significant drop in the mining reward, which could otherwise force them out of business. Finally, an important factor to consider is how volatility of mining reward affects mining centralization. Cryptocurrencies are a highly volatile asset class, and, as mentioned above, fluctuations in mining revenue may present greater challenges for smaller miners. We leave these questions for future research. }

\appendix 

\section{Accuracy of Approximation Formulas in Section \ref{secEffectBeta}}\label{secNumInvest} 

We verify the accuracy of the approximation formulas used to analyze the effect of investment on equilibrium hash rates and mining profits. Figure \ref{fig_eta12} confirms our analytical findings from % from Proposition \ref{propDeltah} and \ref{propDeltaPi} 
Section \ref{secEffectBeta}, i.e., that smaller miners gain hash rate shares and increase their profits, while the opposite is true for larger miners. It is also evident from the figure that these trends magnified for smaller values of $\eta$, i.e., lower adjustment costs, which leads to larger investment levels. 

\begin{figure}[ht!]
	\centering
	%\hspace{-1 cm} %[width=20.0cm,height=6.0cm]
	\includegraphics[width=\textwidth]{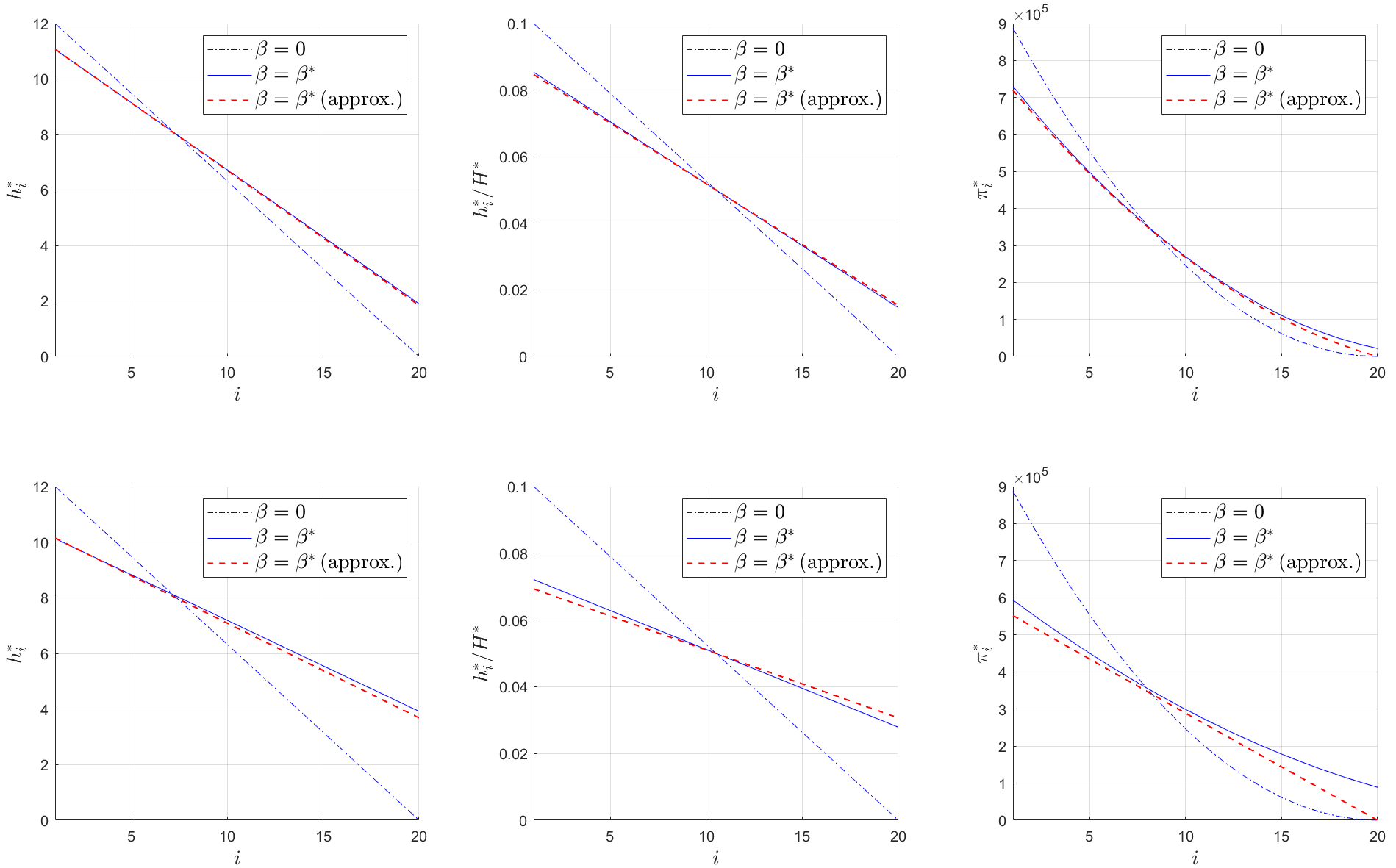}
	\caption{\small The left and middle panels plot the hash rates and hash rate shares of miners, indexed by $i$, with and without investment (i.e., $\beta=\beta^*$ and $\beta=0$). The red {dotted} lines show the approximations in Proposition \ref{propDeltah}. The right panels plot the profits of miners with and without investment. The red {dotted} lines plot the approximation in Proposition \ref{propDeltaPi}. The model parameters are set as described in Section \ref{secParams}, with $\eta=2$ (upper panels) and $\eta=1$ (lower panels).}\label{fig_eta12}
\end{figure} 

Noticeably, the approximation formulas for both hash rates and profits, given in Propositions \ref{propDeltaH} and \ref{propDeltaPi},  closely resemble their exact counterparts. {The accuracy of the approximations increases as $\eta$ gets larger, {but it remains good for both hash rates and hash rate shares if $\eta$ is small, and within acceptable levels for mining profits.} It is worth observing that $\eta=1$ corresponds to a reasonable lower bound for the value of $\eta$ (see Section \ref{secParams}). 
	
\section{Proofs of Propositions}\label{appProofs}

\paragraph{Proof of Proposition \ref{propH}:} 
%\paragraph{Proof of Proposition \ref{propExist}:} 
Sufficient conditions for the existence of a pure-strategy equilibrium are (i) concavity of $\pi_i$ in $h_i$, (ii) continuity of $\pi_i$ in $h_i$ and $h_{-i}$, and (iii) the strategy space of miner $i$ being compact and convex. %\footnote{See: \url{https://ocw.mit.edu/courses/electrical-engineering-and-computer-science/6-254-game-theory-with-engineering-applications-spring-2010/lecture-notes/MIT6_254S10_lec05.pdf}.} 
For the concavity of $\pi_i$ we have
\begin{align}
\frac{\partial\pi_i}{\partial h_i} = \frac{H-h_i}{H^2}R - c_i - \gamma_ih_i, \qquad 
\frac{\partial^2\pi_i}{\partial h_i^2} = -2\frac{H-h_i}{H^3}R  - \gamma_i < 0. 
\end{align}
For the continuity properties it is easy to see that for any given $h_{-i}$ (resp., $h_i$), the objective function $\pi_i$ is continuous in $h_i$ (resp., $h_{-i}$). %, and for any given $h_i$, $\pi_i$ is continuous in $h_{-i}$. 
The strategy space of miner $i$ is $[0,\infty)$, which is unbounded. However, since $\pi_i<0$ for $h_i>R/c_i$, the strategy space of miner $i$ can be restricted to the compact and convex set $[0,R/c_i]$. 

To show uniqueness, first observe that any equilibrium will consist of at least two active miners. Namely, if only a single miner applies a positive hash rate, then a marginal reduction in its hash rate will increase the value of its objective function. Similarly, if no miner applies a positive hash rate, then each miner has an incentive to marginally increase its hash rate. In both cases we have a contradiction. Next, observe that if miner $k+1$ is active in equilibrium, then miner $k$ cannot be inactive, because in that case we would have
\begin{align}
MG_k := \frac{R}{H} > \frac{R}{H}\Big(1-\frac{h_{k+1}}{H}\Big)  = MG_{k+1} = MC_{k+1} = c_{k+1} + \gamma_{k+1}h_{k+1} > c_k =: MC_k,
\end{align}
which is a contradiction. Using these two facts, it is easy to show inductively that in a game consisting of a total of $N_0$ miners, where $2\leq N_0\leq N$, the equilibrium must be unique. When showing the result for $N_0+1$, given that the results holds for $N_0$, we must consider two cases. First, if miner $N_0$ is not active in the unique equilibrium with a total of $N_0$ miners, then the same equilibrium is also unique in a game with a total of $N_0+1$ miners. Second, if miner $N_0$ is active, then miner $N_0+1$ is active if and only if $c_{N_0+1}<R/H_{N_0}$, where $H_{N_0}$ is the overall hash rate in the equilibrium with a total of $N_0$ miners. It follows that the equilibrium is unique in a game of $N_0+1$ miners. 

We now derive the equations for the unique equilibrium hash rates. %The first-order condition for miner $i$ is
%\begin{align}
%0 = \frac{\partial\pi_i}{\partial h_i} = \frac{1}{H}R - \frac{h_i}{H^2}R - c_i - \gamma h_i 
%= \frac{1}{H}R\Big(1 - \frac{h_i}{H}\Big) - c_i - \gamma_ih_i 
%= \frac{\sum_{j\neq i}h_j}{H^2}R - c_i - \gamma_ih_i,  
%\end{align}
%and by using $\sum_{j\neq i}h_j=H-h_i$ we have
%\begin{align}
%0 = \frac{H-h_i}{H^2}R - c_i - \gamma_ih_i %= \frac{R}{H} - \frac{h_i}{H^2}R - c_i - \gamma_ih_i 
%= \frac{R}{H} - h_i\Big(\frac{R}{H^2} + \gamma \Big) - c_i,
%\end{align}
%and 
Summing over the first-order condition (\ref{1st}) of active miners gives
\begin{align}\label{H2nd}
0 %= \sum_i\frac{\sum_{j\neq i}h_j}{H^2}R - c^{(n)} - \sum_{i}\gamma_ih_i
= \frac{n-1}{H}R - c^{(n)} - \gamma H
= \frac{1}{H}\big((n-1)R - c^{(n)}H - \gamma H^2\big).
\end{align}
The aggregate equilibrium hash rate $H^*$ is then given by (\ref{optH}), where the case $\gamma>0$ requires solving a second order polynomial which can be shown to have a unique positive solution. The equation for the hash rate $h_i^*$ then follows from the first-order condition (\ref{1st}) of miner $i$. 
\hfill\qed

\paragraph{Proof of Proposition \ref{propProfit}:} The profit-per-hash of an active miner $i$ is
\begin{align}
\frac{\pi_i^*}{h_i^*} = \frac{R}{H^*}-c_i-\frac{\gamma}{2}h_i^* = \frac{R}{H^*}-(c_i + \gamma h_i^*) + \frac{\gamma}{2}h_i^*,
\end{align}
which is decreasing in $i$ because $c_i+\gamma h_i^*$ is increasing in $i$, by Proposition \ref{propMC}, and $h_i^*$ is decreasing in $i$, by Proposition \ref{propH}. It follows that the profit $\pi_i^*$ is also decreasing in $i$.
\hfill\qed

\paragraph{Proof of Proposition \ref{propN}:} From Proposition \ref{propH} it follows that miner $i$ is active in equilibrium if and only if $R/H^*>c_i$. Thus, 
\begin{align}
h_i^*>0 \quad\Longleftrightarrow\quad R-c_iH^* > 0 &\quad\Longleftrightarrow\quad 2\frac{R\gamma}{c_i} + c^{(i)} > \sqrt{(c^{(i)})^2+4(i-1)R\gamma} 
\quad\Longleftrightarrow\quad c_i < \frac{c^{(i)}+R\gamma/c_i}{i-1}.
\end{align} 
The number of active miners is the largest value of $i$ satisfying this equation, since, in equilibrium, miner $i+1$ cannot be active if miner $i$ is not active. \hfill\qed

\paragraph{Proof of Proposition \ref{propDeltaC} and Proposition \ref{propHdelta}:} We split the proof up into three parts. 

\smallskip\noindent{\em (i) Aggregate hash rate:} For $H^*$ we use equation (\ref{optH}) to obtain
\begin{align}\label{derivsH}
\frac{\partial H^*}{\partial c_i} &= \frac{1}{2\gamma}\Big(\frac{c^{(n)}}{\sqrt{(c^{(n)})^2+4(n-1)R\gamma}}-1\Big) 
%=-\frac{H^*}{\sqrt{(c^{(n)})^2+4(n-1)R\gamma}} 
=-\frac{H^*}{c^{(n)} + 2\gamma H^*} < 0, \\
\frac{\partial H^*}{\partial \gamma} %&= \frac{1}{4\gamma^2}\Big(2\gamma\frac{4(n-1)R}{2\sqrt{(c^{(n)})^2+4(n-1)R\gamma}} - 2\big(\sqrt{(c^{(n)})^2+4(n-1)R\gamma}-c^{(n)}\big)\Big)\\
%&=\frac{1}{2\gamma^2}\Big(\frac{2(n-1)R\gamma}{\sqrt{(c^{(n)})^2+4(n-1)R\gamma}} - \big(\sqrt{(c^{(n)})^2+4(n-1)R\gamma}-c^{(n)}\big)\Big)\\
%&=\frac{1}{2\gamma^2}\Big(\frac{2(n-1)R\gamma- (c^{(n)})^2-4(n-1)R\gamma+c^{(n)}\sqrt{(c^{(n)})^2+4(n-1)R\gamma}}{\sqrt{(c^{(n)})^2+4(n-1)R\gamma}} \Big)\\
&=-\frac{(c^{(n)})^2 + 2(n-1)R\gamma - c^{(n)}\sqrt{(c^{(n)})^2+4(n-1)R\gamma}}{2\gamma^2\sqrt{(c^{(n)})^2+4(n-1)R\gamma}} 
%&=-\frac{1}{2\gamma^2}\Big(\frac{ 2(n-1)R\gamma - 2\gamma c^{(n)}H}{\sqrt{(c^{(n)})^2+4(n-1)R\gamma}} \Big) 
%=-\frac{1}{\gamma}\frac{ (n-1)R - c^{(n)}H^*}{c^{(n)} + 2\gamma H^*} 
=-\frac{(H^*)^2}{c^{(n)} + 2\gamma H^*}  <0, \\
%&\leq-\frac{1}{2\gamma^2}\Big(\frac{(c^{(n)})^2 + 2(n-1)R\gamma - (c^{(n)})^2-2c^{(n)}\sqrt{(n-1)R\gamma}}{\sqrt{(c^{(n)})^2+4(n-1)R\gamma}} \Big)\\
%&\leq-\frac{\sqrt{(n-1)R\gamma}}{\gamma^2}\Big(\frac{ \sqrt{(n-1)R\gamma} -c^{(n)}}{\sqrt{(c^{(n)})^2+4(n-1)R\gamma}} \Big),\\
\frac{\partial H^*}{\partial R} &= \frac{n-1}{\sqrt{(c^{(n)})^2+4(n-1)R\gamma}} = \frac{n-1}{c^{(n)} + 2\gamma H^*} >0, 
\end{align}
where we used (\ref{H2nd}) and that $\sqrt{(c^{(n)})^2+4(n-1)R\gamma}=c^{(n)} + 2\gamma H^*$.
%where the inequality for the $\gamma$-derivative follows from the fact that $H<(n-1)R/c^{(n)}$, which follows from a first-order Taylor expansion. %$\sqrt{a^2+x}\leq a+x/(2a)$, for any $a,x\geq 0$.\footnote{If we can show that $H$ increases as $\gamma$ goes down, it also follows from the fact that {\Red$H<(n-1)R/c^{(n)}$}.} 

\medskip\noindent{\em (ii) Individual hash rates:} From Proposition \ref{propH}, it follows that for an active miner $i$ we can write
\begin{align}\label{fig}
h_i^* &= f_i(c_1,\dots,c_n,\gamma,R)
:= g\frac{R-c_ig}{R+\gamma g^2}, \\ 
g &:= g(c_1,\dots,c_n,\gamma,R) := \frac{\sqrt{(c^{(n)})^2+4(n-1)R\gamma}-c^{(n)}}{2\gamma},
\end{align}
and we now show that the derivatives of $h_i^*$ satisfy
\begin{align}\label{derivsh}
\frac{\partial h_i^*}{\partial c_i} &= \frac{\partial f_i}{\partial c_i} + \frac{\partial f_i}{\partial g}\frac{\partial g}{\partial c_i} =: \Delta_{i,1} + \Delta_{i,2} = -\frac{g^2}{R+\gamma g^2}\Big(1 + \frac{\frac{R}{g}-2(c_i + \gamma f_i)}{c^{(n)}+2\gamma g}\Big), \\
\frac{\partial h_i^*}{\partial c_j} &= \frac{\partial f_i}{\partial g}\frac{\partial g}{\partial c_j} = \Delta_{i,2} 
= -\frac{g^2}{R+\gamma g^2}\frac{\frac{R}{g}-2(c_i + \gamma f_i)}{c^{(n)} + 2\gamma g}, \\
\frac{\partial h_i^*}{\partial \gamma} &= \frac{\partial f_i}{\partial \gamma} + \frac{\partial f_i}{\partial g}\frac{\partial g}{\partial \gamma} =: \Delta_{i,1}^{(\gamma)} + \Delta_{i,2}^{(\gamma)} = -\frac{f_ig^2}{R+\gamma g^2}\Big(1 + \frac{ g}{ f_i}\frac{\frac{R}{g}-2(c_i + \gamma f_i)}{c^{(n)} + 2\gamma g}\Big), \\
\frac{\partial h_i^*}{\partial R} &= \frac{\partial f_i}{\partial R} + \frac{\partial f_i}{\partial g}\frac{\partial g}{\partial R} =: \Delta_{i,1}^{(R)} + \Delta_{i,2}^{(R)} = \frac{g^2(c_i + \gamma f_i)}{R(R+\gamma g^2)}\Big(1 + \frac{\frac{R}{g}-2(c_i + \gamma f_i)}{c_i+\gamma f_i}\frac{c^{(n)}+\gamma g}{c^{(n)}+2\gamma g}\Big).
\end{align} 
To obtain the condition for the sign of $\Delta_{i,2}$, we use Proposition \ref{propMC} to write
\begin{align}\label{condHalf}
\Delta_{i,2} > 0 \quad\Longleftrightarrow\quad c_i+\gamma f_i = \frac{R}{g}\Big(1-\frac{f_i}{g}\Big) > \half\frac{R}{g} 
\quad\Longleftrightarrow\quad \frac{f_i}{g} < \half.
\end{align}
The signs of $\Delta_{i,2}^{(\gamma)}$ and $\Delta_{i,2}^{(R)}$ are analyzed in the same way.

To show (\ref{derivsh}), first note that simple calculations give 
\begin{align}
\frac{\partial f_i}{\partial c_i} &= -\frac{g^2}{R+\gamma g^2}<0,\qquad
\frac{\partial f_i}{\partial \gamma} = -g^3\frac{R-c_ig}{(R+\gamma g^2)^2}<0,\qquad
\frac{\partial f_i}{\partial R} %=  H\frac{R+\gamma H^2 -(R-c_iH)}{(R+\gamma H^2)^2} 
= g^2\frac{c_i + \gamma g }{(R+\gamma g^2)^2}>0,
\end{align} 
and %For the $H$-derivative of $f$ we have 
%and the sign of the derivatives may seem ambiguous because $\partial f/\partial H$ is generally negative,
\begin{align}
\frac{\partial f_i}{\partial g} %&= \frac{R-c_iH}{R+\gamma H^2} + H\frac{-c_i(R+\gamma H^2)-2\gamma(R-c_iH)H}{(R+\gamma H^2)^2} \\
%&= \frac{R-c_iH}{R+\gamma H^2} + H\frac{-c_iR-\gamma c_i H^2 -2\gamma RH+ 2\gamma c_iH^2}{(R+\gamma H^2)^2}\\
= \frac{R-c_ig}{R+\gamma g^2} + g\frac{-c_iR -2\gamma Rg+ \gamma c_ig^2}{(R+\gamma g^2)^2} 
%&= \frac{1}{R+\gamma H^2}\Big(R-c_iH + H\frac{-c_iR -2\gamma RH+ \gamma c_iH^2}{R+\gamma H^2}\Big)\\ 
&= \frac{1}{R+\gamma g^2}\Big(R-c_ig + g\frac{-\gamma g(R-c_ig)-c_iR -\gamma Rg}{R+\gamma g^2}\Big) \\
&= \frac{g}{R+\gamma g^2}\Big(\frac{R}{g}-(c_i + \gamma f_i) - R\frac{c_i +\gamma g}{R+\gamma g^2}\Big) \\
&= \frac{g}{R+\gamma g^2}\Big(\frac{R}{g}-2(c_i + \gamma f_i)\Big),
\end{align} 
where the final equality uses that
\begin{align}\label{eqhmc}
R\frac{c_i +\gamma g}{R+\gamma g^2} = c_i - \gamma g\frac{R - c_ig}{R+\gamma g^2} = c_i + \gamma f_i.
\end{align}
%and $\partial H/\partial c_j<0$. 
%For the $H$-derivative of $f$ we also have 
%\begin{align}
%\frac{\partial}{\partial c_i}\frac{\partial f}{\partial H} &= \frac{-H}{R+\gamma H^2} + H\frac{-R + \gamma H^2}{(R+\gamma H^2)^2} = \frac{-H}{R+\gamma H^2}\Big(1 - \frac{-R + \gamma H^2}{R+\gamma H^2}\Big) < 0, 
%\end{align}
%so ${\partial f}/{\partial H}$ is increasing is 
\begin{comment}
We also have, 
\begin{align}
\frac{\partial f}{\partial H} %&= \frac{R-c_iH}{R+\gamma H^2} + H\frac{-c_i(R+\gamma H^2)-2\gamma(R-c_iH)H}{(R+\gamma H^2)^2} \\
%&= \frac{R-c_iH}{R+\gamma H^2} + H\frac{-c_iR-\gamma c_i H^2 -2\gamma RH+ 2\gamma c_iH^2}{(R+\gamma H^2)^2}\\
&= \frac{R-c_iH}{R+\gamma H^2} + H\frac{-c_iR -2\gamma RH+ \gamma c_iH^2}{(R+\gamma H^2)^2} >0
\quad\Longleftrightarrow\quad \frac{R}{H} > 2(c_i + \gamma h_i).
\end{align}
In terms of the model, this condition means that the reward-per-hash should be greater than the average cost-per-hash when including the quadratic cost. In general this condition is not satisfied.
\end{comment}
%and 
%\begin{align}
%\frac{\partial f}{\partial H}= \frac{R-c_iH}{R+\gamma H^2} + H\frac{-c_i(R+\gamma H^2)-2\gamma(R-c_iH)H}{(R+\gamma H^2)^2} > 0 \\
%R-c_iH > H\frac{c_i(R+\gamma H^2)+2\gamma(R-c_iH)H}{R+\gamma H^2}  \\
%R-c_iH > c_iH + 2\gamma H^2\frac{R-c_iH}{R+\gamma H^2}  \\
%\frac{R}{H} > 2(c_i + \gamma h_i).
%\end{align}
For the $c_i$-derivative of $h_i^*$, we have 
\begin{align}\label{derivC}
\frac{\partial h_i^*}{\partial c_i} &= -\frac{g^2}{R+\gamma g^2} - \frac{g^2}{R+\gamma g^2}\frac{\frac{R}{g}-2(c_i + \gamma f_i)}{\sqrt{(c^{(n)})^2+4(n-1)R\gamma}} 
= -\frac{g^2}{R+\gamma g^2}\Big(1 + \frac{\frac{R}{g}-2(c_i + \gamma f_i)}{c^{(n)}+2\gamma g}\Big) < 0.
\end{align}
The sign of the derivative is negative because, using $R/g>c_i+\gamma f_i$ from Proposition \ref{propMC}, we have
\begin{align}
\frac{\frac{R}{g}-2(c_i + \gamma f_i)}{c^{(n)}+2\gamma g} 
&= \frac{\frac{R}{g}-(c_i + \gamma f_i)}{c^{(n)}+2\gamma g} - \frac{c_i + \gamma f_i}{c^{(n)}+2\gamma g} > - \frac{c_i + \gamma f_i}{c^{(n)}+2\gamma g} > -1.
\end{align}
The equation for ${\partial h_i^*}/{\partial c_j}$ is obtained in the same way.
Next, for the $\gamma$-derivative of $h_i^*$ we have
\begin{align}\label{derivGamma}
\frac{\partial h_i^*}{\partial \gamma} %= \frac{\partial f_i}{\partial \gamma} + \frac{\partial f_i}{\partial g}\frac{\partial g}{\partial \gamma} 
&= -g^3\frac{R-c_ig}{(R+\gamma g^2)^2} + \frac{g}{R+\gamma g^2}\Big(\frac{R}{g}-2(c_i + \gamma f_i)\Big)\Big(-\frac{1}{\gamma}\frac{ (n-1)R - c^{(n)}g}{c^{(n)} + 2\gamma g}\Big) \\
&= -g^3\frac{R-c_ig}{(R+\gamma g^2)^2}\Big(1 + \frac{R+\gamma g^2}{H^2(R-c_ig)}\Big(\frac{R}{g}-2(c_i + \gamma f_i)\Big)\frac{1}{\gamma}\frac{ (n-1)R - c^{(n)}g}{c^{(n)} + 2\gamma g}\Big) \\
&= -g^3\frac{R-c_ig}{(R+\gamma g^2)^2}\Big(1 + \frac{1}{\gamma}\frac{1}{f_ig}\Big(\frac{R}{g}-2(c_i + \gamma f_i)\Big)\frac{ (n-1)R - c^{(n)}g}{c^{(n)} + 2\gamma g}\Big) \\
%&= -H^3\frac{R-c_iH}{(R+\gamma H^2)^2}\Big(1 + \frac{(n-1)R/H - c^{(n)}}{\gamma h_i}\frac{\frac{R}{H}-2(c_i + \gamma h_i)}{c^{(n)} + 2\gamma H}\Big)\\
%&= -g^3\frac{R-c_ig}{(R+\gamma g^2)^2}\Big(1 + \frac{ g}{ g_i}\frac{\frac{R}{g}-2(c_i + \gamma g_i)}{c^{(n)} + 2\gamma g}\Big) \\
&= -\frac{f_ig^2}{R+\gamma g^2}\Big(1 + \frac{ g}{ f_i}\frac{\frac{R}{g}-2(c_i + \gamma f_i)}{c^{(n)} + 2\gamma g}\Big),
\end{align}
where we have used (\ref{H2nd}).
Finally, for the $R$-derivative of $h_i^*$, we have
\begin{align}\label{derivR}
\frac{\partial h^*_i}{\partial R} %= \frac{\partial f_i}{\partial R} + \frac{\partial f_i}{\partial g}\frac{\partial g}{\partial R} 
&= g^2\frac{c_i + \gamma g }{(R+\gamma g^2)^2} + \frac{g}{R+\gamma g^2}\Big(\frac{R}{g}-2(c_i + \gamma f_i)\Big)\frac{n-1}{c^{(n)}+2\gamma g} \\
&= g^2\frac{c_i + \gamma g }{(R+\gamma g^2)^2}\Big(1 + \frac{1}{g}\frac{R+\gamma g^2}{c_i+\gamma g}\Big(\frac{R}{g}-2(c_i + \gamma f_i)\Big)\frac{n-1}{c^{(n)}+2\gamma g}\Big)\\
&= g^2\frac{c_i + \gamma g }{(R+\gamma g^2)^2}\Big(1 + \frac{n-1}{g}\frac{R+\gamma g^2}{c_i+\gamma g}\frac{\frac{R}{g}-2(c_i + \gamma f_i)}{c^{(n)}+2\gamma g}\Big) \\
%&= H^2\frac{c_i + \gamma H }{(R+\gamma H^2)^2}\Big(1 + (n-1)\frac{R}{H}\frac{1}{c_i+\gamma_ih_i}\frac{\frac{R}{H}-2(c_i + \gamma h_i)}{c^{(n)}+2\gamma H}\Big) \\
%&= H^2\frac{c_i + \gamma H }{(R+\gamma H^2)^2}\Big(1 + (c^{(n)}+\gamma H)\frac{1}{c_i+\gamma_ih_i}\frac{\frac{R}{H}-2(c_i + \gamma h_i)}{c^{(n)}+2\gamma H}\Big)\\
%&= g^2\frac{c_i + \gamma g}{(R+\gamma g^2)^2}\Big(1 + \frac{\frac{R}{g}-2(c_i + \gamma g_i)}{c_i+\gamma g_i}\frac{c^{(n)}+\gamma g}{c^{(n)}+2\gamma g}\Big) \\
&= \frac{g^2(c_i + \gamma f_i)}{R(R+\gamma g^2)}\Big(1 + \frac{\frac{R}{g}-2(c_i + \gamma f_i)}{c_i+\gamma f_i}\frac{c^{(n)}+\gamma g}{c^{(n)}+2\gamma g}\Big) > 0,
\end{align}
where we used (\ref{H2nd}) and (\ref{eqhmc}).
The sign of the derivative is positive because
\begin{align}
\frac{n-1}{g}\frac{R+\gamma g^2}{c_i+\gamma g}\frac{\frac{R}{g}-2(c_i + \gamma f_i)}{c^{(n)}+2\gamma g}
&= \frac{1}{g}\frac{R+\gamma g^2}{c_i+\gamma g}\Big(\frac{R}{g}-(c_i + \gamma f_i)\Big)\frac{n-1}{c^{(n)}+2\gamma g} 
- \frac{n-1}{g}\frac{R+\gamma g^2}{c_i+\gamma g}\frac{c_i + \gamma f_i}{c^{(n)}+2\gamma g} \\
&= \frac{1}{g}\frac{R+\gamma g^2}{c_i+\gamma g}\Big(\frac{R}{g}-(c_i + \gamma f_i)\Big)\frac{n-1}{c^{(n)}+2\gamma g} 
- (n-1)\frac{R}{g}\frac{1}{c^{(n)}+2\gamma g} \\
&= \frac{1}{g}\frac{R+\gamma g^2}{c_i+\gamma g}\Big(\frac{R}{g}-(c_i + \gamma f_i)\Big)\frac{n-1}{c^{(n)}+2\gamma g} 
- \frac{c^{(n)}+\gamma g}{c^{(n)}+2\gamma g} > -1,
\end{align}
where, again, we have used (\ref{H2nd}) and (\ref{eqhmc}). The final inequality follows from $R/g>c_i+\gamma f_i$. 
%\begin{align}
%\frac{n-1}{g}\frac{R+\gamma g^2}{c_i+\gamma g}\frac{c_i + \gamma g_i}{c^{(n)}+2\gamma g} 
%= (n-1)\frac{R}{g}\frac{1}{c_i+\gamma g_i}\frac{c_i + \gamma g_i}{c^{(n)}+2\gamma g} 
%= (n-1)\frac{R}{g}\frac{1}{c^{(n)}+2\gamma g} = \frac{c^{(n)}+\gamma g}{c^{(n)}+2\gamma g} < 1.
%\end{align}
\begin{comment} %big-O part 3
We also have
\begin{align}
\frac{\partial f}{\partial H} %&= \frac{R-c_iH}{R+\gamma H^2} + H\frac{-c_iR -2\gamma RH+ \gamma c_iH^2}{(R+\gamma H^2)^2} \\
&= \frac{1}{R+\gamma H^2}\Big(R-c_iH + H\frac{-\gamma H(R-c_iH)-c_iR -\gamma RH}{R+\gamma H^2}\Big) \\
&= H^2\frac{c_i + \gamma H }{(R+\gamma H^2)^2}\Big(\frac{(R-c_iH)(R+\gamma H^2)}{H^2(c_i+\gamma H)} + \frac{-\gamma H(R-c_iH)-c_iR -\gamma RH}{H(c_i+\gamma H)}\Big) \\
&= \frac{\partial f}{\partial R}O\Big(\frac{1}{H}\Big),
\end{align} 
and it follows that\note{Can make this $O(1/H^2)$}
\begin{align}
\frac{\partial h_i}{\partial R} &= \frac{\partial f}{\partial R}\Big(1 + O\Big(\frac{1}{H}\Big)\frac{\partial H}{\partial R}\Big) 
= \frac{\partial f}{\partial R}\Big(1 + O\Big(\frac{1}{H}\Big)\frac{n-1}{c^{(n)}+2\gamma H}\Big) 
= \frac{\partial f}{\partial R}\Big(1 + O\Big(\frac{1}{H}\Big)\Big).
\end{align} 
\end{comment}

\medskip\noindent{\em (iii) Hash rate shares:} We will show that the hash rate share of miner $i$ satisfies
\begin{align}\label{derivsHh}
\frac{\partial }{\partial c_i}\frac{h_i^*}{H^*} &= -\frac{g}{R+\gamma g^2}\Big(1 - \frac{c_i + 2\gamma f_i }{c^{(n)} + 2\gamma g}\Big)<0, \\
\frac{\partial}{\partial c_j}\frac{h_i^*}{H^*} &= \frac{g}{R+\gamma g^2}\frac{c_i +2\gamma f_i}{c^{(n)}+2\gamma g}>0, \\
\frac{\partial }{\partial \gamma}\frac{h_i^*}{H^*} &= -\frac{1}{R+\gamma g^2}\Big(f_ig  - \frac{g^2}{c^{(n)} + 2\gamma g}(c_i + 2\gamma f_i)\Big), \\
\frac{\partial }{\partial R}\frac{h_i^*}{H^*} &= \frac{1}{R+\gamma g^2}\Big(\frac{g}{R}(c_i + \gamma f_i) - (n-1)\frac{c_i + 2\gamma f_i}{c^{(n)} + 2\gamma g}\Big),
\end{align}
where $f_i$ and $g$ are defined in (\ref{fig}). First, for the $c_i$-derivative we have
\begin{align}
\frac{\partial }{\partial c_i}\frac{h_i^*}{H^*} = \frac{\partial }{\partial c_i}\frac{f_i}{g} &= \frac{1}{g}\frac{\partial f_i}{\partial c_i} - \frac{f_i}{g^2}\frac{\partial g}{\partial c_i} = \frac{1}{g}\Big(\frac{\partial f_i}{\partial c_i} + \frac{\partial f_i}{\partial g}\frac{\partial g}{\partial c_i}\Big) - \frac{f_i}{g^2}\frac{\partial g}{\partial c_i}  
=\frac{1}{g}\Big(\frac{\partial f_i}{\partial c_i} + \frac{\partial g}{\partial c_i}\Big(\frac{\partial f_i}{\partial g}-\frac{f_i}{g}\Big)\Big).
\end{align}
Using our previous results for the partial derivatives, we then obtain
\begin{align}\label{derivHh}
\frac{\partial }{\partial c_i}\frac{h_i^*}{H^*} &= \frac{1}{g}\Big(-\frac{g^2}{R+\gamma g^2} - \frac{g}{2\gamma g+c^{(n)}}\Big(\frac{g}{R+\gamma g^2}\Big(\frac{R}{g}-2(c_i + \gamma f_i)\Big)-\frac{f_i}{g}\Big)\Big) \\
&= \frac{1}{g}\Big(-\frac{g^2}{R+\gamma g^2} - \frac{g}{2\gamma g+c^{(n)}}\frac{1}{R+\gamma g^2}\Big(R-2g(c_i + \gamma f_i)-(R-c_ig)\Big)\Big) \\
&= \frac{1}{g}\Big(-\frac{g^2}{R+\gamma g^2} - \frac{g^2}{2\gamma g+c^{(n)}}\frac{1}{R+\gamma g^2}(-2(c_i + \gamma f_i) + c_i)\Big) \\
&= -\frac{g}{R+\gamma g^2}\Big(1 - \frac{2\gamma f_i + c_i}{2\gamma g+c^{(n)}}\Big)<0.
\end{align}
We can also prove the following inequality: 
\begin{align}\label{derivhHj}
\frac{\partial}{\partial c_j}\frac{h_i^*}{H^*} = \frac{\partial}{\partial c_j}\frac{f_i}{g} = \frac{1}{g}\frac{\partial f_i}{\partial c_j} - \frac{f_i}{g^2}\frac{\partial g}{\partial c_j} 
&= -\frac{1}{g}\frac{g^2}{R+\gamma g^2}\frac{\frac{R}{g}-2(c_i + \gamma f_i)}{c^{(n)} + 2\gamma g} +\frac{f_i}{g^2}\frac{g}{c^{(n)}+2\gamma g} \\
&= \frac{1}{c^{(n)}+2\gamma g}\Big(-\frac{g}{R+\gamma g^2}\Big(\frac{R}{g}-2(c_i + \gamma f_i)\Big) +\frac{f_i}{g}\Big) \\
&= \frac{1}{R+\gamma g^2}\frac{1}{c^{(n)}+2\gamma g}(-(R-2g(c_i + \gamma f_i)) +R-c_ig)\\
%&= \frac{1}{R+\gamma H^2}\frac{1}{c^{(n)}+2\gamma H}( c_iH +2\gamma h_iH) \\
&= \frac{g}{R+\gamma g^2}\frac{c_i +2\gamma f_i}{c^{(n)}+2\gamma g}>0.
\end{align}
In the same way, we deduce 
\begin{align}
\frac{\partial }{\partial \gamma}\frac{h_i^*}{H^*} %&= \frac{1}{g}\frac{\partial f_i}{\partial \gamma} - \frac{f_i}{g^2}\frac{\partial g}{\partial \gamma} = \frac{1}{g}\Big(\frac{\partial f_i}{\partial \gamma} + \frac{\partial f_i}{\partial \gamma}\frac{\partial \gamma}{\partial \gamma}\Big) - \frac{f_i}{g^2}\frac{\partial g}{\partial \gamma}  
=\frac{1}{g}\Big(\frac{\partial f_i}{\partial \gamma} + \frac{\partial g}{\partial \gamma}\Big(\frac{\partial f_i}{\partial g}-\frac{f_i}{g}\Big)\Big),
\qquad 
\frac{\partial }{\partial R}\frac{h_i^*}{H^*} %&= \frac{1}{g}\frac{\partial f_i}{\partial R} - \frac{f_i}{g^2}\frac{\partial g}{\partial R} = \frac{1}{g}\Big(\frac{\partial f_i}{\partial R} + \frac{\partial f_i}{\partial g}\frac{\partial g}{\partial R}\Big) - \frac{f_i}{g^2}\frac{\partial g}{\partial R}  
=\frac{1}{g}\Big(\frac{\partial f_i}{\partial R} + \frac{\partial g}{\partial R}\Big(\frac{\partial f_i}{\partial g}-\frac{f_i}{g}\Big)\Big).
\end{align}
Computation of the $\gamma$-derivative yields
\begin{align}\label{derivHh}
\frac{\partial }{\partial \gamma}\frac{h_i^*}{g} 
&= \frac{1}{g}\Big(-g^3\frac{R-c_ig}{(R+\gamma g^2)^2}  -\frac{g^2}{c^{(n)} + 2\gamma g}\Big(\frac{g}{R+\gamma g^2}\Big(\frac{R}{g}-2(c_i + \gamma f_i)\Big)-\frac{f_i}{g}\Big)\Big) \\
&= \frac{1}{g}\Big(-g^3\frac{R-c_ig}{(R+\gamma g^2)^2}  -\frac{g^2}{c^{(n)} + 2\gamma g}\frac{1}{R+\gamma g^2}\Big(R-2g(c_i + \gamma f_i)-(R-c_ig)\Big)\Big) \\
&= -\frac{1}{R+\gamma g^2}\Big(g^2\frac{R-c_ig}{R+\gamma g^2}  - \frac{g^2}{c^{(n)} + 2\gamma g}(c_i + 2\gamma f_i)\Big) \\
&= -\frac{1}{R+\gamma g^2}\Big(f_ig  - \frac{g^2}{c^{(n)} + 2\gamma g}(c_i + 2\gamma f_i)\Big) \\
&= -\frac{1}{R+\gamma g^2}\Big(f_i\Big(g  - \frac{\gamma g^2}{c^{(n)} + 2\gamma g}\Big) - \frac{g^2}{c^{(n)} + 2\gamma g}(c_i + \gamma f_i)\Big).
\end{align}
The above expression is decreasing in $i$ because $f_i$ is increasing in $i$, $c_i+\gamma f_i$ is increasing in $i$ by Prop.\ \ref{propMC}, and the factor multiplying $f_i$ is positive by (\ref{H2nd}). 
For the $R$-derivative, we have
\begin{align}\label{derivHh}
\frac{\partial }{\partial R}\frac{h_i^*}{H^*} 
&= \frac{1}{g}\Big(g^2\frac{c_i + \gamma g }{(R+\gamma g^2)^2} + \frac{n-1}{c^{(n)} + 2\gamma g}\Big(\frac{g}{R+\gamma g^2}\Big(\frac{R}{g}-2(c_i + \gamma f_i)\Big)-\frac{f_i}{g}\Big)\Big) \\
%&= \frac{1}{g}\Big(g^2\frac{c_i + \gamma g }{(R+\gamma g^2)^2} + \frac{n-1}{c^{(n)} + 2\gamma H^*}\frac{1}{R+\gamma g^2}\Big(R-2g(c_i + \gamma f_i)-(R-c_ig)\Big)\Big) \\
&= \frac{1}{g}\Big(g^2\frac{c_i + \gamma g }{(R+\gamma g^2)^2} + \frac{n-1}{c^{(n)} + 2\gamma g}\frac{g}{R+\gamma g^2}(-2(c_i + \gamma f_i) + c_i)\Big) \\
&= \frac{1}{R+\gamma g^2}\Big(g\frac{c_i + \gamma g }{R+\gamma g^2} - (n-1)\frac{c_i + 2\gamma f_i}{c^{(n)} + 2\gamma g}\Big) \\
&= \frac{1}{R+\gamma g^2}\Big(\frac{g}{R}(c_i + \gamma f_i) - (n-1)\frac{c_i + 2\gamma f_i}{c^{(n)} + 2\gamma g}\Big) \\
&= \frac{1}{R+\gamma g^2}\Big((c_i + \gamma f_i)\Big(\frac{g}{R} - \frac{n-1}{c^{(n)} + 2\gamma g}\Big) - (n-1)\frac{\gamma f_i}{c^{(n)} + 2\gamma g}\Big),
\end{align}
where we have used (\ref{eqhmc}). The above expression is decreasing in $i$ because $c_i+\gamma f_i$ is increasing in $i$, $f_i$ is decreasing in $i$, and the factor multiplying $c_i+\gamma f_i$ is positive by (\ref{H2nd}).  \hfill\qed

%\begin{align} 
%\frac{\partial f_i}{\partial g} < 0 \quad\Longleftrightarrow\quad \frac{R}{g} < 2(c_i + \gamma f_i) 
%\quad\Longleftrightarrow\quad \frac{f_i}{g} < \frac{1}{2},
%\end{align}
%where the second equivalence follows from the first-order condition for miner $i$, which can be written as
%\[
%\frac{R}{g}\Big(1-\frac{f_i}{g}\Big) - (c_i+\gamma f_i) = 0.
%\frac{R}{H} = c_i +\gamma_ih_i + \frac{R}{H^2}h_i,
%\]

\medskip  

\begin{prop}\label{propPi}
	For active miners $i$ and $j$, the sensitivities of miner $i$'s profit to $c_i$ and $c_j$ satisfy
	\begin{align}
	\frac{\partial\pi^*_i}{\partial c_i}  < 0, \qquad 
	\frac{\partial\pi_i^*}{\partial c_j}>  0.
	%&= \frac{H}{R+\gamma (H^*)^2}\frac{c_i +2\gamma h_i^*}{c^{(n)}+2\gamma H^*} > 0.
	\end{align} 
\end{prop}

\paragraph{Proof:} Using the notation in (\ref{fig}), we have the following inequality for the $c_i$-derivative of the profit: %, given by $\pi_i = R(f_i/g) - c_if_i - \gamma f_i^2/2$, we have
\begin{align}\label{derivPi}
\frac{\partial\pi^*_i}{\partial c_i} = \frac{\partial}{\partial c_i}\Big(\frac{R}{g}f_i-c_if_i-\frac{\gamma}{2}f_i^2\Big) &= R\Big(\frac{1}{g}\frac{\partial f_i}{\partial c_i} - \frac{f_i}{g^2}\frac{\partial g}{\partial c_i}\Big) - f_i - c_i\frac{\partial f_i}{\partial c_i} - \gamma f_i\frac{\partial f_i}{\partial c_i} \\
&= f_i\Big(- \frac{R}{g^2}\frac{\partial g}{\partial c_i} - 1\Big) + \frac{\partial f_i}{\partial c_i}\Big(\frac{R}{g} - c_i - \gamma f_i\Big) \\
&= f_i\Big( \frac{R}{g}\frac{1}{2\gamma g+c^{(n)}} - 1\Big) + \frac{\partial f_i}{\partial c_i}\Big(\frac{R}{g} - c_i - \gamma f_i\Big) < 0. 
\end{align}
This sign of the derivative is negative because ${\partial f_i}/{\partial c_i}<0$, $R/g>c_i+\gamma f_i$, and, using (\ref{H2nd}),
\begin{align}
2\gamma g+c^{(n)} > \gamma g + c^{(n)} = (n-1)\frac{R}{g} \geq \frac{R}{g}.
\end{align}
For the $c_j$-derivative, we have 
\begin{align}\label{derivPi2}
\frac{\partial\pi^*_i}{\partial c_j} = \frac{\partial}{\partial c_j}\Big(\frac{R}{g}f_i-c_if_i-\frac{\gamma}{2}f_i^2\Big)
&= R\Big(\frac{1}{g}\frac{\partial f_i}{\partial c_j} - \frac{f_i}{g^2}\frac{\partial g}{\partial c_j}\Big) - c_i\frac{\partial f_i}{\partial c_j} - \gamma f_i\frac{\partial f_i}{\partial c_j} \\
&= f_i\Big(- \frac{R}{g^2}\frac{\partial g}{\partial c_j}\Big) + \frac{\partial f_i}{\partial c_j}\Big(\frac{R}{g} - c_i - \gamma f_i\Big) \\
%&= f_i\frac{R}{g}\frac{1}{2\gamma g+c^{(n)}} + \frac{\partial f_i}{\partial c_j}\Big(\frac{R}{g} - c_i - \gamma f_i\Big) \\
&= f_i\frac{R}{g}\frac{1}{2\gamma g+c^{(n)}} -\frac{g^2}{R+\gamma g^2}\frac{\frac{R}{g}-2(c_i + \gamma f_i)}{c^{(n)} + 2\gamma g}\Big(\frac{R}{g} - c_i - \gamma f_i\Big).
\end{align} 
The above expression can be further rewritten as 
\begin{align} 
\frac{\partial\pi^*_i}{\partial c_j}&= \frac{R}{g}\frac{1}{2\gamma g+c^{(n)}}\Big(f_i -\frac{g^2}{R+\gamma g^2}\Big(\frac{R}{g} - c_i - \gamma f_i\Big)\Big)
+\frac{g^2}{R+\gamma g^2}\frac{2(c_i + \gamma f_i)}{c^{(n)} + 2\gamma g}\Big(\frac{R}{g} - c_i - \gamma f_i\Big) > 0,
\end{align}
which is positive because $R/g-c_i-\gamma f_i>0$, and 
\begin{align}
f_i -\frac{g^2}{R+\gamma g^2}\Big(\frac{R}{g} - c_i - \gamma f_i\Big) > 0 
&\quad\Longleftrightarrow\quad \frac{f_i(R+\gamma g^2)}{g^2} = \frac{R-c_ig}{g}  = \frac{R}{g} - c_i > \frac{R}{g} - c_i - \gamma f_i,
\end{align}
which is  trivially satisfied.
\begin{comment}
For the profit we further have
\begin{align}
\frac{\partial\pi^*_i}{\partial\gamma} &= R\Big(\frac{1}{g}\frac{\partial f_i}{\partial\gamma} - \frac{f_i}{g^2}\frac{\partial g}{\partial\gamma}\Big)  - c_i\frac{\partial f_i}{\partial\gamma} - \gamma f_i\frac{\partial f_i}{\partial\gamma} - \frac{f_i^2}{2} \\
&= f_i\Big(- \frac{R}{g^2}\frac{\partial g}{\partial\gamma} - \frac{f_i}{2}\Big) + \frac{\partial f_i}{\partial\gamma}\Big(\frac{R}{g} - c_i - \gamma f_i\Big) \\
&= f_i\Big( \frac{R}{c^{(n)} + 2\gamma H^*} - \frac{f_i}{2}\Big) + \frac{\partial f_i}{\partial\gamma}\Big(\frac{R}{g} - c_i - \gamma f_i\Big) < 0. 
\end{align}
The first term is a bit problematic, and plugging the second derivative in gets messy.
\end{comment}
\hfill\qed

\paragraph{Proof of Proposition \ref{propBeta}:} It follows from Proposition \ref{propPi} that the profit of an active miner $i$ is decreasing in its cost parameter $c_i$, irrespective of the cost parameters of other miners. Hence, for any value of $\beta_{-i}$, i.e., the investment of other miners, the optimal investment of an active miner $i$ is obtained by minimizing the cost $c_i(\beta_i)$. This implies that any potential equilibrium investment $\beta^*$ is such that $\beta_i^*=\min\{1/\eta,1\}$ for $i\in A(\beta_i^*)$, and $\beta_i^*=0$ for $i\notin A(\beta_i^*)$. 

We begin by showing that if all miners have a positive investment level, then the order of miners in terms of their mining efficiency remains the same as without investment. From (\ref{Ieq}), we have that the cost-per-hash of miner $i$ with $\beta_i^*=\min\{1/\eta,1\}$ is
\begin{align}
c_i(\beta_i^*) &=
\left\{
\begin{array}{ll}   
%c_{i,0} - \beta_i^*( c_{i,0}-\tilde c_0) + \frac{\eta_i}{2}(\beta_i^*)^2 
%=  c_{i,0} - \frac{u_i^2}{\eta_i} + \frac{\eta_i}{2}\frac{u_i^2}{\eta_i^2} 
\tilde c_{i} - \frac{1}{2}\frac{u_i}{\eta},& \; \eta>1, \\
\tilde c_{i} - u_i\big(1 - \frac{\eta }{2}\big),& \; \eta\leq 1.
\end{array}
\right.
\end{align}
If $\eta>1$, we have 
\begin{align}
c_{i+1}(\beta_{i+1}^*) \leq c_i(\beta_i^*) 
\quad\Longleftrightarrow\quad  \tilde c_{i+1} - \tilde c_{i} \leq \frac{u_{i+1}-u_i}{2\eta}  = \frac{ \tilde c_{i+1}- \tilde c_{i}}{2\eta} \quad\Longleftrightarrow\quad \eta \leq \half, 
\end{align}
which cannot hold if $\eta>1$. If $\eta\leq 1$, then 
\begin{align}
c_{i+1}(\beta_{i+1}^*) \leq c_i(\beta_i^*) %\quad\Longleftrightarrow\quad 
%\tilde c_{i+1}-u_{i+1}\Big(1-\frac{\eta}{2}\Big) \leq \tilde c_{i}-u_{i}\Big(1-\frac{\eta}{2}\Big)
\quad\Longleftrightarrow\quad  \tilde c_{i+1} -  \tilde c_{i} \leq (u_{i+1} - u_i)\Big(1-\frac{\eta}{2}\Big) = ( \tilde c_{i+1} -  \tilde c_{i})\Big(1-\frac{\eta}{2}\Big) \quad\Longleftrightarrow\quad \eta \leq 0,
\end{align} 
which can only hold if $\eta=0$, in which case the cost of all miners becomes $\tilde c_0$. It follows from the above argument that the order of miners remains the same if all miners have a positive investment level. Using this results, it is then easy to see that for $i$ and $i'$ such that $\tilde c_i\leq\tilde c_{i'}$, miner $i'$ cannot have a positive equilibrium investment if miner $i$ has zero investment.

%Hence, for $i$ and $i'$ such that $\tilde c_i\leq\tilde c_{i'}$, miner $i'$ cannot have a positive equilibrium investment if miner $i$ has zero investment. Namely, if that is the case, then miner $i$ is inactive and has an incentive to deviate from the equilibrium. First, if $i\leq|A(0)|$, then, by investing, miner $i$ has a lower cost-per-hash than miner $i'$, and a positive profit. If $i>|A(0)|$, then miner $i$ additionally faces a fixed cost, but since $i'>i$, miner $i'$ faces the same fixed cost, and miner $i$ again has an incentive to deviate from the equilibrium. 

The unique equilibrium $\beta^*$ is therefore obtained by finding the largest $1\leq n\leq N$ such that if $\beta_i^*=\min\{1/\eta,1\}$ for $i\leq n$, and $\beta_i^*=0$ for $n<i\leq N$, then $\pi_n>0$. To show that this value is at least equal to $|A(0)|$, it is sufficient to show that the participation condition in Proposition \ref{propN} is satisfied for miner $i=|A(0)|$, assuming that miners $j=1,\dots,i$ have positive investment levels. 
%We now show that active miners who are active without investment remain active with investment. First note that since the mining equilibrium contains at least two active miners, the two most efficient miners (with initial costs $\tilde c_1$ and $\tilde c_2$) invest and remain the two most efficient active miners. We can then show inductively that for miners $i=3,\dots,|A(0)|$, the participation condition in Proposition \ref{propN} is satisfied with investment. 
In what follows we write $c_i$ for $c_i(\beta_i^*)$. The participation condition can then be rewritten as
\begin{align}
{c_i} < \frac{c^{(i)} + \frac{R\gamma}{c_i}}{i-1} 
\quad\Longleftrightarrow\quad 
c_i < \frac{i-1}{i-2}\frac{c^{(i-1)}}{i-1} + \frac{R\gamma}{(i-2)c_i},
\end{align}
where, from (\ref{Ieq}), the cost of miner $i$ is
\begin{align}
c_i = \left\{
\begin{array}{ll}
\tilde c_i\big(1-\frac{1}{2\eta}\big) + \tilde c_0\frac{1}{2\eta}, &\; \eta>1, \\
\tilde c_0\frac{\eta}{2} + \tilde c_0\big(1-\frac{\eta}{2}\big), &\; \eta\leq 1.
\end{array}
\right.
\end{align}
Hence, $c_i=\kappa \tilde c_i+(1-\kappa)\tilde c_0$, for some $0<\kappa<1$, and the participation above becomes
\begin{align}
\kappa\tilde c_i + (1-\kappa)\tilde c_0 
%&< \frac{i-1}{i-2}\Big(\kappa\frac{\tilde c^{(i-1)}}{i-1} + (1-\kappa)\tilde c_0\Big) + \frac{R\gamma}{(i-2)c_i}  \\
&< \kappa\frac{i-1}{i-2}\frac{\tilde c^{(i-1)}}{i-1}  + \frac{i-1}{i-2} (1-\kappa)\tilde c_0 + \frac{R\gamma}{(i-2)c_i}.
\end{align}
This is satisfied because  
\begin{align}
\kappa\tilde c_i < \kappa \frac{i-1}{i-2}\frac{\tilde c^{(i-1)}}{i-1} + \kappa  \frac{R\gamma}{(i-2)\tilde c_i}
< \kappa \frac{i-1}{i-2}\frac{\tilde c^{(i-1)}}{i-1} +   \frac{R\gamma}{(i-2) c_i}, 
\qquad (1-\kappa)\tilde c_0 <  \frac{i-1}{i-2}(1-\kappa)\tilde c_0,
\end{align} 
The first {inequality} uses that the participation condition is satisfied for miner $i$ before investment, in addition to using $\kappa<1$ and $c_i\leq\tilde c_i$. The second {inequality follows from the fact} that $(i-1)/(i-2)>1$. 
\hfill\qed

\paragraph{Proof of Proposition \ref{propDeltaH}:}  We begin by introducing the notation 
\begin{align}\label{notation}
H^* &:= H^*(\beta^*), \mkern-36mu& h_i^* &:=  h_i^*(\beta^*), \mkern-36mu& H^*_0 &:= H^*(0), \mkern-36mu & h^*_{i,0} &:= h_i^*(0), \\
c^{(n)} &:= c^{(n)}(\beta^*), \mkern-36mu&  c^{(n)}_0 &:= c^{(n)}(0), \mkern-36mu& K&:=4(n-1)R\gamma. \mkern-36mu&
\end{align}
The derivative of $H^*$ with respect to $c^{(n)}$ is given by 
\begin{align}
\frac{\partial H^*}{\partial\bar I} = \frac{1}{2\gamma}\Big(\frac{c^{(n)}}{\sqrt{( c^{(n)})^2+K}}  -1\Big) = -\frac{1}{2\gamma}\Big(1 - \frac{ c^{(n)}}{\sqrt{( c^{(n)})^2+K}}\Big) = -\frac{ H^*}{\sqrt{( c^{(n)})^2+K}}.
\end{align}
Using that $c^{(n)}=c^{(n)}_0-\bar I$ we then obtain the first-order Taylor expansion 
\begin{align}\label{Hstar}
H^*
%&= \frac{1}{2\gamma}\Big(\sqrt{( c^{(n)})^2+K} - c^{(n)} + \Big(1-\frac{c^{(n)}}{\sqrt{( c^{(n)})^2+K}}\Big)\bar\beta\Big) + O(\bar\beta^2) 
=  H_0^* + \frac{H_0^*}{\sqrt{( c_0^{(n)})^2+K}}\bar I + O(\bar I^2) 
=  H_0^* + \frac{H_0^*}{2\gamma H_0^* + c_0^{(n)}}\bar I + O(\bar I^2)
=: H_0^* + a\bar I+ O(\bar\beta^2).
\end{align}
Using $a = {H_0^*}/{\sqrt{(c_0^{(n)})^2+K}}$, and that $\bar I=uc^{(n)}_0-v$ for some $u,v>0$ (see (\ref{Ieq})), simple calculations show that $a\bar I$ is increasing in $c_0^{(n)}$. %, because the function $x\mapsto x/\sqrt{\kappa+x^2}$, where $\kappa>0$ is a constant, is increasing for $x\geq 0$.
Moreover, $a\bar I$ is clearly decreasing in $\gamma$ since $\bar I$ is independent of $\gamma$.  \hfill\qed 

\paragraph{Proof of Proposition \ref{propDeltah}:} 
Using the relation $c_i=c_{i,0}-I_i$, we can decompose the hash rate share of miner $i$ as
\begin{align}
\frac{ h_i^*}{ H^*} =  \frac{R- c_{i} H^*}{R+\gamma  (H^*)^2}
= \frac{R- c_{i,0} H^*}{R+\gamma  (H^*)^2} + \frac{ H^*}{R+\gamma  (H^*)^2}I_i.
\end{align}
Using (\ref{Hstar}), the first term can be written as
\begin{align}
\frac{R- c_{i,0} H^*}{R+\gamma  (H^*)^2}
%&= \frac{R- c_i H_0}{R+\gamma H_0^2} + \frac{- c_i(R+\gamma H_0^2) - 2\gamma H_0(R- c_i H_0)}{(R+\gamma H_0^2)^2}\frac{ H_0}{\sqrt{( c^{(n)})^2+K}} \bar\eta + O(\bar\eta^2) \\
&= \frac{h_{i,0}^*}{H_0^*} - \frac{ c_{i,0}(R+\gamma (H_0^*)^2) + 2\gamma H_0^*(R- c_i H_0^*)}{(R+\gamma (H_0^*)^2)^2}\frac{ H_0^*}{\sqrt{( c_0^{(n)})^2+K}} \bar I + O(\bar I^2),
\end{align}
and the second one as
\begin{align}
\frac{ H^*}{R+\gamma  (H^*)^2} %&= \frac{H_0}{R+\gamma H_0^2} + \frac{R+\gamma H_0^2 - 2\gamma H_0^2}{(R+\gamma H_0^2)^2}\frac{ H_0}{\sqrt{( c^{(n)})^2+K}}\bar\eta + O(\bar\eta^2)\\
&= \frac{H_0^*}{R+\gamma (H_0^*)^2} + \frac{R - \gamma (H_0^*)^2}{(R+\gamma (H_0^*)^2)^2}\frac{ H_0^*}{\sqrt{( c_0^{(n)})^2+K}}\bar I + O(\bar I^2).
\end{align}
It follows that
\begin{align}
\frac{ h_i^*}{ H^*} 
&= \frac{h_{i,0}^*}{H_0^*} + \frac{H_0^*}{R+\gamma (H_0^*)^2}I_i - \frac{ c_{i,0}(R+\gamma (H_0^*)^2) + 2\gamma H_0^*(R- c_{i,0} H_0^*)}{(R+\gamma (H_0^*)^2)^2}\frac{ H_0^*}{\sqrt{( c_0^{(n)})^2+K}} \bar I  + O(\bar I^2)  \\
%&= \frac{h_{i,0}}{H_0} + \frac{H_0}{R+\gamma H_0^2}u_i\beta_i - \frac{H_0}{R+\gamma H_0^2}\frac{ c_i(R+\gamma H_0^2) + 2\gamma H_0(R- c_i H_0)}{R+\gamma H_0^2}\frac{1}{\sqrt{( c^{(n)})^2+K}} \bar\beta  + O(\bar\beta^2)\\
%&= \frac{h_{i,0}}{H_0} + \frac{H_0}{R+\gamma H_0^2}u_i\beta_i - \frac{H_0}{R+\gamma H_0^2}\frac{c_i + 2\gamma h_{i,0}}{\sqrt{( c^{(n)})^2+K}} \bar\beta  + O(\bar\beta^2) \\
%&= \frac{h_{i,0}}{H_0} + \frac{H_0}{R+\gamma H_0^2}u_i\beta_i - \frac{H_0}{R+\gamma H_0^2}\frac{c_i + 2\gamma h_{i,0}}{c^{(n)}+2\gamma H_0} \bar\beta  + O(\bar\beta^2) \\
&= \frac{h_{i,0}^*}{H_0^*} + \frac{H_0^*}{R+\gamma (H_0^*)^2}\Big(I_i - \frac{c_{i,0} + 2\gamma h^*_{i,0}}{c_0^{(n)}+2\gamma H^*_0} \bar I\Big)  + O(\bar I^2) \\
&= \frac{h_{i,0}^*}{H_0^*} + \alpha_0\big((1-\alpha_i)I_i - \alpha_i \bar I_{-i}\big)  + O(I^2), %\\
%&\approx \frac{h_{i,0}}{H_0} + \frac{H_0}{R+\gamma H_0^2}\Big(\Big(1-\frac{h_{i,0}}{H_0}\Big)\eta_i - \frac{h_{i,0}}{H_0}\sum_{j\neq i}\eta_j\Big)  + O(\eta^2),
\end{align}
where
\begin{align}\label{alpha_eq}
\alpha_0 := \frac{H_0^*}{R+\gamma (H_0^*)^2}, \qquad 
\alpha_i := \frac{c_{i,0} + 2\gamma h_{i,0}^*}{c_0^{(n)}+2\gamma H_0^*}. %\in (0,1), \qquad \sum_{i=1}^n\alpha_i=1.
\end{align}
Let $u_i:=\tilde c_i-\tilde c_0$ and $u^{(n)}:=\sum_{i=1}^nu_i$. We then have 
\begin{align}\label{x1}
(1-\alpha_i)I_i - \alpha_i \bar I_{-i} = I_i - \alpha_i \bar I 
&= \half\frac{u_i}{\eta} - \half\frac{u^{(n)}}{\eta} \frac{c_{i,0} + 2\gamma h^*_{i,0}}{c_0^{(n)}+2\gamma H^*_0} \\
&= \frac{1}{2\eta}\Big(u_i - u^{(n)} \frac{c_{i,0} + 2\gamma h^*_{i,0}}{c_0^{(n)}+2\gamma H^*_0}\Big) \\
%&= \frac{1}{2\eta}\Big(c_{i,0} - \tilde c_0 - u^{(n)} \frac{c_{i,0} + 2\gamma h^*_{i,0}}{c_0^{(n)}+2\gamma H^*_0}\Big) \\
&= \frac{1}{2\eta}\Big(\tilde c_{i}\Big(1- \frac{u^{(n)}}{c_0^{(n)}+2\gamma H^*_0}\Big) - \tilde c_{0} - u^{(n)} \frac{ 2\gamma h^*_{i,0}}{c_0^{(n)}+2\gamma H^*_0}\Big).
\end{align}
The above expression is increasing in $i$, because $\tilde c_{i}$ is increasing in $i$, $h_{i,0}^*$ is decreasing in $i$, and 
\[
\frac{u^{(n)}}{c_0^{(n)}+2\gamma H^*_0} < \frac{c_0^{(n)}}{c_0^{(n)}+2\gamma H^*_0}<1.
\] 
\begin{comment}
Furthermore, 
\begin{align}
(1-\alpha_i)I_i - \alpha_i \bar I_{-i} > 0 \quad\Longleftrightarrow\quad 
I_i > \alpha \bar I  \quad\Longleftrightarrow\quad 
\frac{u_{i}}{c_{i,0} + 2\gamma h^*_{i,0}} > \frac{u^{(n)}}{c_0^{(n)}+2\gamma H^*_0},
\end{align}
and it follows from Proposition \ref{propMC} and the fact that $h^*_{i,0}$ is decreasing in $c_{i,0}$, that the left-hand side of the final inequality is increasing in $i$. 
\end{comment}
\begin{comment}
The approximated hash rate share of all miners sums to one, 
\begin{align}
\sum_{i=1}^n\Big(\frac{h_{i,0}^*}{H_0^*} + \frac{H_0^*}{R+\gamma (H_0^*)^2}\big((1-\alpha_i)I_i - \alpha_i\bar I_{-i}\big)\Big)
&= 1 + \frac{H_0^*}{R+\gamma (H_0^*)^2}\Big(\bar I - \sum_{i=1}^n\alpha_i(I_i+\bar I_{-i}) \Big) 
%&= 1 + \frac{H_0^*}{R+\gamma (H_0^*)^2}\Big(\bar I - \sum_{i=1}^n\alpha_i\bar I\Big) \Big) \\
= 1,
\end{align}
%\begin{align}
%\sum_{i=1}^n\frac{h_{i,0}^*}{H^*}  &= \sum_{i=1}^n\Big(\frac{h_{i,0}^*}{H_0^*} + \frac{H_0^*}{R+\gamma (H_0^*)^2}\Big((1-\alpha_i)I_i - \alpha_i\bar I_{-i}\Big) + O(\bar I^2) \Big) \\
%&= 1 + \frac{H_0^*}{R+\gamma (H_0^*)^2}\Big(\bar I - \sum_{i=1}^n\alpha_iI_i - \sum_{i=1}^n\alpha_i\bar I_{-i}\Big) + O(\bar I^2) \Big)\\
%&= 1 + \frac{H_0^*}{R+\gamma (H_0^*)^2}\Big(\bar I - \sum_{i=1}^n\alpha_i\bar I\Big) + O(\bar I^2) \Big) \\
%&= 1 + O(\bar I^2),
%\end{align}
where we used that $I_i+\bar I_{-i}=\bar I$, and $\sum_{i=1}^n\alpha_i=1$.
\end{comment}
To analyze the value of $h_i^*$, we introduce the notation
\begin{align}
\kappa_1 := \frac{H_0^*}{\sqrt{( c_0^{(n)})^2+K}} = \frac{H_0^*}{2\gamma H_0^* + c_0^{(n)}}, \qquad 
\kappa_2 := \frac{H_0^*}{R+\gamma (H_0^*)^2}, \qquad 
\xi_i := \frac{c_{i,0} + 2\gamma h^*_{i,0}}{c_0^{(n)}+2\gamma H_0^*},
\end{align}
so the expansions for $H^*$ and $h_i^*/H^*$ become
\begin{align}
H^*  %&=  H^*_0 + \frac{H_0^*}{\sqrt{( c_0^{(n)})^2+K}}\bar I + O(\bar I^2) 
&=  H_0^* + \kappa_1\bar I + O(\bar I^2), \qquad 
\frac{ h_i^*}{ H^*} %&= \frac{h_{i,0}^*}{H_0^*} + \frac{H_0^*}{R+\gamma (H_0^*)^2}I_i - \frac{H_0^*}{R+\gamma (H_0^*)^2}\frac{c_{i,0} + 2\gamma h_{i,0}^*}{c_0^{(n)}+2\gamma H_0^*} \bar I  + O(\bar I^2) 
= \frac{h_{i,0}^*}{H_0^*} + \kappa_2I_i - \kappa_2\xi_i \bar I  + O(\bar I^2).
\end{align} 
For the hash rate of miner $i$, we then have
\begin{align}\label{hi_exp}
h_i^* = \frac{h_{i}^*}{H^*}H^*
&= \frac{h_{i,0}^*}{H_0^*}(H_0^* + \kappa_1\bar I) + \kappa_2H_0^*I_i - \kappa_2H_0^*\xi_i \bar I + O(\bar I^2) \\
&= h_{i,0}^* + \frac{h_{i,0}^*}{2\gamma H_0^* + c_0^{(n)}}\bar I + \frac{(H_0^*)^2}{R+\gamma (H_0^*)^2}I_i - \frac{(H_0^*)^2}{R+\gamma (H_0^*)^2}\xi_i \bar I + O(\bar I^2) \\
%&= h_{i,0} + \frac{H_0^2}{R+\gamma H_0^2}u_i\beta_i + \frac{h_{i,0}}{2\gamma H_0 + c^{(n)}}\bar\beta  - \frac{H_0^2}{R+\gamma H_0^2}\xi_i \bar\beta + %O(\bar\beta^2) \\
%&= h_{i,0} + \frac{H_0^2}{R+\gamma H_0^2}u_i\beta_i + \Big(\frac{h_{i,0}}{2\gamma H_0 + c^{(n)}}  - \frac{H_0^2}{R+\gamma H_0^2}\frac{c_i + 2\gamma h_{i,0}}{c^{(n)}+2\gamma H_0}\Big) \bar\beta + O(\bar\beta^2) \\
&= h_{i,0}^* + \frac{(H_0^*)^2}{R+\gamma (H_0^*)^2} I_i + \frac{h_{i,0}^*}{c_0^{(n)} + 2\gamma H_0^* }\Big(1 - \frac{(H_0^*)^2}{R+\gamma (H_0^*)^2}\frac{c_{i,0} + 2\gamma h_{i,0}^*}{h_{i,0}^*}\Big) \bar I + O(\bar I^2)\\
&= h_{i,0}^* + a_{i}I_i + a_{-i}\bar I_{-i} + O(\bar I^2),
\end{align}
where 
\begin{align}\label{a_eq}
a_{i} &:= \frac{(H_0^*)^2}{R+\gamma (H_0^*)^2} + a_{-i} > 0, \\ 
a_{-i} &:=  \frac{h_{i,0}^*}{c_0^{(n)} + 2\gamma H_0^* }\Big(1 - \frac{(H_0^*)^2}{R+\gamma (H_0^*)^2}\frac{c_{i,0} + 2\gamma h_{i,0}^*}{h_{i,0}^*}\Big) < 0
\quad\Longleftrightarrow\quad \frac{h_{i,0}^*}{H_0^*} < \half.
\end{align}
The equivalence relation for $a_{-i}$ follows from 
\begin{align}
\frac{(H_0^*)^2}{R+\gamma (H_0^*)^2}\frac{c_{i,0} + 2\gamma h^*_{i,0}}{h^*_{i,0}} > 1
%&\quad\Longleftrightarrow\quad \frac{c_iH_0 + 2\gamma h_{i,0}H_0}{R+\gamma H_0^2} > \frac{h_{i,0}}{H_0} \\
%&\quad\Longleftrightarrow\quad \frac{R-c_iH_0 + 2c_iH_0 + 2\gamma h_{i,0}H_0-R}{R+\gamma H_0^2} > \frac{h_{i,0}}{H_0} \\
&\quad\Longleftrightarrow\quad \frac{R-c_{i,0}H_0^*}{R+\gamma (H_0^*)^2} + \frac{2H_0^*(c_{i,0} + \gamma h^*_{i,0})-R}{R+\gamma (H_0^*)^2} > \frac{h^*_{i,0}}{H^*_0} \\
&\quad\Longleftrightarrow\quad \frac{h^*_{i,0}}{H^*_0} + \frac{2H^*_0\frac{R}{H^*_0}\Big(1-\frac{h^*_{i,0}}{H^*_0}\Big)-R}{R+\gamma (H_0^*)^2} > \frac{h^*_{i,0}}{H^*_0} \\
%&\quad\Longleftrightarrow\quad  \frac{2\Big(1-\frac{h_{i,0}}{H_0}\Big)-1}{R+\gamma H_0^2} > 0 \\
&\quad\Longleftrightarrow\quad \frac{h^*_{i,0}}{H^*_0} <\frac{1}{2},
\end{align}
where we have used (\ref{1st}) in the second equivalence relation. 
\hfill\qed 

\paragraph{Proof of Proposition \ref{propDeltaPi}:} Using the expression (\ref{hi_exp}), we have
%the proof of Proposition \ref{propDeltaH} we have
%\begin{align}
% h_i %= \frac{ h_i}{ H} H %&= \Big(\frac{h_{i,0}}{H_0} + a\eta_i - a\xi_i \bar\eta\Big)(H_0 + \alpha\bar\eta) + O(\bar\eta^2) 
%&= \frac{h_{i,0}}{H_0}(H_0 + \alpha\bar\beta) + au_i\beta_i(H_0 + \alpha\bar\beta) - a\xi_i \bar\beta(H_0 + \alpha\bar\beta) + O(\bar\beta^2) \\
%= \frac{h_{i,0}}{H_0}(H_0 + \alpha\bar\eta) + aH_0\eta_i - aH_0\xi_i \bar\eta + O(\bar\eta^2),
%\end{align}
%so
\begin{align}
(h_i^*)^2 %&= \Big(\frac{h_{i,0}}{H_0}(H_0 + \alpha\bar\beta) + aH_0u_i\beta_i - aH_0\xi_i \bar\beta\Big)^2 + O(\bar\beta^2) \\
%&= \Big(\frac{h_{i,0}}{H_0}(H_0 + \alpha\bar\beta)\Big)^2 + 2\frac{h_{i,0}}{H_0}(H_0 + \alpha\bar\beta)\Big(aH_0u_i\beta_i - aH_0\xi_i \bar\beta\Big) + O(\bar\beta^2) \\
&= \Big(\frac{h^*_{i,0}}{H^*_0}\Big)^2((H_0^*)^2 + 2H_0^*\kappa_1\bar I) + 2h^*_{i,0}\kappa_2H^*_0(I_i - \xi_i \bar I) + O(\bar I^2).
\end{align}
Using the expressions for $h_i^*/H^*$, $h_i^*$, and $(h_i^*)^2$, the profit of miner $i$ becomes
\begin{align}
\pi_i^* := \pi_i^*(\beta^*) &= \frac{h_i^*}{ H^*} R - c_i h_i^* - \frac{\gamma}{2} (h_{i,0}^*)^2 \\
&= R\Big(\frac{h^*_{i,0}}{H_0^*} + \kappa_2 I_i - \kappa_2\xi_i \bar I\Big) - (c_{i,0}-I_i)\Big(\frac{h^*_{i,0}}{H^*_0}(H^*_0 + \kappa_1\bar I) + aH^*_0I_i - \kappa_2H_0^*\xi_i \bar I\Big) \\
&\quad  - \frac{\gamma}{2}\Big(\Big(\frac{h^*_{i,0}}{H^*_0}\Big)^2((H_0^*)^2 + 2H_0^*\kappa_1\bar I) + 2h^*_{i,0}\kappa_2H^*_0(I_i - \xi_i \bar I)\Big) + O(\bar I^2)\\
%&= R\frac{h_{i,0}}{H_0} - c_ih_{i,0} - \frac{\gamma_i}{2}h_{i,0}^2
%+ R(au_i\beta_i - a\xi_i \bar\beta)  
%- c_i\Big(\frac{h_{i,0}}{H_0}\alpha\bar\beta + aH_0u_i\beta_i - aH_0\xi_i \bar\beta\Big) 
%+ \beta_iu_ih_{i,0}\\
%&\quad  
%- \frac{\gamma_i}{2}\Big(\Big(\frac{h_{i,0}}{H_0}\Big)^2 2H_0\alpha\bar\beta + 2h_{i,0}aH_0(u_i\beta_i - \xi_i \bar\beta)\Big) \\
&= R\frac{h_{i,0}^*}{H_0^*} - c_{i,0}h_{i,0}^* - \frac{\gamma}{2}(h_{i,0}^*)^2
+ I_ih_{i,0}^* + R\kappa_2(I_i - \xi_i \bar I)\\
&\quad   - c_{i,0}\Big(\frac{h^*_{i,0}}{H^*_0}\kappa_1\bar I + \kappa_2H_0^*(I_i - \xi_i \bar I)\Big)   
- \gamma\Big(\Big(\frac{h_{i,0}^*}{H_0^*}\Big)^2 H_0^*\kappa_1\bar I + \kappa_2h_{i,0}^*H_0^*(I_i - \xi_i \bar I)\Big) + O(\bar I^2).
\end{align}
\begin{comment} 
We have, besides the $\beta_iu_ih_{i,0}$ term,
\begin{align}
&Ra(u_i\beta_i - \xi_i \bar\beta) - c_iaH_0(u_i\beta_i - \xi_i \bar\beta) - \gamma_iah_{i,0}H_0(u_i\beta_i - \xi_i \bar\beta)  \\
=\;& aH_0(u_i\beta_i - \xi_i \bar\beta)\Big(\frac{R}{H_0} - c_i - \gamma_ih_{i,0}\Big),
\end{align}
and the first term is negative for the ``big guys'', while the second term is positive for everyone. 
The leftover terms are always negative,
\begin{align}
- c_i\frac{h_{i,0}}{H_0}\alpha\bar\beta 
- \gamma_i\Big(\frac{h_{i,0}}{H_0}\Big)^2 H_0\alpha\bar\beta
= -\frac{h_{i,0}}{H_0}\alpha\bar\beta(c_i + \gamma_ih_{i,0}).
\end{align}
\end{comment}
Writing $\pi_{i,0}^*:=\pi_i^*(0)$, the change in profit due to investment is 
\begin{align}
\pi_i^* - \pi_{i,0}^*
&= I_ih_{i,0}^* +\kappa_2H_0^*(I_i - \xi_i \bar I)\Big(\frac{R}{H_0^*} - c_{i,0} - \gamma h_{i,0}^*\Big) -\frac{h_{i,0}^*}{H_0^*}\kappa_1\bar I(c_{i,0} + \gamma h_{i,0}^*) + O(\bar I^2) \\
&= I_i\Big(h_{i,0}^*+\kappa_2H_0^*\Big(\frac{R}{H_0^*} - c_{i,0} - \gamma h_{i,0}^*\Big)\Big) -\bar I\Big(\kappa_2H_0^*\xi_i\Big(\frac{R}{H_0^*} - c_{i,0} - \gamma h_{i,0}^*\Big)
+\frac{h_{i,0}^*}{H_0^*}\kappa_1(c_{i,0} + \gamma h_{i,0}^*)\Big) + O(\bar I^2) \\
&= I_i\Big(h_{i,0}^*+\kappa_2H_0^*\frac{h_{i,0}^*}{H_0^*}\frac{R}{H_0^*}\Big) -\bar I\Big(\frac{(H_0^*)^2}{R+\gamma (H_0^*)^2}\frac{c_{i,0}+2\gamma h_{i,0}^*}{c_0^{(n)}+2\gamma H_0^*}\frac{h_{i,0}^*}{H_0^*}\frac{R}{H_0^*} 
+ \frac{h_{i,0}^*}{c_0^{(n)}+2\gamma H_0^*}(c_{i,0} + \gamma h_{i,0}^*)\Big) + O(\bar I^2)\\
&= I_ih_{i,0}^*\Big(1+\frac{R}{R+\gamma (H_0^*)^2}\Big) -\bar I h^*_{i,0}\frac{c_{i,0}+2\gamma h^*_{i,0}}{c_0^{(n)}+2\gamma H^*_0}\Big(\frac{R}{R+\gamma (H_0^*)^2}
+ \frac{c_{i,0} + \gamma h^*_{i,0}}{c_{i,0} + 2\gamma h^*_{i,0}}\Big) + O(\bar I^2),
\end{align}
where we used the first-order condition (\ref{1st}) for miner $i$. 
\begin{comment}
Combining the $\beta_i$ terms and using the first-order condition gives
\begin{align}
\eta_i\Big(h_{i,0}+aH_0\Big(\frac{R}{H_0} - c_i - \gamma h_{i,0}\Big)\Big)
= \eta_i\Big(h_{i,0}+aH_0\frac{h_{i,0}}{H_0}\frac{R}{H_0}\Big) 
%&= \beta_iu_ih_{i,0}\Big(1+\frac{H_0}{R+\gamma H_0^2}\frac{R}{H_0}\Big) \\
&= \eta_ih_{i,0}\Big(1+\frac{R}{R+\gamma H_0^2}\Big).
\end{align}
Combining the $\bar\eta$ terms and again the first-order condition gives
\begin{align}
%&-aH_0\xi_i \bar\beta\Big(\frac{R}{H_0} - c_i - \gamma_ih_{i,0}\Big)-\frac{h_{i,0}}{H_0}\alpha\bar\beta(c_i + \gamma_ih_{i,0}) \\
& -\bar\eta\Big(aH_0\xi_i\Big(\frac{R}{H_0} - c_i - \gamma h_{i,0}\Big)
+\frac{h_{i,0}}{H_0}\alpha(c_i + \gamma h_{i,0})\Big) \\
%=\;& -\bar\beta\Big(H_0\frac{H_0}{R+\gamma H_0^2}\frac{c_i+2\gamma h_{i,0}}{c^{(n)}+2\gamma H_0}\Big(\frac{R}{H_0} - c_i - \gamma_ih_{i,0}\Big)
%+\frac{h_{i,0}}{H_0}\frac{H_0}{c^{(n)}+2\gamma H_0}(c_i + \gamma_ih_{i,0})\Big) \\
=\;& -\bar\eta\Big(\frac{H_0^2}{R+\gamma H_0^2}\frac{c_i+2\gamma h_{i,0}}{c^{(n)}+2\gamma H_0}\frac{h_{i,0}}{H_0}\frac{R}{H_0} 
+ \frac{h_{i,0}}{c^{(n)}+2\gamma H_0}(c_i + \gamma h_{i,0})\Big) \\
%=\;& -\bar\beta \frac{h_{i,0}}{c^{(n)}+2\gamma H_0}\Big(\frac{c_i+2\gamma h_{i,0}}{R+\gamma H_0^2}R+ c_i + \gamma h_{i,0}\Big) \\
=\;& -\bar\eta h_{i,0}\frac{c_i+2\gamma h_{i,0}}{c^{(n)}+2\gamma H_0}\Big(\frac{R}{R+\gamma H_0^2}
+ \frac{c_i + \gamma h_{i,0}}{c_i + 2\gamma h_{i,0}}\Big) \\
\approx \;& -\bar\eta \frac{h_{i,0}^2}{H_0}\Big(\frac{R}{R+\gamma H_0^2}
+ \half\Big).
\end{align}
\end{comment}
This can also be written as
\begin{align}
{\pi^*_i - \pi^*_{i,0}}
&= h^*_{i,0}(b_i I_i + b_{-i} \bar I_{-i}) + O(\bar I^2), 
\end{align}
where 
\begin{align}\label{b_eq}
b_i &:= \Big(1+\frac{R}{R+\gamma (H^*_0)^2}\Big) + b_{-i} > 0, \\
b_{-i} &:= -\frac{c_{i,0}+2\gamma h^*_{i,0}}{c_{0}^{(n)}+2\gamma H_0^*}\Big(\frac{R}{R+\gamma (H_0^*)^2}
+ \frac{c_{i,0} + \gamma h^*_{i,0}}{c_{i,0} + 2\gamma h^*_{i,0}}\Big) < 0.
\end{align}
Using the notation $u_i:=\tilde c_i-\tilde c_0$ and $u^{(n)}:=\sum_{i=1}^nu_i$, the profit of miner $i$ can be written as 
\begin{align}\label{pi_decomp}
\frac{\pi_i^* - \pi^*_{i,0}}{h^*_{i,0}}
&= \Big(1+\frac{R}{R+\gamma (H_0^*)^2}\Big)I_i - \frac{c_{i,0}+2\gamma h^*_{i,0}}{c_0^{(n)}+2\gamma H_0^*}\Big(\frac{R}{R+\gamma (H_0^*)^2}
+ \frac{c_{i,0} + \gamma h^*_{i,0}}{c_{i,0} + 2\gamma h^*_{i,0}}\Big)\bar I + O(\bar I^2) \\
%&= \Big(1 - \frac{c_i+\gamma h_i}{c^{(n)}+2\gamma H_0} + \frac{R}{R+\gamma H_0^2}\Big(1-\frac{c_i+2\gamma h_i}{c^{(n)}+2\gamma H_0}\Big)\Big)\tilde\eta_i \\
%&\quad - \frac{c_i+2\gamma h_{i,0}}{c^{(n)}+2\gamma H_0}\Big(\frac{R}{R+\gamma H_0^2}
%+ \frac{c_i + \gamma h_{i,0}}{c_i + 2\gamma h_{i,0}}\Big)(\bar\eta-\tilde\eta_i) + O(\bar\eta^2)\\
%&= \Big(1 - \frac{c_{i,0}+\gamma h^*_{i,0}}{c_0^{(n)}+2\gamma H_0^*}\Big)I_i 
%- \frac{c_{i,0} + \gamma h^*_{i,0}}{c_0^{(n)}+2\gamma H_0^*}(\bar I-I_i) \\
%&\quad + \frac{R}{R+\gamma (H_0^*)^2}\Big(\Big(1-\frac{c_{i,0}+2\gamma h_{i,0}^*}{c_0^{(n)}+2\gamma H_0^*}\Big)I_i
%- \frac{c_{i,0}+2\gamma h^*_{i,0}}{c_0^{(n)}+2\gamma H_0^*}(\bar I-I_i)\Big)
%+ O(\bar\eta^2) \\
&= \frac{1}{2\eta}\Big(u_i - \frac{c_{i,0}+\gamma h^*_{i,0}}{c_0^{(n)}+2\gamma H_0^*}u^{(n)}   + \frac{R}{R+\gamma (H_0^*)^2}\Big(u_i-\frac{c_{i,0}+2\gamma h_{i,0}^*}{c_0^{(n)}+2\gamma H_0^*}u^{(n)}\Big)\Big)  + O(\bar I^2) \\
&=: \frac{1}{2\eta}\Big(x_{i,1}   + \frac{R}{R+\gamma (H_0^*)^2}x_{2,i}\Big)  + O(\bar I^2).
\end{align}
%If $\eta_i=\eta u_i$, we have $I_i=u_i/(2\eta)$ and $\bar I=u^{(n)}/(2\eta)$, and 
In the  proof of Proposition \ref{propDeltah} (see (\ref{x1})), we have shown that $x_{2,i}$ is increasing in $i$. Equivalently, $x_{i,2}$ is increasing in $c_{i,0}$. We also have
\begin{align}
x_{2,i} > 0
%\Big(1-\frac{c_{i,0}+2\gamma h_i^*}{c_0^{(n)}+2\gamma H_0^*}\Big)I_i
%- \frac{c_{i,0}+2\gamma h^*_{i,0}}{c_0^{(n)}+2\gamma H_0^*}(\bar I-I_i) > 0 
%&\quad\Longleftrightarrow\quad 
%\Big(1 - \frac{c_{i,0}+2\gamma h_{i,0}^*}{c_0^{(n)}+2\gamma H_0^*}\Big)u_{i} >\frac{c_i^* + 2\gamma h^*_{i,0}}{c_0^{(n)}+2\gamma H_0^*}u^{(n)}_{-i} \\
%&\quad\Longleftrightarrow\quad 
%(c_0^{(n)}+2\gamma H_0^*)u_{i} - (c_{i,0}+2\gamma h_i^*)u_{i} >(c_{i,0} + 2\gamma h^*_{i,0})u^{(n)}_{-i} \\
%&\quad\Longleftrightarrow\quad 
%(c_0^{(n)}+2\gamma H^*_0)u_{i}  >(c_{i,0} + 2\gamma h^*_{i,0})u_0^{(n)} \\
&\quad\Longleftrightarrow\quad 
\frac{u_{i}}{c_{i,0} + 2\gamma h^*_{i,0}}  >\frac{u_0^{(n)}}{c_0^{(n)}+2\gamma H_0^*}.
\end{align}
The right-hand side is independent of $i$, but the left-hand side of the inequality is increasing in $i$, because $u_i=c_{i,0}-c_0$, $c_{i,0}$ is increasing in $i$, and $h_{i,0}^*$ is decreasing in $i$. Furthermore, 
\begin{align} 
\frac{u_{i}}{c_{i,0} + 2\gamma h^*_{i,0}}  >\frac{u_0^{(n)}}{c_0^{(n)}+2\gamma H_0^*}
\quad\Longleftrightarrow\quad 
{u_{i}(c_0^{(n)}+2\gamma H_0^*)}  > {u_0^{(n)}}(c_{i,0} + 2\gamma h^*_{i,0}),
\end{align} 
and summing the left- and right-hands sides of the inequality for $i=1,\dots,n$ yields the same result. From the above, it follows that $x_{2,i}>0$ only holds for large enough values of $i$. Similar steps can be used to show that $x_{1,i}$ is increasing in $i$, and can be negative for small enough values of $i$. 
\begin{comment}
\begin{align}
\Big(1 - \frac{c_{i,0}+\gamma h^*_i}{c_0^{(n)}+2\gamma H_0^*}\Big)I_i 
- \frac{c_{i,0} + \gamma h^*_{i,0}}{c_0^{(n)}+2\gamma H_0^*}(\bar I-I_i) > 0
&\quad\Longleftrightarrow\quad 
\Big(1 - \frac{c_{i,0}+\gamma h_i^*}{c_0^{(n)}+2\gamma H_0^*}\Big)u_{i} >\frac{c_{i,0} + \gamma h^*_{i,0}}{c_0^{(n)}+2\gamma H_0^*}u^{(n)}_{-i}  \\
&\quad\Longleftrightarrow\quad 
(c_0^{(n)}+2\gamma H_0^*)u_i - (c_{i,0}+\gamma h_{i,0}^*)u_{i} >(c_{i,0} + \gamma h^*_{i,0})u^{(n)}_{0,-i} \\
&\quad\Longleftrightarrow\quad 
(c_0^{(n)}+2\gamma H_0^*)u_i  > (c_{i,0} + \gamma h^*_{i,0})u^{(n)} \\
&\quad\Longleftrightarrow\quad 
\frac{u_{i}}{c_{i,0} + \gamma h^*_{i,0}}  >\frac{u^{(n)}}{c_0^{(n)}+2\gamma H_0^*}.
\end{align}
The left-hand side of the final inequality is increasing in $i$, i.e., it is larger for larger values of $c_{i,0}$. This follows from the fact that $u_i=c_{i,0}-\tilde c_0$ and that $h_{i,0}^*$ is decreasing in $c_{i,0}$.
%Can we know that the final inequality does not hold for all $i$? I guess we can at least see it for $c_{i,0}\to 0$.
\end{comment}
\hfill\qed 

\paragraph{Proof of Proposition \ref{cor1}:}
We use that $I_i$ is independent of $i$, $c_0^{(n)}=nc_{i,0}$, and $H_0^*=nh_{i,0}^*$, to write
\begin{align}\label{homogPi}
\pi_i^*-\pi_{i,0}^*
&= {h_{i,0}^*} I_i \Big(1+\frac{R}{R+\gamma (H_0^*)^2}\Big) -{h_{i,0}^*} I_i \Big(\frac{R}{R+\gamma (H_0^*)^2}
+ \frac{c_{0}^{(n)} + \gamma H^*_{0}}{c_{0}^{(n)} + 2\gamma H^*_{0}}\Big) + O(\bar I^2) \\
%&= {h_{i,0}^*} I_i \Big(1 - \frac{c^{(n)}_0 + \gamma H^*_{0}}{c^{(n)}_{0} + 2\gamma H^*_{0}}\Big) + O(\bar I^2) \\
&= {h_{i,0}^*}\frac{1}{n}\Big(1 - \frac{c^{(n)}_0 + \gamma H^*_{0}}{c^{(n)}_{0} + 2\gamma H^*_{0}}\Big)\bar I + O(\bar I^2) \\
&=: {h^*_{i,0}}b\bar I + O(\bar I^2).
\end{align}
We have that $b$ is decreasing in $n$ because $1/n$ is decreasing in $n$, and 
\begin{align}\label{b}
\frac{c^{(n)}_0 + \gamma H^*_{0}}{c^{(n)}_{0} + 2\gamma H^*_{0}}
&= \frac{nc + \half(\sqrt{(nc)^2+4(n-1)R\gamma}-nc)}{nc + \sqrt{(nc)^2+4(n-1)R\gamma}-nc}
%&= \half\frac{nc + \sqrt{(nc)^2+4(n-1)R\gamma}}{\sqrt{(nc)^2+4(n-1)R\gamma}} \\
%&= \half\frac{1 + \sqrt{1+4\frac{n-1}{n^2}\frac{R\gamma}{c^2}}}{\sqrt{1+4\frac{n-1}{n^2}\frac{R\gamma}{c^2}}} \\
= \half\Big(1+\frac{1}{\sqrt{1+4\frac{n-1}{n^2}\frac{R\gamma}{c^2}}}\Big),
\end{align}
is decreasing in $n$. It also follows that $b$ is increasing in $\gamma$, and that $b\to 0$ as $n\to\infty$ or $\gamma\to 0$. \hfill\qed

\section{Additional Proofs}\label{appOtherProofs} 

\paragraph{Proof related to Equation (\ref{1st}):} %The first-order condition (\ref{1st}) is trivially obtained from the objective function (\ref{pi_i}). 
From the first-order condition (\ref{1st}) it can be seen directly that the equilibrium marginal gain of an active miner is smaller than the equilibrium reward-per-hash. Intuitively, this is because each additional unit of hash contributes to the aggregate hash rate, and thus reduces the reward-per-hash. The underlying mathematical reason is that, given a hash rate profile $h$, 
\begin{align}
\frac{\partial}{\partial h_i}\frac{h_i}{H} %&= \frac{\partial}{\partial h_i}\frac{h_i}{h_i+h_{-i}} 
= \frac{H-h_i}{H^2} > 0, \qquad 
\frac{\partial^2}{\partial h_i^2}\frac{h_i}{H} &= -2\frac{H-h_i}{H^3} < 0. 
\end{align}
That is, %all else being equal, 
the probability $h_i/H$ of a miner earning the reward is an increasing but concave function of the miner's hash rate. In other words, the marginal probability of earning the reward is decreasing. \hfill\qed

\paragraph{Proof related to Section \ref{secDeltaC}:} We show that the effect of an increase in the cost $c_j$ on the hash rate $h_i^*$ depends on the value of $h_i^*/H^*$. Define the marginal gain and marginal cost functions 
\begin{align}
MG_i(h_i,H):=\frac{R}{H}\Big(1-\frac{h_i}{H}\Big), \qquad MC_i(h_i,H):=c_i+\gamma h_i. 
\end{align}
It is then easy to see that
\begin{align}\label{monotone}
\text{(i)}\;\;\frac{\partial MG_i}{\partial h_i} < 0, \quad \frac{\partial MC_i}{\partial h_i} > 0, \qquad
\text{(ii)}\;\;\frac{\partial MG_i}{\partial H} > 0 \;\;\Longleftrightarrow\;\; \frac{h_i}{H} > \half.
\end{align}
In equilibrium, marginal gain equals marginal cost for each active miner. Denote the ``initial'' equilibrium hash rate of miner $i$ by $h_{i,0}^*$ and the ``initial'' aggregate hash rate by $H^*_0$. From equation (\ref{derivsH}) it follows that an increase in the cost $c_j$ results in aggregate hash rate $H^*$ such that $H^*<H^*_0$. From (\ref{monotone})-(ii), $MG_i(h_{i,0}^*,H_0^*)=MC_i(h_{i,0}^*,H_0^*)$, and $MC_i(h_{i,0}^*,H^*)=MC_i(h_{i,0}^*,H_0^*)$, it then follows that
\[
MG_i(h_{i,0}^*,H^*) < MC_i(h_{i,0}^*,H^*) \quad\Longleftrightarrow\quad \frac{h_{i,0}^*}{H_0^*}>1/2.
\] 
Using (\ref{monotone})-(i), we then see that if $h_{i,0}^*/H_0^*>1/2$, the equilibrium value $h_i^*$ needs to be smaller than the initial value $h_{i,0}^*$. The opposite happens if $h_{i,0}^*/H_0^*<1/2$, in which case $h_i^*$ needs to be greater than $h_{i,0}^*$. Finally, if $h_{i,0}^*/H_0^*=1/2$, then $h_i^*$ equals $h_{i,0}^*$. \hfill\qed

\bibliographystyle{plainnat}

\end{document}